\documentclass{aa}
\usepackage{natbib}
\usepackage{pdflscape} % or {lscape}
\usepackage{wasysym}
\usepackage[varg]{txfonts}
\usepackage{graphicx}
\usepackage{hyperref}
\bibpunct{(}{)}{;}{a}{}{,} % to follow the A&A style
\usepackage{longtable}
\usepackage{amssymb}

\usepackage{afterpage}

\def\ms{\hbox{\,m\,s$^{-1}$}}         %m.s -1
\def\m2s2{\hbox{\,m$^{2}$\,s$^{-2}$}} %m2.s -2
\def\logrhk{$\log$(R$^{\prime}_{HK}$)}

\begin{document}

\title{The HARPS search for southern extra-solar planets. }
\subtitle{
XLII. Eight HARPS multi-planet systems hosting 20 super-Earth and Neptune-mass companions   
%      \object{HD\,20003},  \object{HD\,20781}, \object{HD\,21693}, \object{HD\,31527}, 
%      \object{HD\,45184}, \object{HD\,51608}, \object{HD\,134060},  and \object{HD\,136352}.
\thanks{Based on observations made with the {\footnotesize HARPS} instrument on the ESO 3.6 m telescope at La Silla Observatory under the GTO program 072.C-0488 and Large program 193.C-0972/193.C-1005/.}
\thanks{The analysis of the radial-velocity measurements were performed using the Data and Analysis Center for Exoplanets (DACE) developed in the frame of the Swiss NCCR {\it PlanetS} and available for the community at the following address: \url{https://dace.unige.ch/}}
\thanks{The {\footnotesize HARPS} RV measurements discussed in this paper are available in electronic form at the CDS via anonymous ftp to cdsarc.u-strasbg.fr (130.79.128.5) or via http://cdsweb.u-strasbg.fr/cgi-bin/qcat?J/A+A/.}
}
\author{S.~Udry\inst{1}
        \and X.~Dumusque\inst{1}
        \and C.~Lovis\inst{1}
        \and D.~S\'egransan\inst{1}
        \and R.F.~Diaz\inst{1}
        \and W.~Benz\inst{2} 
        \and F.~Bouchy\inst{1,3}  
        \and A.~Coffinet\inst{1}
        \and G.~Lo Curto\inst{4}   
        \and M.~Mayor\inst{1} 
        \and C.~Mordasini\inst{2}
        \and F.~Motalebi\inst{1}
        \and F.~Pepe\inst{1} 
        \and D.~Queloz\inst{1} 
        \and N.C.~Santos\inst{5,6}  
        \and A.~Wyttenbach\inst{1}  
        \and R.~Alonso\inst{7}
        \and A.~Collier Cameron\inst{8}
        \and M.~Deleuil\inst{3}
        \and P.~Figueira\inst{5}
        \and M.~Gillon\inst{9}
        \and C.~Moutou\inst{3,10}
        \and D.~Pollacco\inst{11}
        \and E.~Pompei\inst{4}
}
\offprints{St\'ephane Udry, \email{Stephane.Udry@unige.ch}}
\institute{Observatoire astronomique de l'Universit\'e de Gen\`eve, 
               51 ch. des Maillettes, CH-1290 Versoix, Switzerland
          \and
          Physikalisches Institut, Universitat Bern, Silderstrasse 5, CH-3012 Bern, Switzerland
          \and
          Aix Marseille Universit\'e, CNRS, LAM (Laboratoire d'Astrophysique de Marseille) UMR 7326, 13388, Marseille, France
          \and
          European Southern Observatory, Karl-Schwarzschild-Str. 2, D-85748 Garching bei M\"unchen, Germany            
          \and 
          Instituto de Astrof\'isica e Ci\^encias do Espa\c{c}o, Universidade do Porto, CAUP, Rua das Estrelas, 4150-762 Porto, Portugal
          \and
          Departamento de F\'isica e Astronomia, Faculdade de Ci\`encias, Universidade do Porto, Rua do Campo Alegre, 4169-007 Porto, Portugal 
          \and
          Instituto de Astrofisica de Canarias, 38025, La Laguna, Tenerife, Spain
          \and
          School of Physics and Astronomy, University of St Andrews, North Haugh, St Andrews, Fife KY16 9SS
          \and
          Institut d'Astrophysique et de G\'eophysique, Universit\'e de Li\`ege, All\'ee du 6 Ao\^ut 17, Bat. B5C, 4000, Li\`ege, Belgium
          \and
          Canada France Hawaii Telescope Corporation, Kamuela, 96743, USA
          \and
          Department of Physics, University of Warwick, Coventry, CV4 7AL, UK
}

\date{Received XXX; accepted XXX}

%"Aims", "Methods", and "Results", are mandatory. When appropriate, the structured abstract may use an introductory paragraph entitled "Context", and a final paragraph entitled "Conclusions".
\abstract
{We present radial-velocity measurements of eight stars observed with the {\footnotesize HARPS} Echelle spectrograph mounted on the 3.6-m telescope in La Silla (ESO, Chile). Data span more than ten years and highlight the long-term stability of the instrument. }
{We search for potential planets orbiting \object{HD\,20003}, \object{HD\,20781}, \object{HD\,21693}, \object{HD\,31527}, \object{HD\,45184}, \object{HD\,51608}, \object{HD\,134060} and  \object{HD\,136352} to increase the number of known planetary systems and thus better constrain exoplanet statistics.}
{After a preliminary phase looking for signals using generalized Lomb-Scargle periodograms, we perform a careful analysis of all signals to separate \emph{bona-fide} planets from signals induced by stellar activity and instrumental systematics. We finally secure the detection of all planets using the efficient MCMC available on the Data and Analysis Center for Exoplanets (DACE web-platform), using model comparison whenever necessary.}
{ In total, we report the detection of twenty new super-Earth to Neptune-mass planets, with minimum masses ranging from  2 to 30 M$_{\rm Earth}$ and periods ranging from 
3 to 1300 days, in multiple systems with two to four planets. Adding {\footnotesize CORALIE} and {\footnotesize HARPS} measurements of \object{HD20782} to the already published data, we also improve the characterization of the extremely eccentric Jupiter orbiting this visual companion of \object{HD\,20781}.}
{}
   

\authorrunning{Udry et al.}
\titlerunning{20 super-Earths and hot Neptunes detected with HARPS}

\maketitle
 \section{Introduction}

The radial velocity (RV) planet search programs with the {\footnotesize HARPS} spectrograph on the ESO 3.6-m telescope \citep{Pepe-2000,Mayor-2003:a} have contributed in a tremendous way to our knowledge of the population of small-mass planets around solar-type stars. The {\footnotesize HARPS} planet-search  program on Guaranteed Time Observations (GTO, PI: M. Mayor) was on-going for 6 years between autumn 2003 and spring 2009. The high-precision part of this {\footnotesize HARPS} GTO survey aimed at the detection of very low-mass planets in a sample of quiet solar-type stars already screened for giant planets at a lower precision with the {\footnotesize CORALIE} Echelle spectrograph mounted on the 1.2-m Swiss telescope on the same site  \citep{Udry-2000:a}.  The GTO was then continued within the ESO Large Programs 183.C-0972, 183.C-1005 and 192.C-0852  (PI: S. Udry), from 2009 to 2016.

Within these programs, {\footnotesize HARPS} has allowed for the detection (or has contributed to the detection) of more than 100 extra-solar planet candidates \citep[see detections in][]{Diaz-2016,Moutou-2015,LoCurto-2013,Dumusque-2011:c,Mayor-2011,Pepe-2011,Moutou-2011,Lovis-2011}. In particular, {\footnotesize HARPS} has unveiled the existence of a large population of low-mass planets including super-Earths and hot Neptunes previous to the launch of the \emph{Kepler} satellite which provided us with an overwhelming sample of thousands of small-size transiting candidates \citep{Coughlin:2016aa,Mullally-2015,Borucki-2011,Batalha-2011}. A preliminary analysis of the {\footnotesize HARPS} data showed at that time that at least 30\,\% of solar-type stars were hosting low-mass planets on short-period orbits with less than 50 days \citep[]{Lovis-2009}. A comprehensive analysis of our high-precision sample combined with 18 years of data from {\footnotesize CORALIE} allowed us later to precise this occurrence rate: about 50\% of the stars surveyed have planets with masses below 50\,M$_{\oplus}$ on short to moderate period orbits \citep{Mayor-2011}. Furthermore, a large fraction of those planets are in multi-planetary systems. This preliminary statistics of hot super-Earth and Neptune frequency is now beautifully confirmed by the impressive results of the \emph{Kepler} mission.

\begin{table*}
\caption{
Observed and inferred stellar parameters for the stars hosting planetary systems described in this paper.
% \object{HD\,20003}, \object{HD\,20781}, \object{HD\,21693}, \object{HD\,31527}, \object{HD\,45184}, \object{HD\,51608}, \object{HD\,134060}, and \object{HD\,136352}. 
}
\label{table:star}
\begin{center}
\begin{small}
\begin{tabular}{lcccccccc}
\hline \hline 
Parameters  &\object{HD\,20003}   &\object{HD\,20781}$^a$   &\object{HD\,21693}    &\object{HD\,31527}  
                    & \object{HD\,45184}  &\object{HD\,51608}   &\object{HD\,134060}  &\object{HD\,136352} \\
\hline                                                       
Sp. Type$^{(1)}$   &G8V    &K0V   &G8V   &G2V   &G1.5V  &G7V    &G3IV   &G4V   \\[1mm]
V$^{(1)}$               &8.39    &8.48   &7.95  &7.49   &6.37     &8.17    &6.29    &5.65  \\[1mm]
$B-V$$^{(1)}$       &0.77    &0.82    &0.76 &0.61   &0.62     &0.77    &0.62    &0.63  \\[1mm]
$\pi$ [mas]$^{(1)}$  &22.83$\pm$0.65    &28.27$\pm$1.08    &30.88$\pm$0.49    &25.93$\pm$0.60    
                                 &45.70$\pm$0.40    &28.71$\pm$0.51    &41.32$\pm$0.45    &67.51$\pm$0.39  \\[1mm] 
\hline
 $M_{V}$$^{(1)}$      &5.22    &5.70   &5.40   &4.87   &4.65   &5.51  &4.38    &4.83    \\[1mm]
 $T_{\rm eff}$ [K]$^{(2)}$      &5494$\pm$27    &5256$\pm$29   &5430$\pm$26   &5898 $\pm$13                  
                                              &5869$\pm$14    &5358$\pm$22   &5966$\pm$14   &5664$\pm$14     \\[1mm]
$[Fe/H]^{(2)}$    &$0.04\pm0.02$    &$-0.11\pm0.02$     &$0.0\pm0.02$     &$-0.17\pm0.01$                
                          &$0.04\pm0.01$    &$-0.07\pm0.01$    &$0.14\pm0.01$   &$-0.34\pm0.01$   \\[1mm]
$\log{(g)}^{(2)}$  &4.41$\pm$0.05    &4.37$\pm$0.05    &4.37 $\pm$0.04    &4.45$\pm0.02$
                          &4.47$\pm$0.02     &4.36$\pm$0.05    &4.43 $\pm$0.03    &4.39$\pm0.02$    \\[1mm]
$M_{\star}$  [M$_{\odot}$]$^{(2)}$    &0.875    &0.70   &0.80   &0.96    &1.03    &0.80   &1.095  &0.81    \\[1mm]
$L_{\star}$   [L$_{\odot}$]$^{(2)}$     &0.72$\pm$0.03   &0.49$\pm$0.04   &0.62$\pm$0.02   &1.20$\pm$0.03  
                                                          &1.13$\pm$0.01   &0.57$\pm$0.02   &1.44$\pm$0.02   &0.99$\pm$0.01   \\[1mm]
%Age$^{(5)}$ [Gyr]     &     &     &     &      &     &     &     &   \\[1mm]
\logrhk$^{(3)}$  &$-4.97\pm$0.05     &$-5.03\pm$0.01    &$-4.91\pm$0.05     &$-4.96\pm$0.01 
                                                            &$-4.91\pm$0.01     &$-4.98\pm$0.02     &$-5.00\pm$0.01    &$-4.95\pm$0.01  \\[1mm]
$v\sin{(i)}$ [km\,s$^{-1}$]$^{(3)}$    &1.9    &1.1   &1.6   &1.9   &2.1   &1.3   &2.6     &$<1$    \\
%$P_{rot}$  [days]$^{(4)}$      &$\sim$33   &$\sim$43   &$\sim$36   &$\sim$19  
%                                             &$\sim$19   &$\sim$36   %&$\sim$21   &$\sim$22 \\[1mm]
$P_{rot}$  [days]$^{(4)}$      &38.9$\pm$4.0   &$46.8\pm$4.4   &35.2$\pm$4.0  
                                             &20.3$\pm$2.9   &21.5$\pm$3.0   &40.0$\pm$4.0   &23.0$\pm$2.9    &23.8$\pm$3.1 \\[1mm]
\hline
\end{tabular}
\end{small}
\end{center}
\tablefoot{
${(1)}$ Astrometric and visual photometric data from the Hipparcos Catalogs  \citep{ESA-1997,vanLeeuwen-2007}.
${(2)}$ From  \citet{Sousa-2008} spectroscopic analysis.
${(3)}$ Parameter derived using { \footnotesize HARPS} spectra or CCF.
${(4)}$ From the calibration of \citet{Mamajek-2008}. \\
%${(5)}$ Parameter derived from  Geneva stellar evolution models.
$^a$ HD\,20782, the stellar visual companion of HD\,20781, hosts a planet as well. The corresponding stellar parameters are given in \citet{Jones-2006}.
}
\end{table*}

With the RV technique, the variation of the velocity of the central star due to the perturbing effect of small-mass planets becomes very small, of the order or even smaller than the uncertainties of individual measurements. The problems to solve and characterize the individual systems are then multi-fold, requiring to disentangle the planetary from the stellar, instrumental, and statistical noise effects. Efficient statistical techniques, mainly based on a Bayesian approach, have been developed to optimise the process and thus the outcome of ongoing RV surveys \citep[see e.g.,][and references therein]{Dumusque-2016b,Diaz:2016ab}. A large number of observations is however paramount for a complete probe of the planetary content of the system, and to take full advantage of the developed technique of analysis. In this context, focusing on the most observed, closest and brightest stars, an on-going {\footnotesize HARPS} LP (198.C-0836) is continuing the original observing efforts and providing us with an unprecedented sample of well observed stars. In this paper we describe 8 planetary systems hosting 20 planets. The detection of these planets has been announced in \citet{Mayor-2011}\footnote{ This paper presents the characterization of half of the planets announced in \citet{Mayor-2011}. The latter paper was then waiting for the detailed analysis to be published before being resubmitted in parallel to the present paper.} studying statistical properties of the systems discovered with {\footnotesize HARPS}. Most of them are super-Earths and Neptunian planets on relatively short periods, and member of a multi-planet system. We present here the orbital solutions for one 4-planet system around {\small HD}\,20781, three 3-planet systems around {\small HD}\,20003, {\small HD}\,31527 and {\small HD}\,136352, and four 2-planet systems around {\small HD}\,21693, {\small HD}\,45184, {\small HD}\,51608, and {\small HD}\,134060. In addition, we give updated parameters for the very eccentric planet orbiting {\small HD}\,20782 \citep[][]{Jones-2006}, derived by combining {\footnotesize HARPS}, {\footnotesize CORALIE} and the published {\footnotesize UCLES} data. The paper is organised as follows. In Sec.\,\ref{sec:star} we discuss the primary star properties. Radial-velocity measurements and  orbital solutions of each system are presented in Sects.\,\ref{sec:meas} and \ref{sec:systems}, while Sec.\,\ref{sec:analysis} describes the framework used for analysing the data. We provide concluding remarks in Sec.\,\ref{sec:concl}.

\section{Stellar characteristics}
\label{sec:star}

This section provides basic information about the stars hosting the planets presented in this paper. Effective temperature, gravity and metallicity are derived from the spectroscopic analysis of {\footnotesize HARPS} spectra and provided in \citet{Sousa-2008}. We used the improved  Hipparcos astrometric  parallaxes re-derived by \citet{vanLeeuwen-2007} to determine the absolute V-band magnitude using the apparent visual magnitude from Hipparcos \citep{ESA-1997}. Metallicities, together with the effective temperatures and $M_{V}$ are then used to estimate basic stellar parameters (ages, masses) using theoretical isochrones from the grid of Geneva stellar evolution models, including a Bayesian estimation method \citep{Mowlavi-2012}.

Individual spectra were also used to derive measurements of the chromospheric activity S-index, \logrhk\,and H$\alpha$, following a similar approach as used by \citet{Santos-2000:b} and \citet{Gomes-da-Silva-2011}. Using \citet{Mamajek-2008}, we estimate rotational periods for our stars from the empirical correlation of stellar rotation and chromospheric activity index \citep[][]{Noyes-1984}.  We also derived  the $v\sin{(i)}$ from a calibration of the Full Width at Half Maximum (FWHM) of the {\footnotesize HARPS} Cross-Correlation Function (CCF) following a standard approach \citep[e.g.][]{Santos-2002}. All those extra indicators are used to disentangle small-amplitude planetary signals from stellar "noise" \citep[see e.g. the cases of CoRoT-7 and $\alpha$\,Centauri\,B;][]{Queloz-2009,Dumusque-2012}. The derived stellar parameters are summarized in Table\,\ref{table:star}.

\section{HARPS RV measurements}
\label{sec:meas}  

Radial velocities presented here have been obtained with the {\footnotesize HARPS} high-resolution spectrograph installed on the 3.6m ESO telescope at La Silla Observatory \citep{Mayor-2003:a}. The long-term \ms RV precision is ensured by nightly ThAr calibrations \citep{Lovis-2008}. On the short timescale of a night, the high precision is obtained using simultaneous ThAr (from 2003 to 2013) or Fabry-Perot \'etalon (since 2013) reference calibrations. The data reduction was performed with the latest version of the HARPS pipeline. In addition to the barycentric radial velocities with internal error bars, the reduction provides the Bisector Inverse Slope (BIS) of the {\footnotesize HARPS} CCF \citep[][]{Pepe-2002:a,Baranne-1996}, as defined by \citet{Queloz-2000:b}, the FWHM and the Contrast of the CCF, as well as the Ca\,II activity indexes S and \logrhk.  

By construction of the {\footnotesize HARPS} high-precision sample from low-activity stars in the  {\footnotesize CORALIE} volume-limited planet-search sample, the eight stars discussed here present low activity levels. The average values of the \logrhk\,activity index, estimated from the chromospheric re-emission in the Ca II H and K lines at $\lambda=3933.66~\AA$ and $3968.47~\AA$, are low, ranging from $-$5.01 to $-$4.87. Activity-induced RV jitter on stellar rotation time scales, due to spots and faculae on the stellar surface, is thus expected to remain at a low level. The potential influence of stellar activity on RVs is nevertheless scrutinised closely when long-term variation of activity indexes are observed (magnetic cycle) or when the planetary and stellar rotation periods (or harmonics) are of similar values.

\begin{table}
\scriptsize
\caption{
\label{table:obsstat}
General statistics of the {\footnotesize HARPS} observations of the planet-host stars presented in this paper, with the number of individual spectra observed, $<$S/N$>$ the average signal-to-noise ratio at 550\,nm of those spectra, the number of measurements obtained after binning the data over 1 hour, $\Delta T$ the time span of the observations, $\sigma_{RV}$ the rms value of the RV set and $< \varepsilon_{RV} >$ the mean RV photon-noise + calibration error.
}
\begin{center}
\begin{tabular}{lcccccc}
\hline\hline
  &$N_{spectr}$  & $<$ SNR $>$  &$N_{meas}$    & $\Delta T$     &$\sigma_{RV}$     &$< \varepsilon_{RV} >$  \\
  &  & 550\,nm  & 1-hour bin  & [days]   &  [m/s] &[m/s] \\[1mm]
\hline   \\[-1mm]    
HD\,20003        &184 & 110 &184   &4063  &5.35   &0.74  \\
HD\,20781        &225 & 112 &216   &4093  &3.41   &0.76  \\
HD\,20782$^a$    &71  & 181 &68    &4111  &36.86   &0.56 \\
HD\,21693        &210 & 141 &210   &4106  &4.72   &0.60  \\
HD\,31527        &256 & 180 &245   &4135  &3.19   &0.64  \\
HD\,45184        &308 & 221 &178   &4160  &4.72   &0.41   \\
HD\,51608        &218 & 133 &216   &4158  &4.07   &0.62  \\
HD\,134060       &335 & 199 &155   &4083  &3.68   &0.40 \\
HD\,136352       &649 & 231 &240   &3993  &2.74   &0.33 \\[1mm]                                                       
\hline       
\end{tabular}
\end{center}
\tablefoot{$^a$ Star already known as a planet host for which we give here an updated orbit including {\footnotesize HARPS} observations.}
\end{table}

\begin{table*}[t!]
\caption{
\label{table:obs}
{\footnotesize HARPS} RVs and parameters inferred from the spectra and cross correlation functions for the 9 planet-host stars discussed in the paper.}
\begin{center}
\begin{tabular}{llrcccccc}
\hline\hline
  & JDB                        &$RV$        &$\varepsilon_{RV}$  &FWHM     &Contrast   &BIS     &\logrhk    &SN50 \\
  & [-2400000 days]    & [km/s]       &[m/s]                               &[km/s]      &[\%]           &[m/s]  &  &                   \\[1mm]
\hline
{\footnotesize HD}\,20003  &53728.576344    &-16.09631  &0.00054    &6.88881   &52.452   &-0.03595 &-4.8864   &141.40 \\
{\footnotesize HD}\,20003  &53763.545747    &-16.09548  &0.00048    &6.88610   &52.481   &-0.03718 &-4.8729   &165.50 \\
{\footnotesize HD}\,20003  &53788.524407    &-16.09975  &0.00052    &6.88607   &52.510   &-0.03588 &-4.8968   &151.20 \\
... & & & & & & & & \\
\hline
\end{tabular}
\end{center}
\end{table*}

Low-mass planets are very often found in multi-planet systems \citep[e.g.][for examples of RV results and \emph{Kepler} findings]{Lovis-2006,Lovis-2011,Udry-2007,Mayor-2009,Vogt-2010,Diaz:2016ab,Motalebi-2015,Latham-2011,Lissauer-2011,Lissauer-2014,Fabrycky-2014}. The several components in the system give rise to often complex, low-amplitude RV signals, not easy to solve for. The optimal observing strategy for a given star is a priori unknown, the relevant planetary periods possibly extending over three orders of magnitude. An adequate and efficient observing strategy has been developed coping with the need to accumulate a large number of measurements and to probe different timescales of variation. We typically follow our targets every night during a few initial observing runs, and after gathering a couple of dozens of observations we compare the observed dispersion with the expected jitter for the considered stellar spectral type. For significant variations we then continue monitoring at the same cadence when high-frequency variations are seen, or we adapt the frequency to possible variations at longer timescales.

The majority of the stars in our {\footnotesize HARPS} high-precision program have been followed since 2003, gathering observations spanning more than 11 years. We are not considering here observations obtained after May 2015 when a major upgrade of the instrument was implemented (change of the optical fibres)\footnote{The change of the fibres induces a small RV offset, different for each target. This change also corresponds to a change of scientific focus (and contributors) of our HARPS programs. After the change, only a few new measurements were obtained for our targets, not enough to precisely characterize the instrument offset and sample long-term signals.}. We also removed measurements obtained on the nights JD=2455115 and 2455399 as an unexplained instrumental systematic produced RV measurements that were off by more than 10 sigma on several stars observed those nights. Each RV measurement corresponds normally to a 15-minute {\footnotesize HARPS} exposure. This long exposure time allows to mitigate the short-timescale variations induced by stellar oscillations and therefore to improve RV precision \citep[][]{Santos-2004:a,Dumusque-2011:a}. For bright stars ($V \le 6.5$), the exposures are split in several sub-exposures within the 15 minutes in order to avoid CCD saturation while keeping the same strategy to mitigate stellar oscillations. The different time series presented in this paper are composed of binned points calculated through a weighted average of all points taken within an hour, so that all the observations taken within 15 minutes are binned together. The stars presented here have between 178 and 245 observations typically spread over $\sim$4000 days and with a sampling allowing the detection of planets with periods from below 1 day to the full span of the measurements. The obtained SNR at 550 nm typically ranges from 100 to 250 (up to 400 for \object{HD\,134060}), depending on the star and weather conditions. The corresponding quantified uncertainties on the RVs range then from 0.33 to 0.76\,\ms, including photon noise, calibration errors and instrumental drift uncertainty (see Table\,\ref{table:obsstat}). This does not include other instrumental systematics like telescope centering and guiding errors, which are expected to be small but difficult to estimate.
Possible additional uncertainties not included in this estimate might originate from the RV intrinsic variability of the star (jitter) due to stellar oscillations, granulation \citep{Dravins-1982,Dumusque-2011:a} and magnetic activity \citep{Dumusque-2014,Dumusque-2011:b,Meunier-2010,Desort-2007,Saar-1997}. Those extra errors induced by instrumental systematics or stellar signals will be taken into account when modeling the RVs (see Sec.~\ref{sec:analysis}).

The analysis of long series of HARPS radial velocities have recently revealed that, at the level of precision of HARPS, stitching of the detector (i.e. difference between the inter-pixel size and the gap between imprinted patches of pixels, see Sect.\,\ref{sec:analysis3}) may introduce a residual signal in the radial velocities at periods close to 6 months or 1 year when a strong stellar line crosses the gaps between pixel patches due to the yearly motion of the Earth around the Sun \citep{Dumusque-2015}. To avoid this effect, new sets of radial velocities for the stars presented in this paper have been obtained by removing the corresponding zone of the spectra in the template used for the cross correlation. 

\begin{figure*}[]
\center
 \includegraphics[angle=0,width=0.28\textwidth,origin=br]{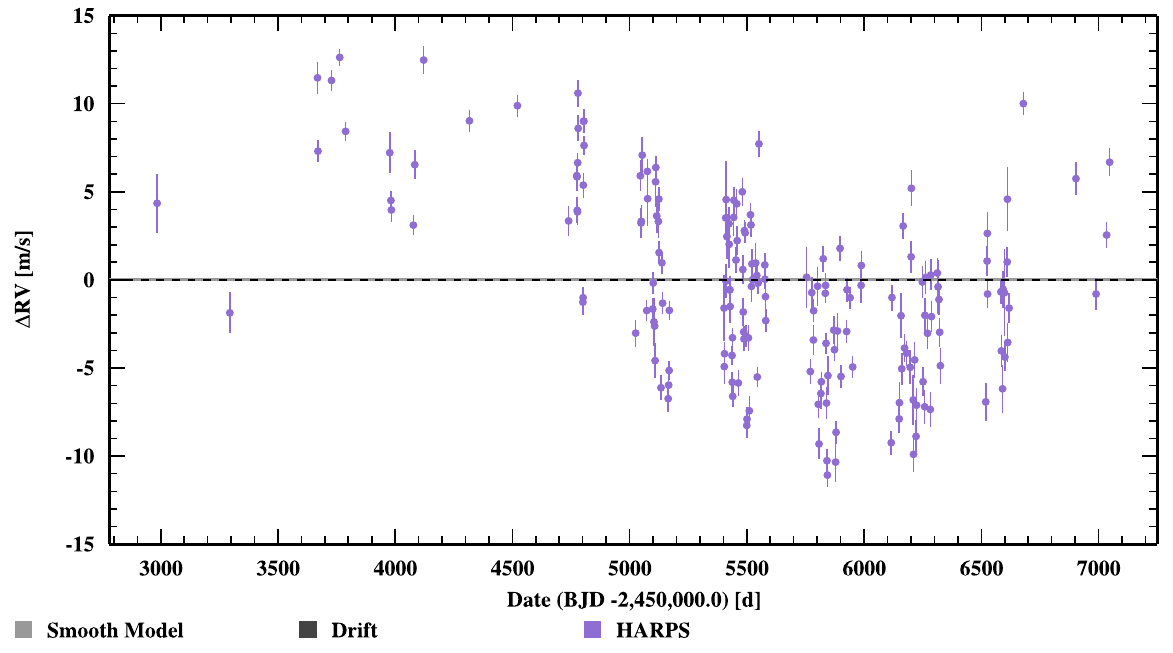}
 \includegraphics[angle=0,width=0.28\textwidth,origin=br]{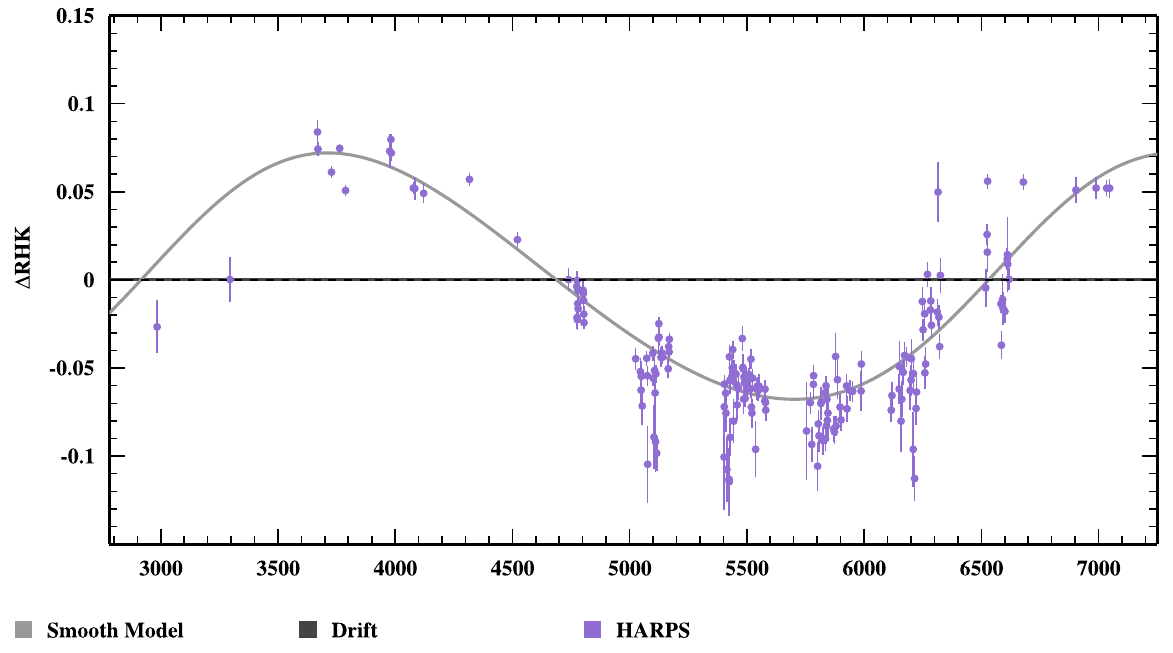}
 \includegraphics[angle=0,width=0.36\textwidth,origin=br]{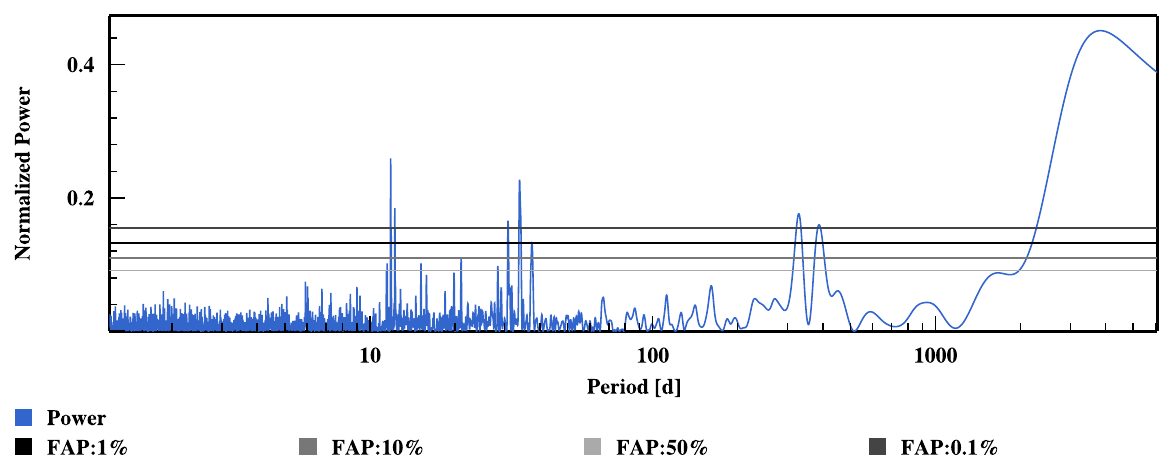}\\
 \includegraphics[angle=0,width=0.28\textwidth,origin=br]{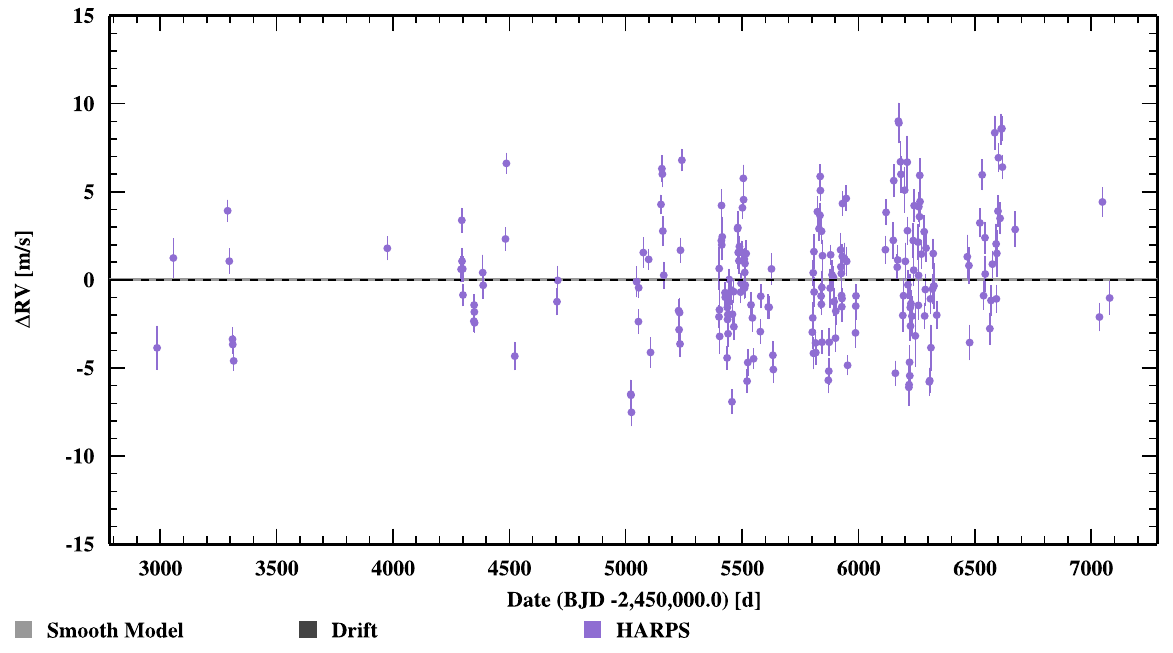}
 \includegraphics[angle=0,width=0.28\textwidth,origin=br]{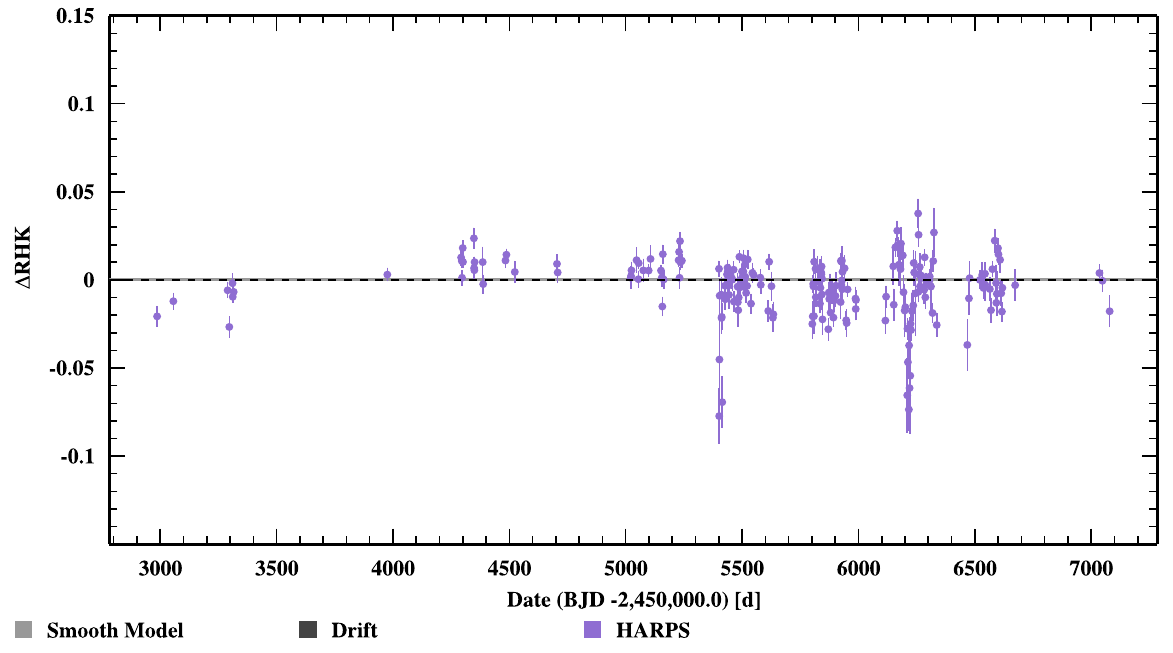}
 \includegraphics[angle=0,width=0.36\textwidth,origin=br]{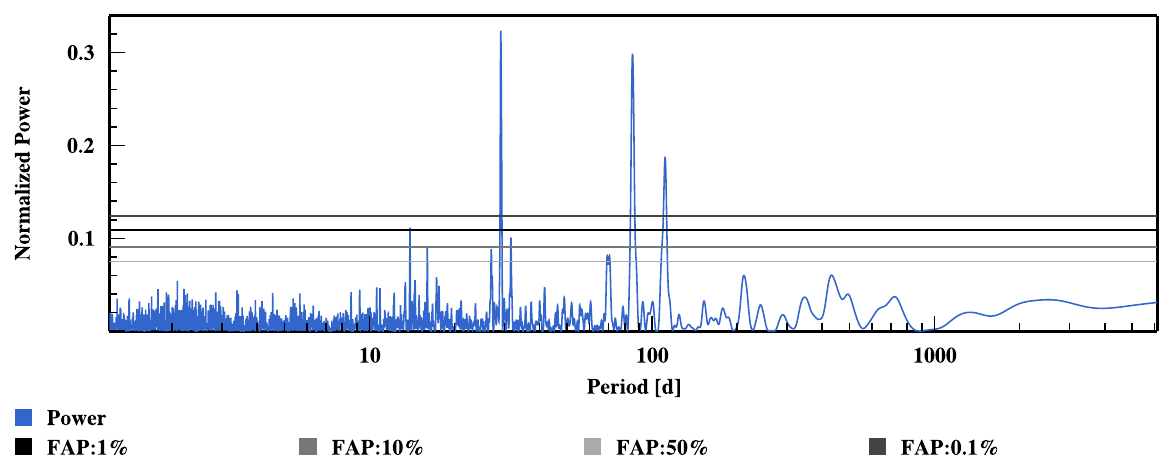}\\
 \includegraphics[angle=0,width=0.28\textwidth,origin=br]{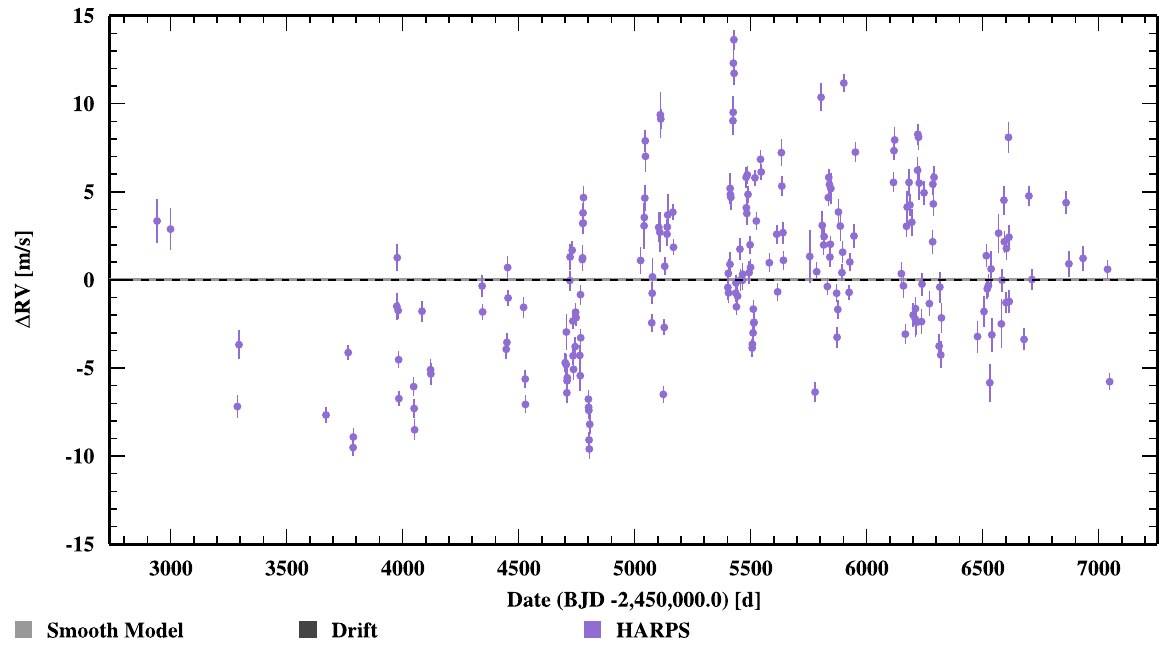}
 \includegraphics[angle=0,width=0.28\textwidth,origin=br]{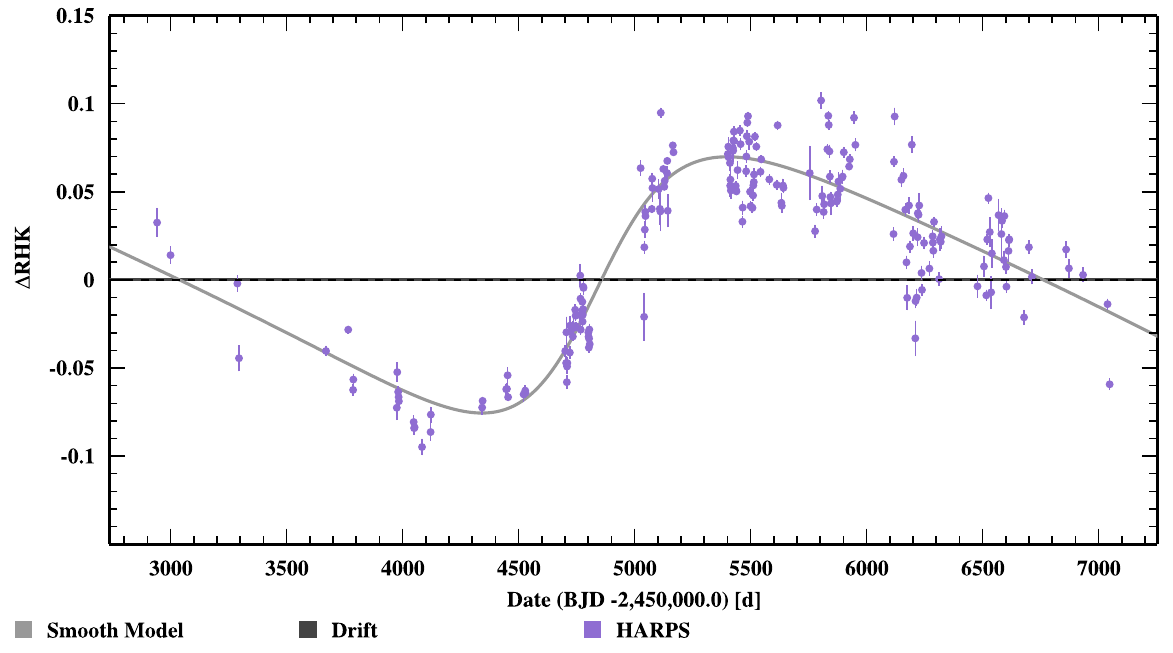}
 \includegraphics[angle=0,width=0.36\textwidth,origin=br]{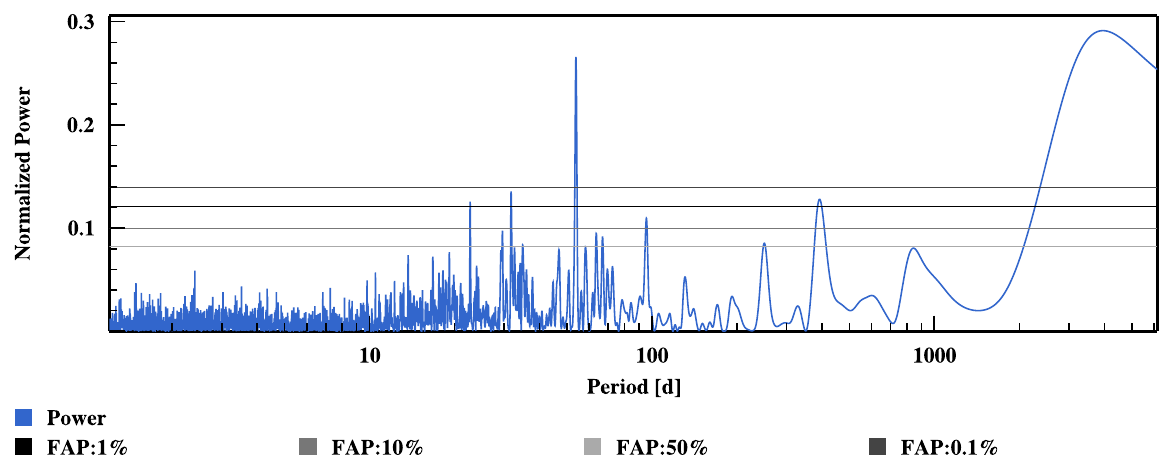}\\
 \includegraphics[angle=0,width=0.28\textwidth,origin=br]{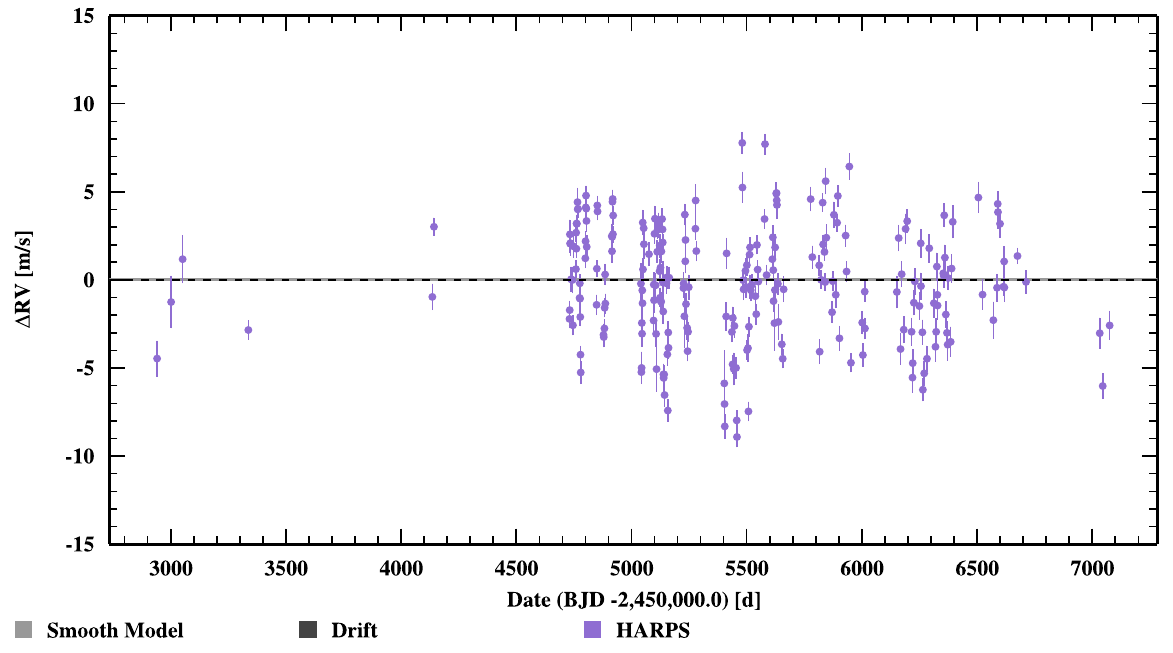}
 \includegraphics[angle=0,width=0.28\textwidth,origin=br]{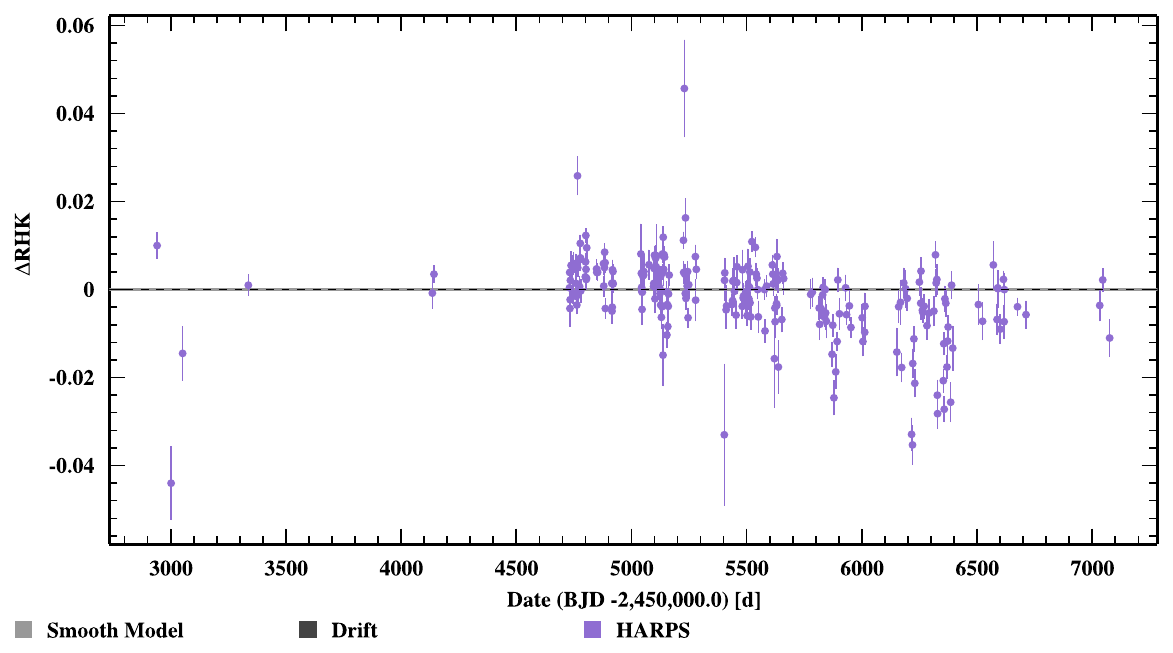}
 \includegraphics[angle=0,width=0.36\textwidth,origin=br]{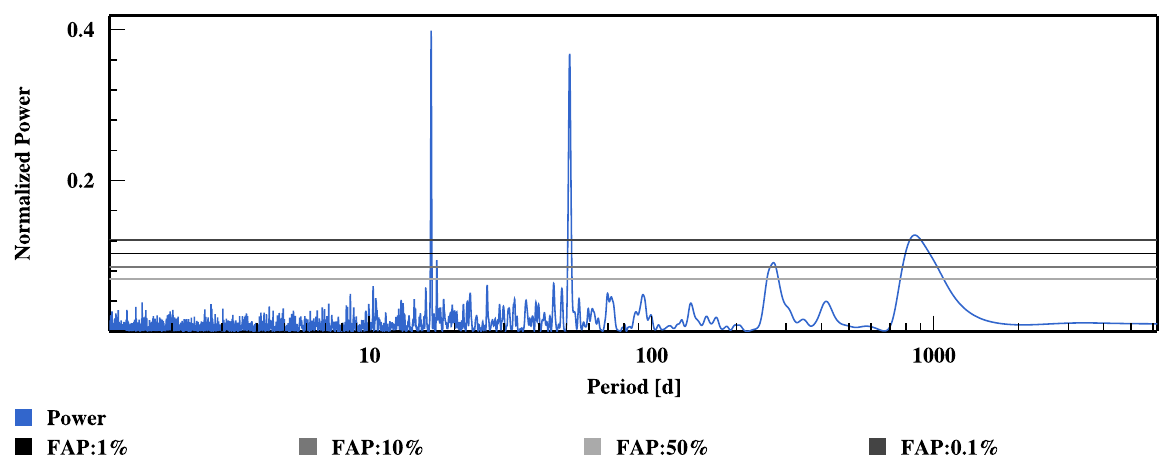}\\
  \includegraphics[angle=0,width=0.28\textwidth,origin=br]{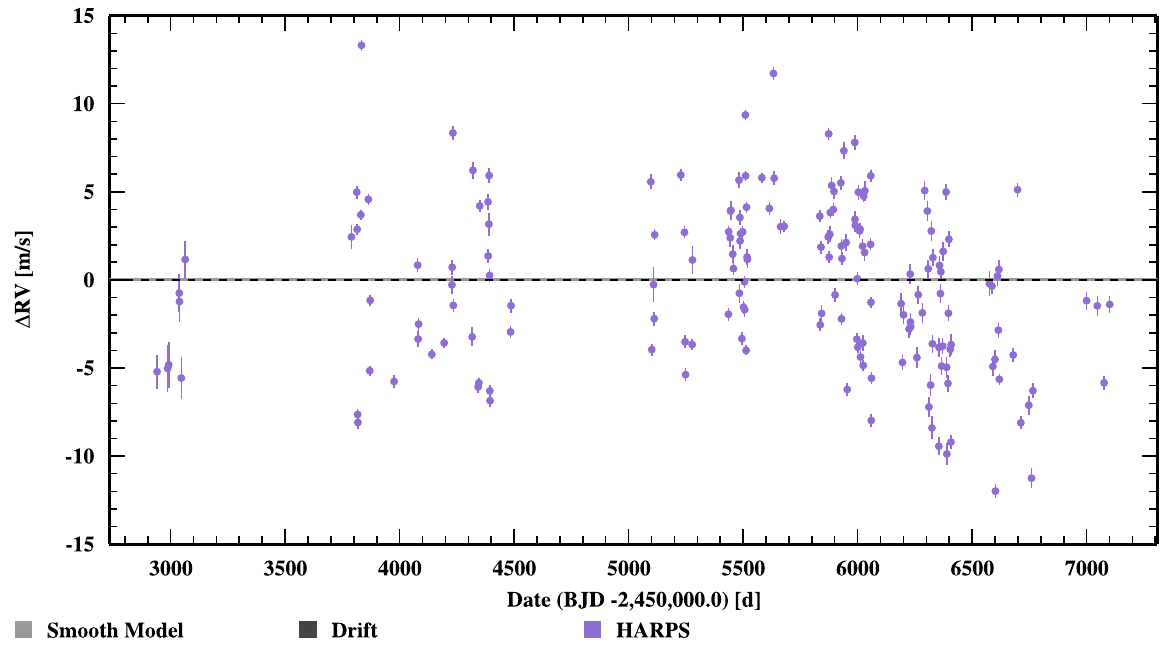}
 \includegraphics[angle=0,width=0.28\textwidth,origin=br]{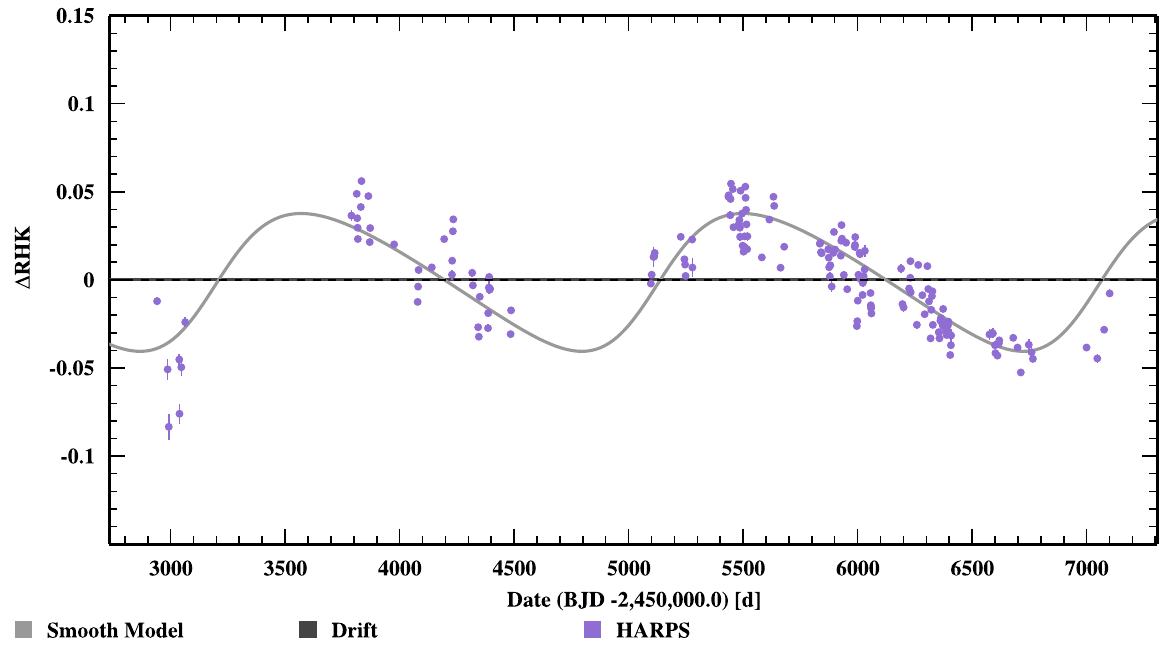}
 \includegraphics[angle=0,width=0.36\textwidth,origin=br]{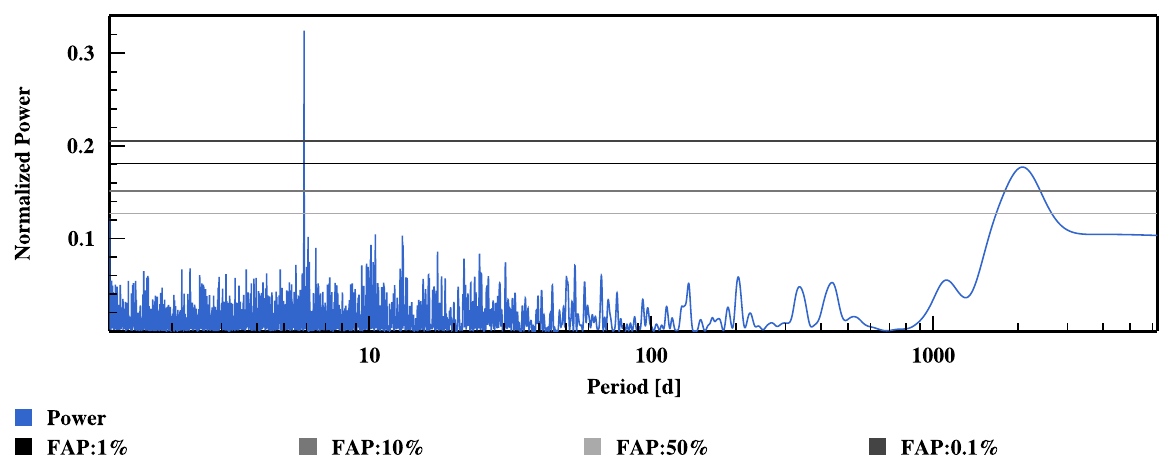}\\
 \includegraphics[angle=0,width=0.28\textwidth,origin=br]{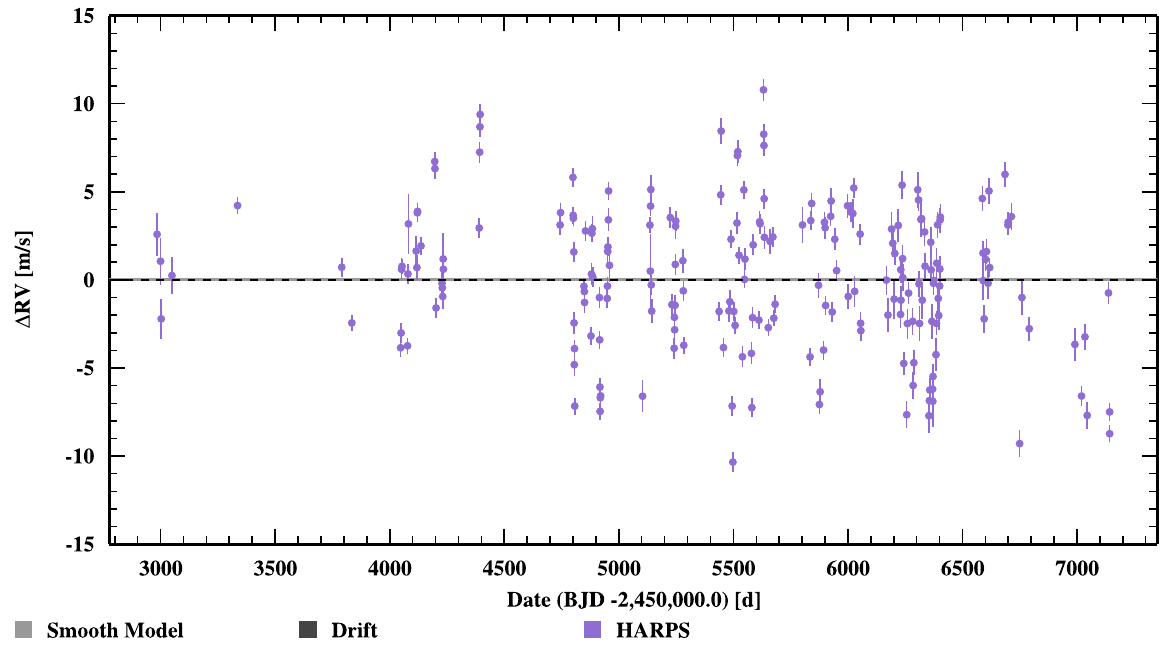}
 \includegraphics[angle=0,width=0.28\textwidth,origin=br]{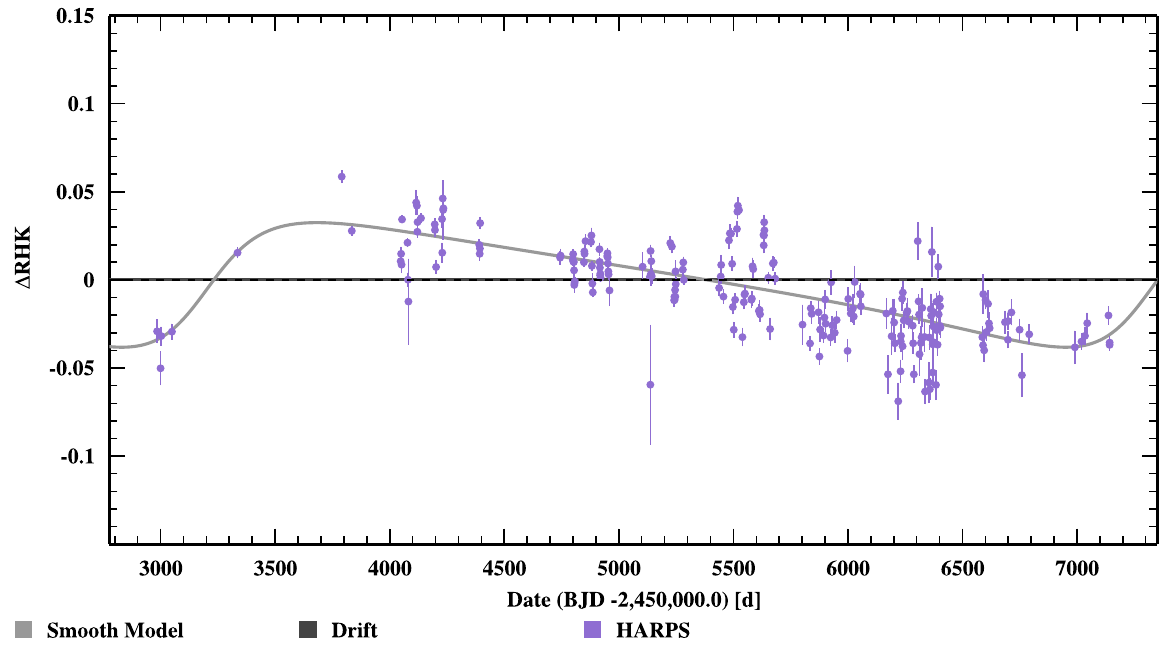}
 \includegraphics[angle=0,width=0.36\textwidth,origin=br]{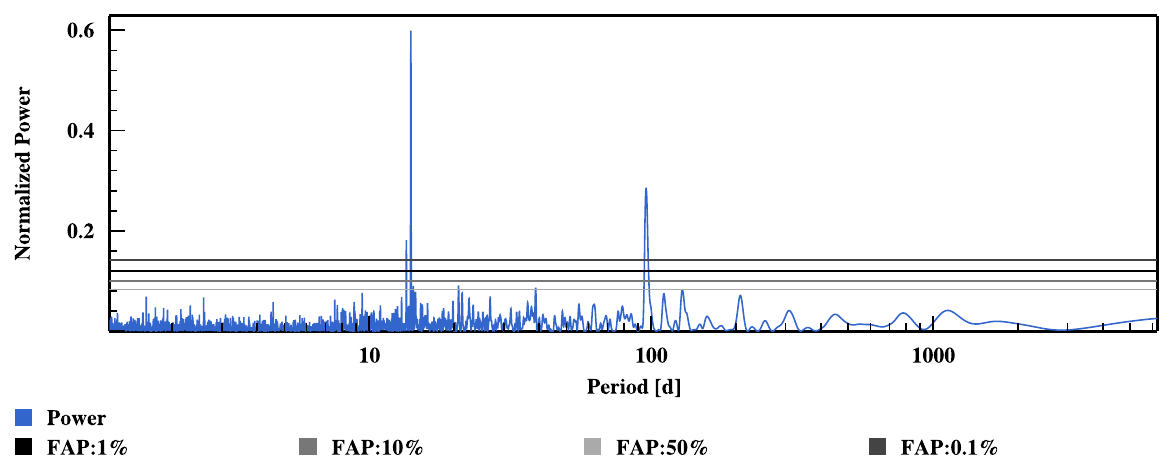}\\
 \includegraphics[angle=0,width=0.28\textwidth,origin=br]{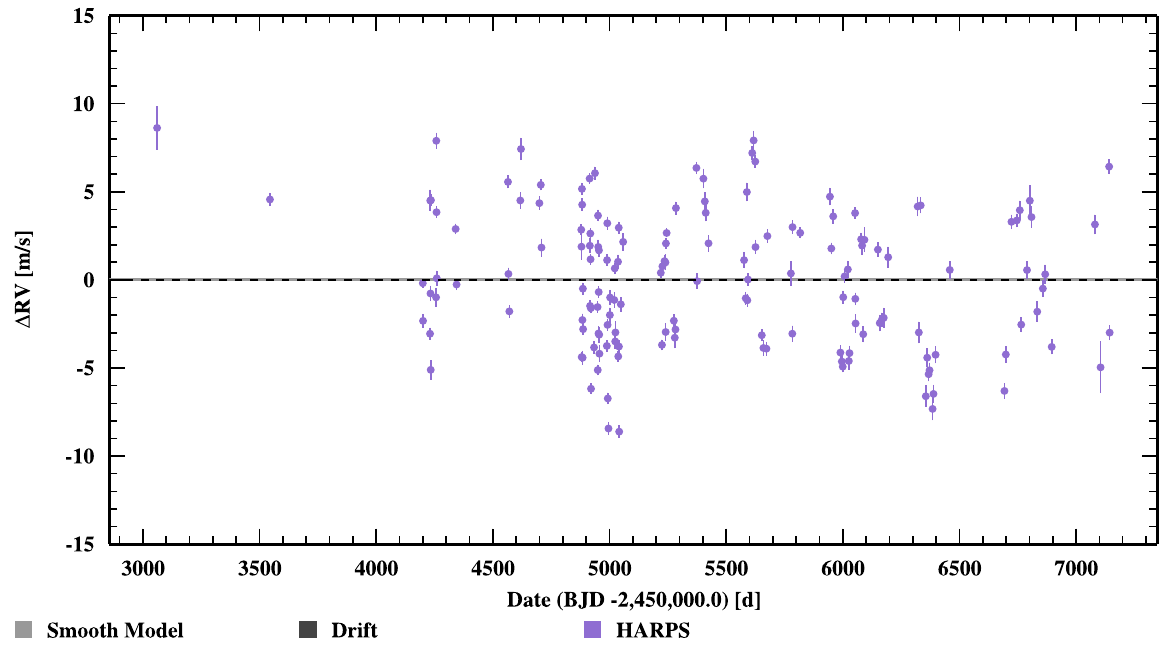}
 \includegraphics[angle=0,width=0.28\textwidth,origin=br]{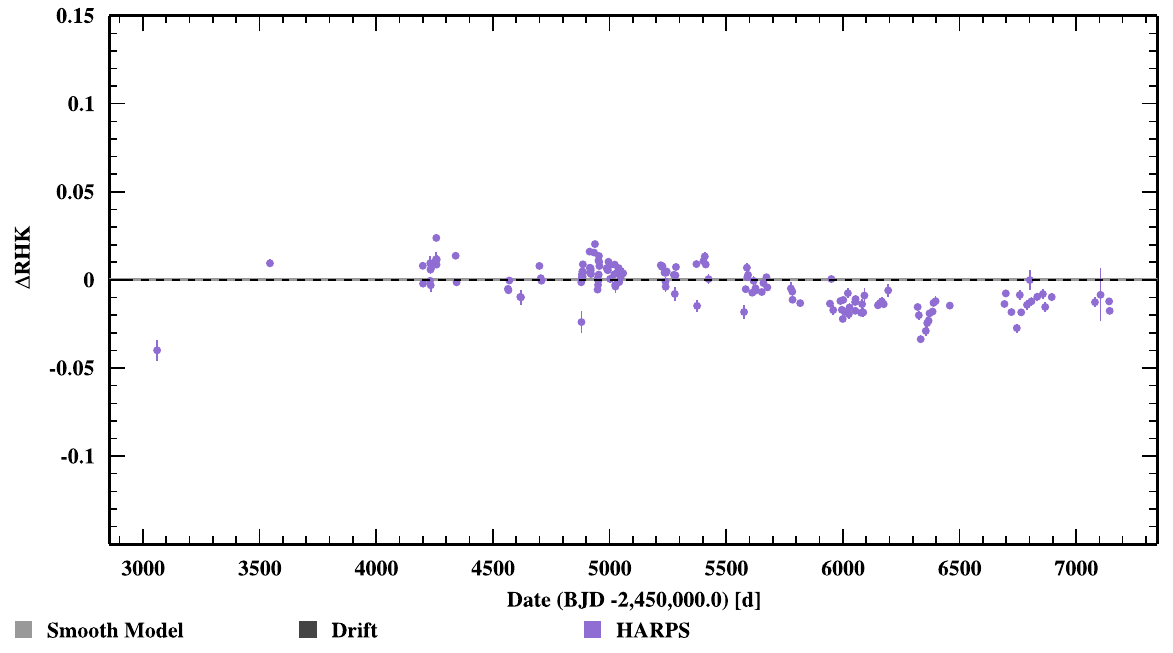}
 \includegraphics[angle=0,width=0.36\textwidth,origin=br]{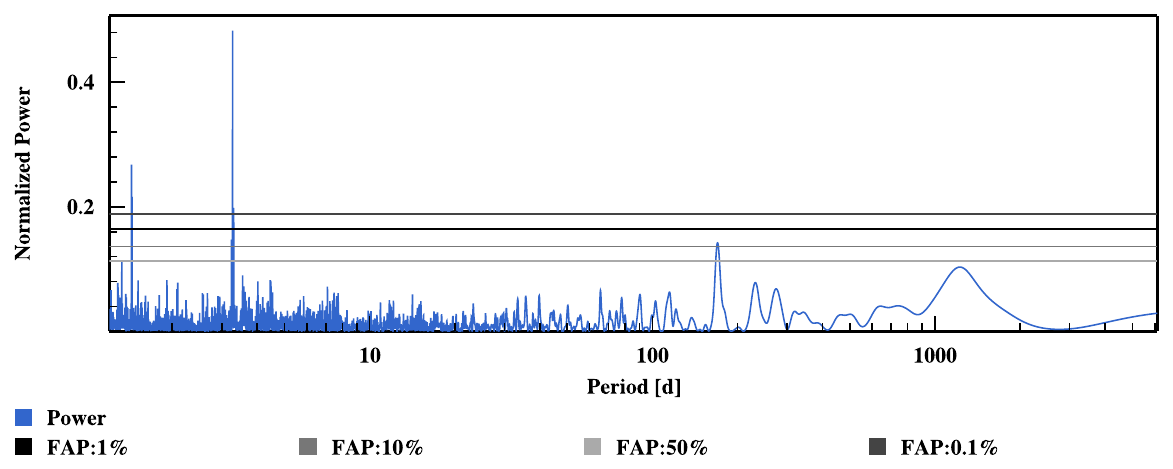}\\
  \includegraphics[angle=0,width=0.28\textwidth,origin=br]{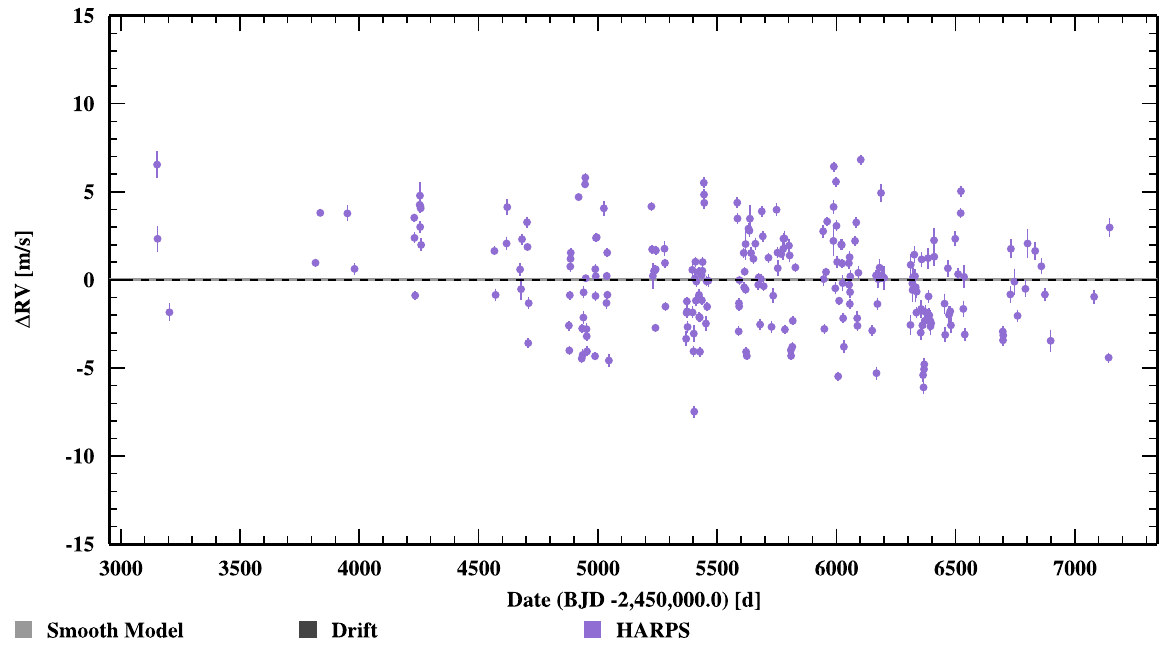}
 \includegraphics[angle=0,width=0.28\textwidth,origin=br]{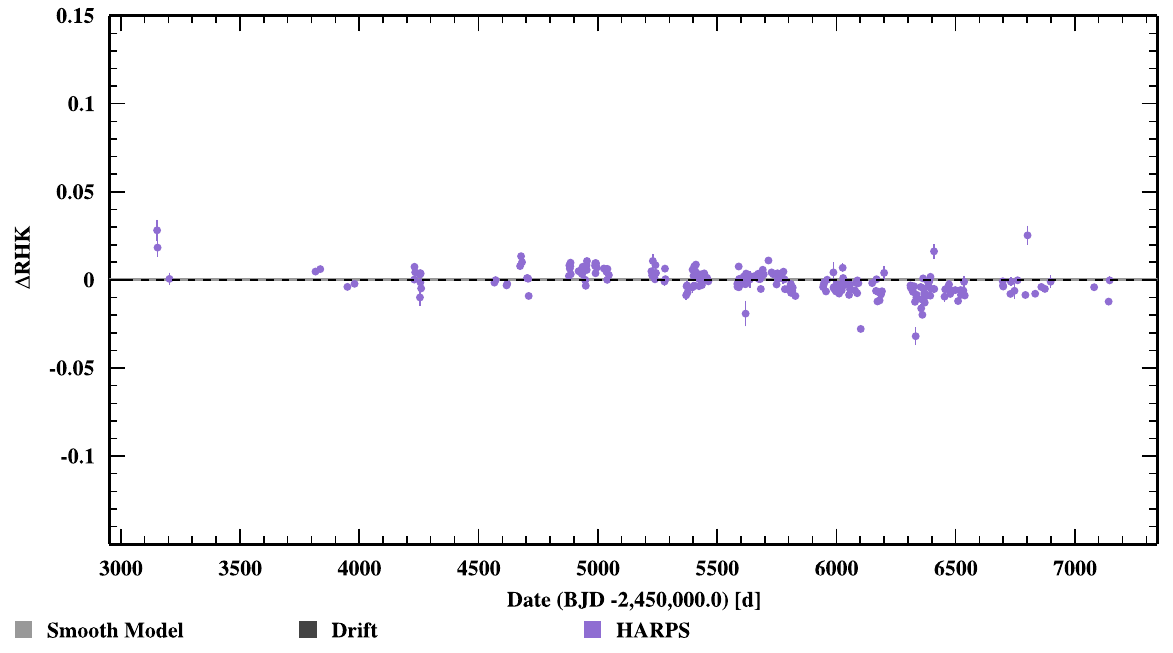}
 \includegraphics[angle=0,width=0.36\textwidth,origin=br]{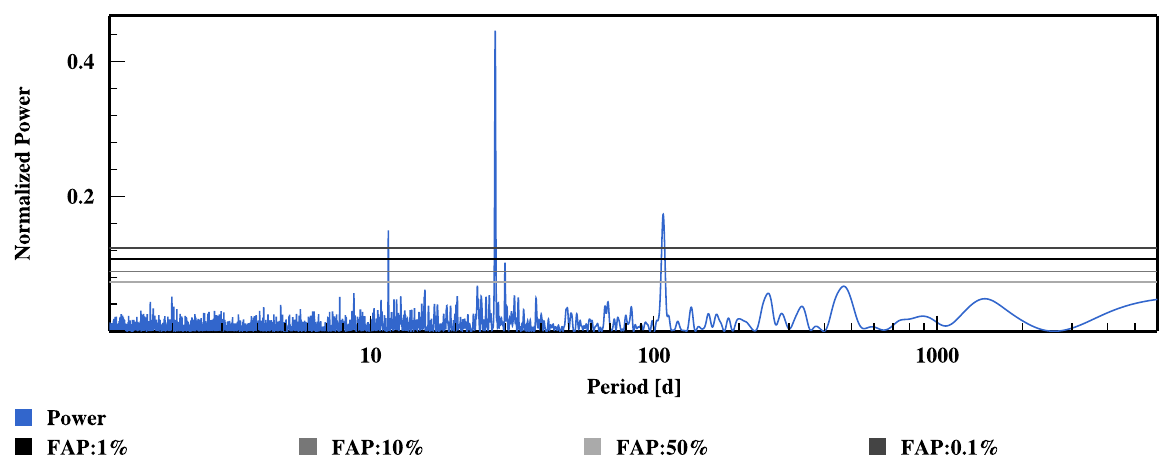}\\
  \caption[]{Left column: From top to bottom, {\footnotesize HARPS} RV measurements as a function of barycentric Julian Date obtained for \object{HD\,20003}, \object{HD\,20781}, \object{HD\,21693}, \object{HD\,31527}, \object{HD\,45184}, \object{HD\,51608}, \object{HD\,134060} and \object{HD\,136352}. Error bars only include photon and calibration noise. Middle column: \logrhk\ activity indicator as a function of time for the same stars. Right column: GLS periodogram of the corresponding RV measurements. }
\label{udry:fig1}
\end{figure*}

The final one-hour binned {\footnotesize HARPS} rsadial velocities and \logrhk\ measurements are displayed in the left column of Fig.~\ref{udry:fig1} for the 8 planet-host stars discussed in the paper. These velocities and the parameters inferred from the spectra and cross correlation functions are provided in electronic form at CDS. A sample of these data is provided in Table\,\ref{table:obs}. The statistics of the RV series are listed in Table\,\ref{table:obsstat}.

\section{General approach of the data analysis}
\label{sec:analysis}  

The data analysis presented in this paper is performed using a set of online tools available from the DACE platform\footnote{The DACE platform is available at http://dace.unige.ch. The online tools used to analyse radial-velocity data can be found in the section Observations $=>$ RVs}. The RV tool on this platform allows to upload any RV measurements and then perform a multiple Keplerian adjustment to the data using the approach described in \citet{Delisle-2016}. Once preliminary Keplerian plus drift parameters are found using this iterative approach, it is possible to run a full Bayesian MCMC analysis using an efficient algorithm, allowing various choices of models for signals (and noise) of different origins \citep{Diaz-2016,Diaz-2014}.

For each system, we performed a full MCMC analysis, probing the following set of variables for planetary signals: $\log{P}$,  $\log{K}$, $\sqrt{e}\,\cos{\omega}$,  $\sqrt{e}\,\sin{\omega}$ and $\lambda_{0}$, each one corresponding to the period, the RV semi-amplitude, the eccentricity, the argument of periastron and the mean longitude at a given reference epoch. We used $\sqrt{e}\,\cos{\omega}$ and$\sqrt{e}\,\sin{\omega}$ as free parameters rather than the eccentricity and the argument of periastron because they translate into a uniform prior in eccentricity \citep{Anderson-2011}. The mean longitude ($\lambda_{0}$) is also preferred as a free parameter (instead of the mean anomaly or the date of passage through periastron) since this quantity is not degenerated at low eccentricities.

When no hint of magnetic activity can be seen in the different activity observables (S-index, \logrhk, H$\alpha$-index), we fitted a RV model composed of Keplerians plus a polynomial up to the second order. The MCMC analysis is performed with uniform priors for all variables, with the exception of the stellar mass for which a Gaussian prior is used based on the information given in Table\,\ref{table:star}. We chose an error in stellar mass of 0.1 M$_{\odot}$ for all systems to propagate the stellar mass error to the estimation of the planet masses. To take into account uncertainty due to instrumental systematics and/or stellar signals not estimated by the reduction pipeline, we included in the MCMC analysis a white-noise jitter parameter, $\sigma_{JIT}$, that is quadratically added to the individual RV error bars.

When a magnetic cycle is detected in the different activity observables (S-index, \logrhk, H$\alpha$-index), we decided, in addition to the model described above, to include two extra components in our RV model to account for the RV variation induced by this magnetic cycle. As explained in \citet{Meunier-2016} and \citet{Dumusque-2011:c}, the variation of the total filling factor of spots and faculae along a magnetic cycle change the total amount of stellar convective blueshift, as it is reduced inside spots and faculae due to strong magnetic fields, which thus change the absolute RV of the star. A positive correlation between the different activity observables and the RV is thus expected. In this paper, we decided to use the method proposed by \citet{Meunier-2013} to mitigate the impact of magnetic cycles, i.e. to adjust a linear correlation between RV and one of the activity index. Here we chose to use the \logrhk\ \citep{Vaughan-1978,Wilson-1968,Noyes-1984}. A larger number of spots and faculae are present on the stellar surface during high-activity phases of the magnetic cycle, which implies a stronger stellar jitter due to those surface features coming in and out of view. To account for this stellar jitter that changes in amplitude along the magnetic cycle, we included in the MCMC analysis two extra white-noise jitter parameters, $\sigma_{JIT\,LOW}$ and $\sigma_{JIT\,HIGH}$, that correspond to the jitter during the minimum and the maximum phases of the magnetic cycle. We therefore replace $\sigma_{JIT}$ described in the precedent paragraph by $\sigma_{JIT\,LOW} + (\sigma_{JIT\,HIGH} - \sigma_{JIT\,LOW}).\mathrm{norm}($ \logrhk), where norm(\logrhk) corresponds to \logrhk\,normalized from 0 to 1. For each RV measurement, a new jitter parameter is derived according to the activity level and is quadratically added to the corresponding RV error bar. This is similar to the approach adopted in \citet{Diaz-2016}. The list of all the parameters probed by our MCMC is given in Table~\ref{tab-mcmc-params}.

\begin{table*}
\caption{Parameters probed by the MCMC. The symbols $\mathcal{U}$ and $\mathcal{N}$ used for the priors definition stands for uniform and normal distributions, respectively.
}
\label{tab-mcmc-params}
\footnotesize
\begin{tabular}{lcclc}
\hline
\hline
Parameters & Units & Priors & Description & Range\\
\hline
 & & & &\\
\multicolumn{5}{l}{\bf Parameters probed by MCMC without magnetic cycle}\\
 & & & &\\
M$_{\star}$ & [M$_{\odot}$]  & $\mathcal{N}$(M$_{\star}$,0.1) & stellar mass (M$_{\star}$ can be found in Table \ref{table:star}) & \\
$\sigma_{X}$ & [\ms] & $\mathcal{U}$ & instrumental jitter of instrument $X$  & ]-$\infty$,$\infty$[\\
$\sigma_{JIT}$ & [\ms] & $\mathcal{U}$ & stellar jitter (If only one instrument is used, this parameter includes $\sigma_{X}$)& ]-$\infty$,$\infty$[\\
$\gamma_{X}$& [km/s]&$\mathcal{U}$ & Constant velocity offset of instrument $X$ & ]-$\infty$,$\infty$[\\
$lin$ &[\ms\,yr$^{-1}$] & $\mathcal{U}$ & Linear drift  & ]-$\infty$,$\infty$[\\
$quad$ &[$m^2s^{-2}\,yr^{-2}$] & $\mathcal{U}$ & Quadratic drift  & ]-$\infty$,$\infty$[\\
%$quad$ & [\ms\,yr$^{-2}$]&$\mathcal{U}$&Quadratic drift  &\\
$\log{(P)}$ & log([days]) & $\mathcal{U}$ & Logarithm of the period  & [0,$\infty$[\\
$\log{(K)}$ & log([\ms]) & $\mathcal{U}$ & Logarithm of the RV semi-amplitude  & [0,$\infty$[\\
$\sqrt{e}\,\cos{\omega}$ & - & $\mathcal{U}$ & & ]-1,1[\\
$\sqrt{e}\,\sin{\omega}$  & - & $\mathcal{U}$ & & ]-1,1[\\
$\lambda_{0} = M_{0}+\omega$ & [deg] & $\mathcal{U}$ & Mean longitude ($M_{0}$ = mean anomaly) & [0,360[\\
 & & & &\\
\multicolumn{5}{l}{{\bf Parameters probed by MCMC with magnetic cycle} (in addition to the previous ones, except $ \sigma_{JIT\,LOW}$ and $\sigma_{JIT\,HIGH}$ that replace $\sigma_{JIT}$}\\
 & & & &\\
$\sigma_{JIT\,LOW}$ & [\ms] & $\mathcal{U}$ & stellar jitter at the minimum of the magnetic cycle (If only one & \\
 & &  & instrument is used, this parameter includes $\sigma_{X}$) & ]-$\infty$,$\infty$[\\
$\sigma_{JIT\,HIGH}$ & [\ms] & $\mathcal{U}$ & stellar jitter at the maximum of the magnetic cycle (If only one & \\
 & &  & instrument is used, this parameter includes $\sigma_{X}$) & ]-$\infty$,$\infty$[\\
\logrhk\,$lin$ & [\ms] & $\mathcal{U}$ & slope of the correlation between the RV and \logrhk\,(The \logrhk\,variation & \\
 & & & is normalized between 0 and 1, thus this parameter is in [\ms]) & ]-$\infty$,$\infty$[\\
 & & & &\\
\multicolumn{5}{l}{{\bf Physical Parameters derived from the MCMC posteriors (not probed)}}\\
 & & & &\\      
 $P$ & [d] & - & Period & \\
 $K$ & [m\,s$^{-1}$] & - & RV semi-amplitude & \\
 $e$ & - & - & Orbit eccentricity & \\
 $\omega$ & [deg] &  - & Argument of periastron & \\
 $T_P$ & [d] &  - & Time of passage at periastron & \\
 $T_C$ & [d] &  - & Time of transit & \\
  $Ar$ & [AU] &  - & Semi-major axis of the relative orbit & \\
 M.$\sin{i}$ & [M$_{\rm Jup}$] &  - & Mass relative to Jupiter & \\
 M.$\sin{i}$ & [M$_{\rm Earth}$] &  - & Mass relative the Earth & \\
 \hline
\end{tabular}
\end{table*}

%\subsection{General approach, non-planetary signals, and validation of the derived solutions}

Before running a full MCMC analysis to obtain reliable posteriors for the orbital parameters of the planets present in the RV measurements, we first iteratively look for significant signals in the data using the approach of \citet{Delisle-2016}, which gives us a good approximation of the orbital parameters that are used as initial conditions for the MCMC analysis. The iterative approach follows several independent but complementary steps.

\subsection{Removing long-term trends} \label{sec:analysis1}
Long-term trends in the data, due either to long-period companions (stellar or planetary) or magnetic cycles \citep[][and \citet{Lovis-2011:b} that provides a broad view of magnetic cycle behavior for the entire {\footnotesize HARPS} high-precision RV survey]{Dumusque-2011:c}, might perturb the detection of planets on shorter-period orbits due to the spread of the energy of the signal at high frequencies in the Generalized Lomb-Scargle periodogram \citep[GLS,][]{Zechmeister-2009,Scargle-1982} and aliases of these signals appearing at shorter periods. We fit a polynomial up to the second order to account for a non-resolved long-period companion. In addition, if a significant long-period signal is seen in the GLS periodogram of the calcium activity index \logrhk, we fit the RVs with a polynomial up to the second order, plus a Keplerian model reproducing the \logrhk\ variation, leaving only the amplitude as a free parameter. Note that when correcting magnetic cycle effect in RVs when adopting a step-by-step approach, we do not consider a simple linear correlation between RV and \logrhk\ as explained above because significant planetary signals not yet removed from the data might destroy any existing correlation. There is no a priori reason why the magnetic cycle should look like a Keplerian, however a Keplerien has more degrees of freedom than a simple sinusoid and can therefore better estimate the long-term variation seen in \logrhk. Regarding signal significance, a signal is considered worth looking into when its $p$-value, which gives the probability that the signal appears just by chance, is smaller or equal to a chosen threshold. Experience has shown us that a  $p$-value of 1\,\% provides a good empirical threshold. Note that in this paper, $p$-values are estimated in a Monte Carlo approach by bootstrapping 10'000 times over the dates of the observations.

\subsection{Detection of periodic signals in the data} \label{sec:analysis2}
\begin{itemize}
\item[--] Planetary signals are searched for in the GLS periodogram of the RV residuals after correcting long-term trends. As for magnetic cycles, we consider a signal as significant if its $p$-value is smaller than 1\,\%\footnote{The goal of our program is to derive reliable statistical distributions of orbital and planet properties that can be used to constrain planet formation models. Signals of less significance are of course of great interest but require more observations to be confirmed as {\it bona fide} planets.}. If the $p$-values of some signals in the GLS periodogram are smaller than 1\%, we adjust a Keplerian orbital solution to the signal presenting the smallest $p$-value following the approach of \citet{Delisle-2016} with the detected period as a guess value. 
\item[--] Additional planets in the systems are then considered if other significant signals are present in the GLS periodogram of the RV residuals. Note that each time a Keplerian signal is added to the model, a global fit including all the previously detected signals is performed.
\item[--] We stopped when no more signals in the residuals present a $p$-value smaller than 1\,\% and we finally visually inspected the solutions. 
\end{itemize}

\subsection{Origin of the different signals found} \label{sec:analysis3}

After the main periodic signals have been recognized in the data, it is important to identify the ones that are very likely not from planetary origin. They may be of different natures: stellar, instrumental or observational. Here is a brief description of some cases encountered:
\begin{itemize}
\item[--] {\it Stellar origin.} The signal is due to stellar intrinsic phenomena. The most common one is the variation of the measured RVs due to spectral line asymmetries induced by spots and faculae coming in and out of view on the surface of the star, modulating the signal over a star rotational period \citep[][]{Haywood-2016,Haywood-2014,Dumusque-2014b,Meunier-2010,Desort-2007,Santos-2000:b,Saar-1997}. In order to recognize RV variations of intrinsic stellar origin, we compare the periods of the derived orbital solutions with the rotational periods of the stars and their harmonics estimated using the \logrhk\,average activity level \citep[see Table\,\ref{table:star},][]{Mamajek-2008,Noyes-1984}. We also compare the orbital periods with the variation timescales of spectroscopic activity indicators derived from the CCF, such as the BIS and the FWHM, and derived from spectral lines sensitive to activity, such as the Ca II H and K lines (S-index and \logrhk) or the H$\alpha$ line (H$\alpha$-index).
%Activity indicators also allow for the detection of long-term magnetic-cycle effects.
\item[--] {\it Instrumental origin.} As already mentioned earlier, spurious signals of small amplitudes with periods close to one year or one of its harmonics (half a year, a third of a year) can be created by a discontinuity in the wavelength calibration introduced by the stitching of the detector. The 4k$\times$4k {\footnotesize HARPS} CCD is composed of 32 blocks of 512$\times$1024 pixels. When each block were imprinted to form the detector mosaic, the technology at the time was not precise enough to ensure that pixels between blocks had the same size than intra-pixels. Therefore, every 512 pixels in the spectral direction, the CCD presents pixels that differs in size \citep[][]{Wilken-2010}. Block stitching may introduce a residual signal in the RVs at periods close to 6 months or 1 year when strong stellar lines crosse block stitchings due to the yearly motion of the Earth around the Sun. As mentioned in Sect.\,\ref{sec:meas}, in order to mitigate this effect new sets of RVs for the stars presented in this paper have been obtained by removing from the correlation masks used the spectral lines potentially affected. For each star, the systemic velocity will shift the stellar spectrum on the CCD, therefore an optimization of the correlation mask is done on a star-by-star basis \citep[for more information, see][]{Dumusque-2015}.
\item[--] {\it Observational limitations.} Aliases have to be taken into account. The ones due to one year or one day sampling effects are well known \citep[e.g.,][]{Dawson-2010}. They apply on all signals and not only on the planetary ones. 
%Other effects are a bit less direct. For example, periods in the range 320 to 420 days can be introduce by aliasing of long-period quadratic or linear drifts that are not taken into account. 
\end{itemize}

\section{Description and analysis of individual systems}
\label{sec:systems}

In this section, we present eight planetary systems including Neptune-mass and super-Earth planets.  Presence of planets around these stars and preliminary system characterizations have been announced at the "Extreme Solar System II" conference held at Grand Teton USA in September 2011 and published in \citet{Mayor-2011}. We present here the detailed analysis of each system with updated data, describing and discussing the derived planetary system characteristics, as well as other (i.e. non-planetary) signals in the data. 

%HD20003
\subsection{{\footnotesize HD}\,20003: Eccentric, close-in Neptune-mass planets close to the 3:1 commensurability}
\label{udry:hd20003}  

\subsubsection{Radial velocity analysis}
An ensemble of 184 high signal-to-noise observations ($<$S/N$>$ of 110 at 550 nm) of \object{HD\,20003} have been gathered covering about 11 years (4063 days). The typical photon-noise plus calibration uncertainty of the observations is 0.74\,ms$^{-1}$, well below the observed dispersion of the RVs at 5.35\,\ms. The RVs with their GLS periodogram and the \logrhk\,time-series are displayed in Fig.\,\ref{udry:fig1}. A long-period variation is observed in the RV data, as well as in \logrhk, indicative of a magnetic cycle effect on the velocities. To correct the velocities for this effect, we modeled the long-term variation by a Keplerian with parameters fixed and determined from the \logrhk, except for the amplitude that was left free to vary.

\begin{figure*}[!h]
\center
  \includegraphics[angle=0,width=0.4\textwidth,origin=br]{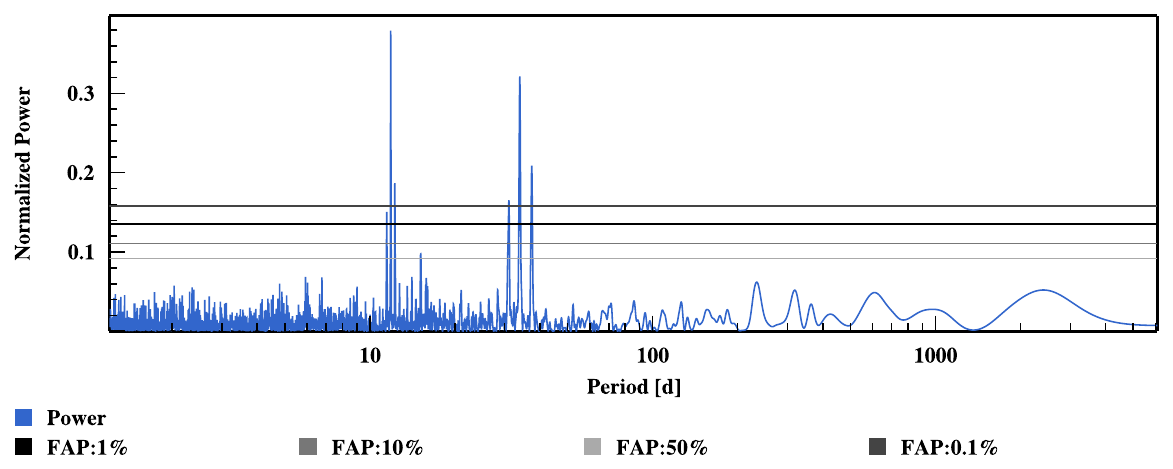} 
  \includegraphics[angle=0,width=0.4\textwidth,origin=br]{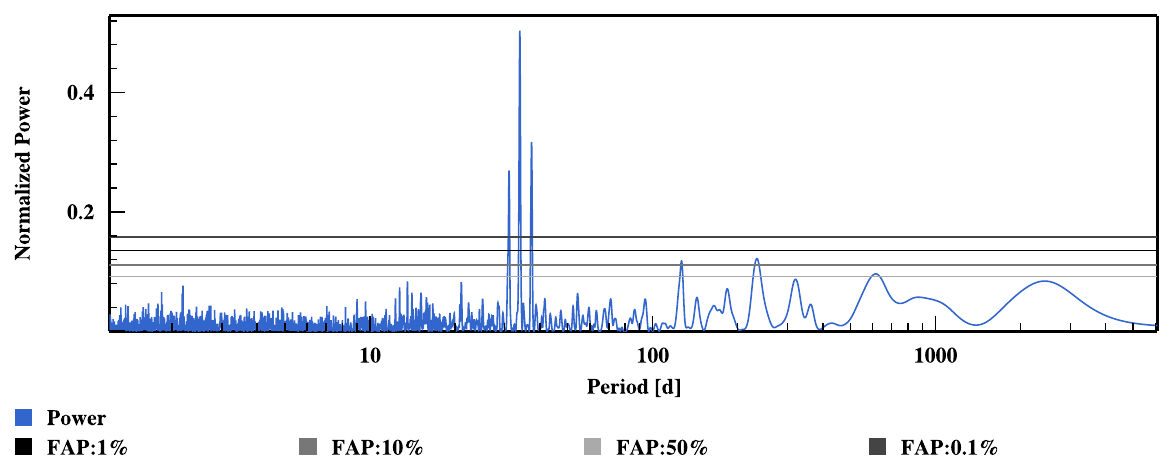} 
  \includegraphics[angle=0,width=0.4\textwidth,origin=br]{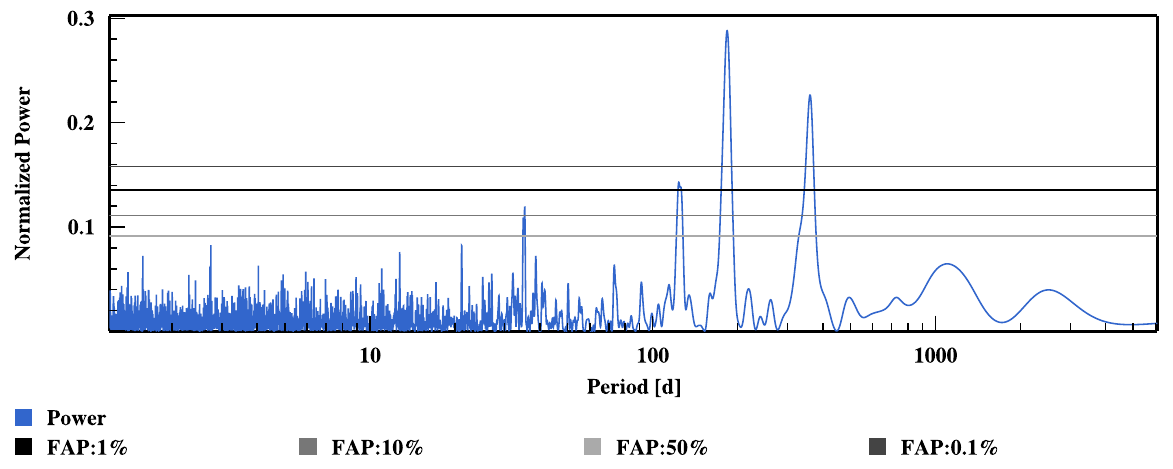} 
    \includegraphics[angle=0,width=0.4\textwidth,origin=br]{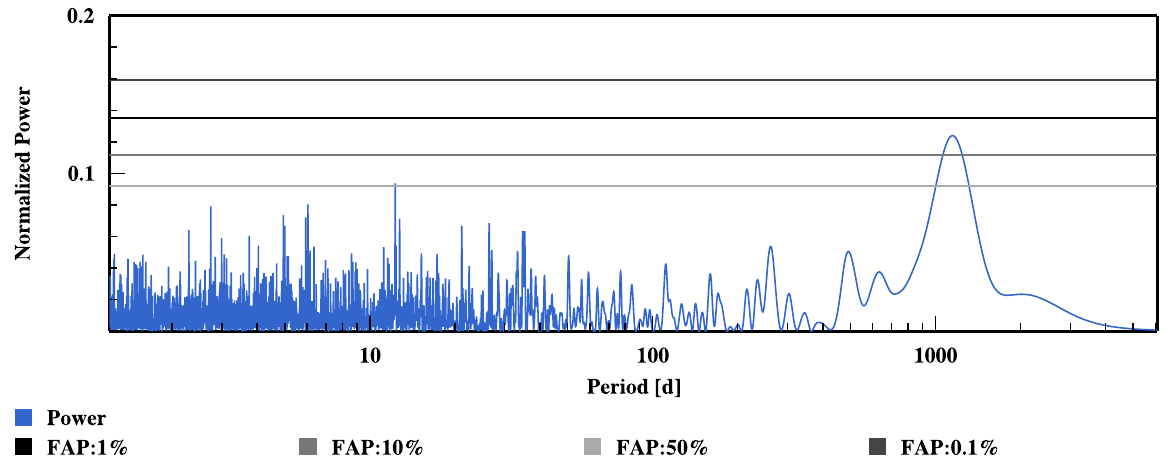} 
\caption[]{GLS Periodogram of the RV residuals of {\footnotesize HD}\,20003 at each step after removing, from top-left to bottom-right, the magnetic cycle effect plus a second-order polynomial drift, then the two planets at 11 and 33 days one after the other, and finally the signal at 184 days. The GLS periodogram of the raw RVs is shown in Fig.\,\ref{udry:fig1}.}
\label{udry:fig3}
\end{figure*}

After correction for the long-period variation due to the magnetic cycle plus fitting a second-order polynomial drift, two peaks very clearly emerge in the periodogram well above the 0.1\,\% $p$-value limit, at periods around 11.9 and 33.9 days (Fig.\,\ref{udry:fig3}). 
%The 1-day aliases of these periods are also clearly visible in the GLS periodogram around 0.91\,d, 0.97\,d, 1.03\,d and 1.09\,d.
Once those two signals are fitted for along with the magnetic cycle and the second-order polynomial drift, a significant signal at 184 days appears in the GLS periodogram of the RV residuals, with strong aliases at 127 and 359 days. A global fit including a second-order polynomial drift, a Keplerian for the magnetic cycle and three Keplerians for signals at 11.9, 33.9 and 184 days allows us to model all the variations seen in the RVs. No significant signal (with $p$-values smaller than 5\,\%) are then present in the RV residuals (Fig.\,\ref{udry:fig3}, bottom-right plot).

\begin{figure}[!h]
\center
  \includegraphics[angle=0,width=0.4\textwidth,origin=br]{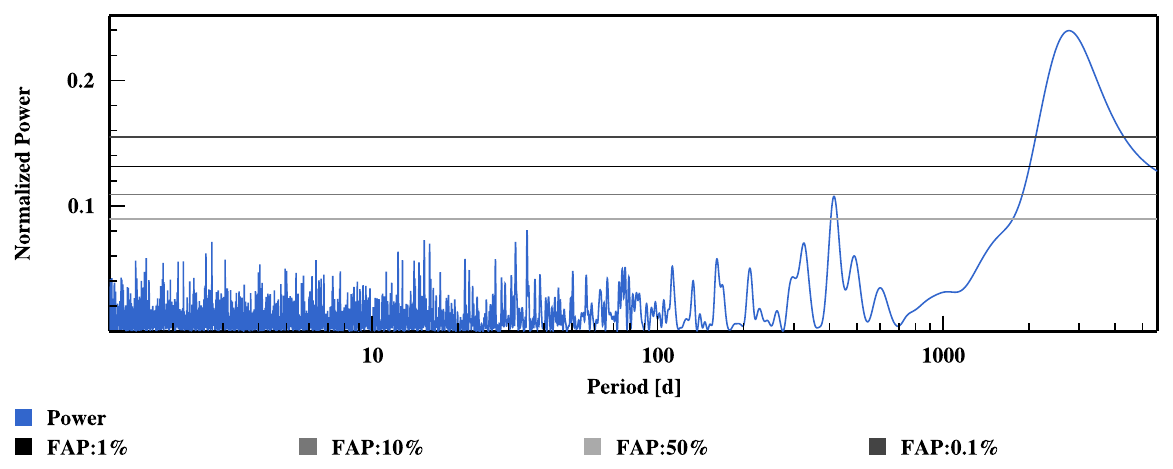} 
\caption[]{GLS Periodogram of the RV residuals of {\footnotesize HD}\,20003 obtained from the MCMC solution including a linear correlation between RV and \logrhk\, to account for the magnetic cycle, a second-order polynomial drift and three Keplerian.}
\label{udry:fig3.1}
\end{figure}
\begin{figure*}[!ht]
\center
  \includegraphics[angle=0,width=0.4\textwidth,origin=br]{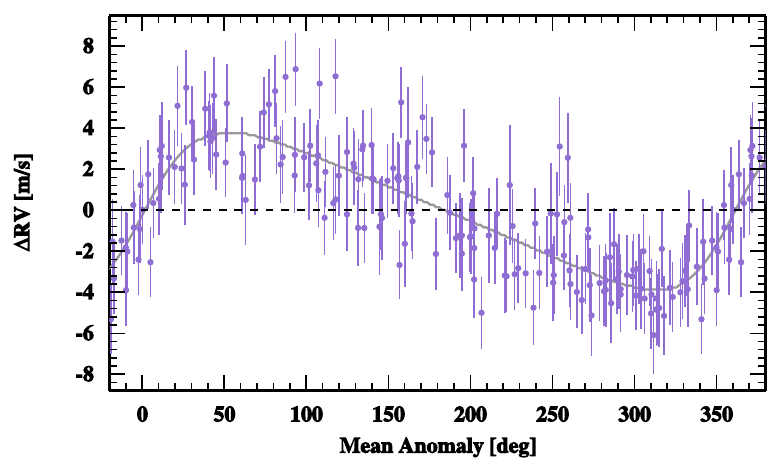} 
  \includegraphics[angle=0,width=0.4\textwidth,origin=br]{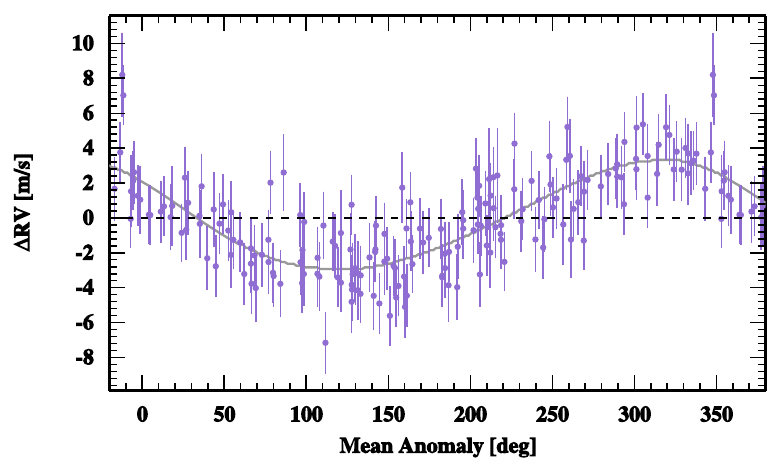} 
    \includegraphics[angle=0,width=0.4\textwidth,origin=br]{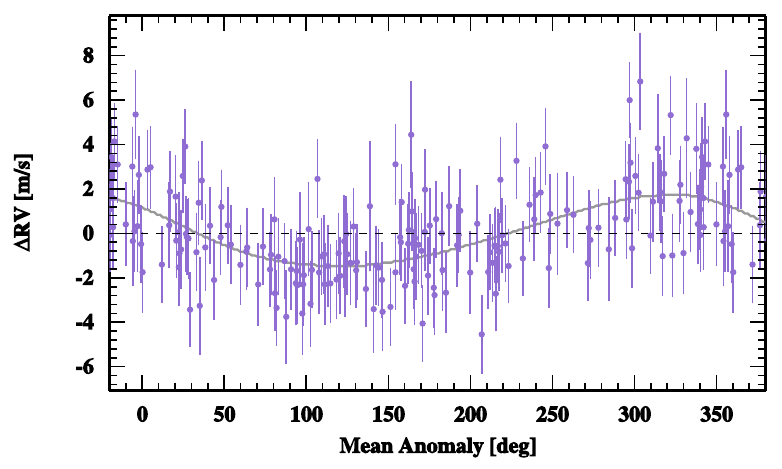} 
  \includegraphics[angle=0,width=0.4\textwidth,origin=br]{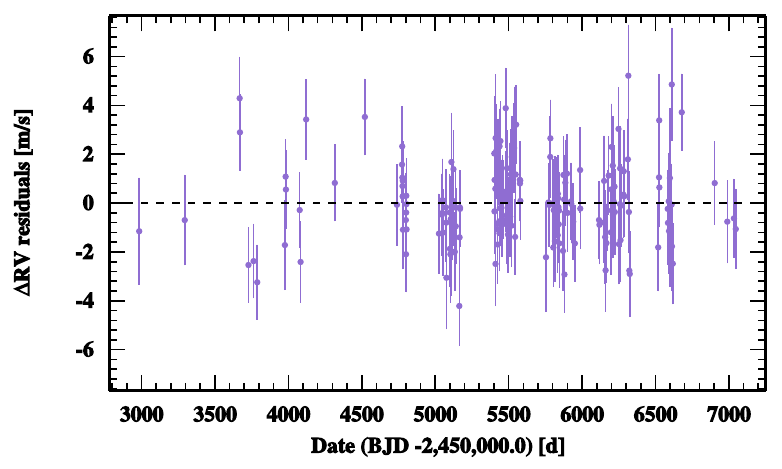} 
\caption[]{Phase-folded RV measurements of {\footnotesize HD}\,20003 with the best Keplerian solution represented as a black curve for each of the signals in the data. From top-left to bottom-right: planet b, planet c and the signal at 184 days. The residuals around the solution are displayed in the lower-right panel. Contrarily to Fig.\,\ref{udry:fig1} error bars include now the instrumental + stellar jitter from the MCMC modeling. Corresponding planetary orbital elements are listed in Table\,\ref{HD20003_tab-mcmc-Summary_params}.}
\label{udry:fig4}
\end{figure*}

\begin{figure*}[!h]
 \center
  \includegraphics[angle=0,width=0.4\textwidth,origin=br]{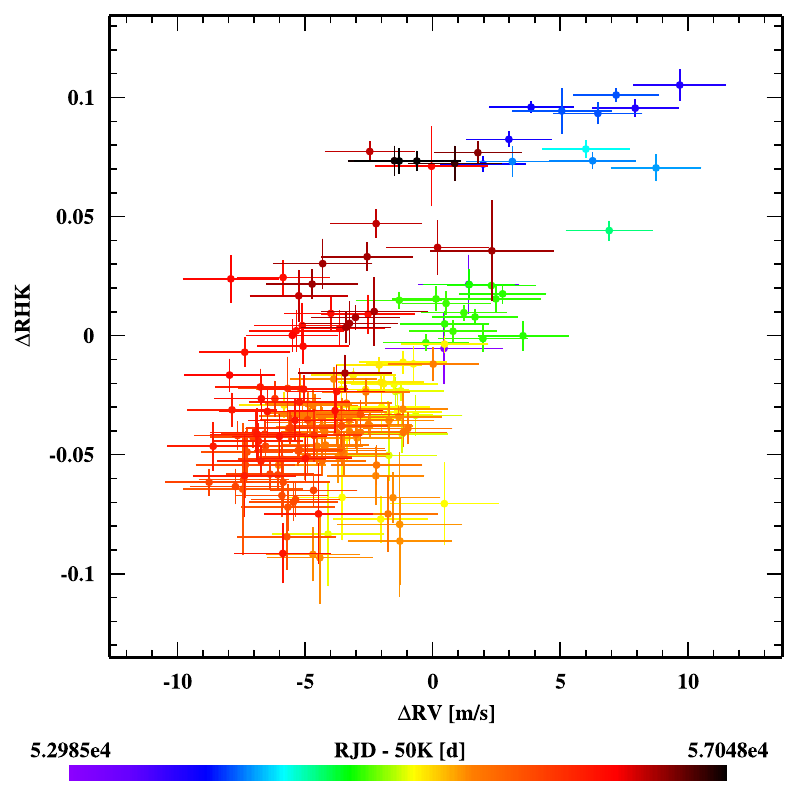} 
    \includegraphics[angle=0,width=0.4\textwidth,origin=br]{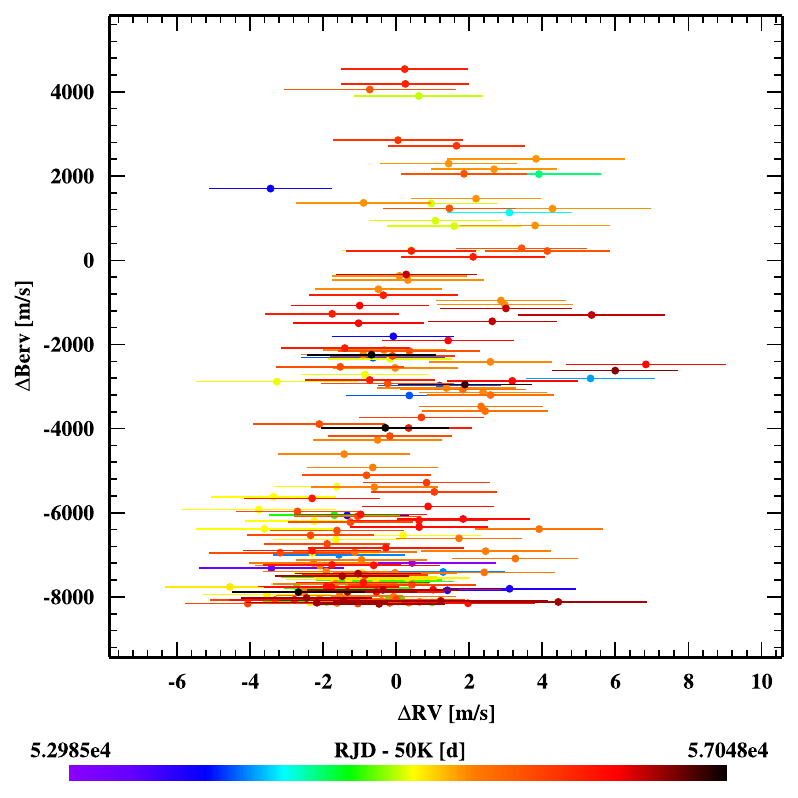} 
    \caption[]{{\emph Left:} Activity index \logrhk\,as a function of the RV residuals when removing all the detected signals except the magnetic cycle effect for {\footnotesize HD}\,20003. The observed correlation indicates that most of the RV residual variation is due to activity-related effects. {\emph Right:} Barycentric Earth velocity projected on the line of sight (BERV) as a function of the RV residuals around the best derived solution without considering the 184-day signal. The fact that no correlation can be observed disfavours the hypothesis that the 184-day signal is due to a discontinuity in the wavelength calibration introduced by tiny gaps between the different quadrants of the detector.}
\label{udry:fig4.1}
\end{figure*}

After this preliminary phase looking for significant signals in the data, we searched for the best-fit parameters with an MCMC, using a model composed of a second-order polynomial drift, a linear correlation of the RVs with \logrhk\ to adjust the magnetic cycle effect, three Keplerians to fit for the signals at 11.9, 33.9 and 184 days, and two jitters that correspond to the instrumental plus stellar noise at the minimum and maximum of the magnetic cycle (see Sec.\,\ref{sec:analysis}). This model converged to a stable solution. However, a significant signal in the residuals was still present near 3000 days (Fig.\,\ref{udry:fig3.1}), a period similar to the one of the magnetic cycle, an indication that our modeling of the long period variation was not satisfactory, very probably because the magnetic cycle of {\footnotesize HD}\,20003 is not fully covered, as seen in  Fig.\,\ref{udry:fig1}. We finally decided then to adjust a model composed of 3 Keplerians at 11.9, 33.9 and 184 days, an extra Keplerian at long-period to absorb the effect of the magnetic signal plus any long-term trend that could be present in the data, and one jitter corresponding to the instrumental plus stellar noise. In this case, the GLS periodogram of the residuals is extremely similar to the last plot in Fig.\,\ref{udry:fig3} and therefore no more significant signal is present in those RV residuals. Note that the period of the signal adjusting the magnetic cycle changed however from 4038 days (\logrhk) to 3298 days (RVs). Although we do not understand fully the long-term signal present in the RVs of {\footnotesize HD}\,20003, adjusting it with different models gives similar orbital parameters for the signals at 11.9, 33.9 and 184 days. We adopted as final solution the model with the magnetic cycle modeled by a long-period Keplerian with free parameters.

The best-fit for each inner planet, the signal at 184 days, and the RV residuals are displayed in Fig.\,\ref{udry:fig4}. The best-fit parameters can be found in Table\,\ref{HD20003_tab-mcmc-Summary_params}. A careful look at the activity indicators, \logrhk, BIS SPAN and FWHM, after removing the long-period signal induced by the stellar magnetic cycle, reveals no significant peaks that match the planetary signals found in our analysis (see Fig.\,\ref{app:figHD20003}).  We are therefore confident that those signals are not induced by stellar activity.

The correlation between the activity index \logrhk\ and the RV residuals when removing all the detected signals except the magnetic cycle effect can be seen in the left panel of Fig.\,\ref{udry:fig4.1}. Note that this dependence in activity is included in our model when adjusting the long-period Keplerian.

%O-C = 1.65
\begin{table*}[!t]
\caption{Best-derived solution for the planetary system orbiting HD20003. For each parameter, the median of the posterior distribution is considered, with error bars computed from the MCMC chains using a 68.3\% confidence interval. $\sigma_{O-C}$ corresponds to the weighted standard deviation of the residuals around this best solution. All the parameters probed by the MCMC can be found in Annex, in Table\,\ref{HD20003_tab-mcmc-Probed_params}}  
\label{HD20003_tab-mcmc-Summary_params}
\def\arraystretch{1.5}
\begin{center}
\begin{tabular}{lccccc}
\hline
\hline
Param. & Units & HD20003b & HD20003c & HD20003d? & magn. cycle \\
\hline
$P$ & [d] & 11.8482$_{-0.0017}^{+0.0016}$  & 33.9239$_{-0.0266}^{+0.0239}$  & 183.6225$_{-1.1057}^{+1.0051}$  & 4037.7931$_{-231.0018}^{+329.9338}$  \\
$K$ & [m\,s$^{-1}$] & 3.84$_{-0.21}^{+0.20}$  & 3.15$_{-0.20}^{+0.22}$  & 1.62$_{-0.20}^{+0.20}$  & 5.72$_{-0.31}^{+0.32}$  \\
$e$ &   & 0.36$_{-0.05}^{+0.05}$  & 0.10$_{-0.07}^{+0.08}$  & 0.13$_{-0.09}^{+0.11}$  & 0.06$_{-0.04}^{+0.06}$  \\
$\omega$ & [deg] & -92.74$_{-8.61}^{+8.52}$  & 53.33$_{-34.56}^{+45.86}$  & 52.14$_{-56.21}^{+55.14}$  & -113.61$_{-54.68}^{+75.59}$  \\
$T_P$ & [d] & 55489.5110$_{-0.2095}^{+0.2090}$  & 55493.2516$_{-3.2265}^{+4.4223}$  & 55426.5836$_{-27.4821}^{+26.7344}$  & 52787.4615$_{-899.1487}^{+1011.0883}$  \\
$T_C$ & [d] & 55483.7622$_{-0.3634}^{+0.3768}$  & 55496.2240$_{-0.8117}^{+0.7332}$  & 55442.1924$_{-5.5073}^{+5.3361}$  & 54968.5034$_{-61.9929}^{+60.5529}$  \\
\hline
$Ar$ & [AU] & 0.0974$_{-0.0039}^{+0.0034}$  & 0.1964$_{-0.0078}^{+0.0068}$  & 0.6053$_{-0.0239}^{+0.0213}$  & 4.7592$_{-0.2678}^{+0.3033}$  \\
M.$\sin{i}$ & [M$_{\rm Jup}$] & 0.0367$_{-0.0033}^{+0.0033}$  & 0.0454$_{-0.0044}^{+0.0046}$  & 0.0406$_{-0.0058}^{+0.0060}$  & 0.4093$_{-0.0388}^{+0.0421}$  \\
M.$\sin{i}$ & [M$_{\rm Earth}$] & 11.66$_{-1.06}^{+1.04}$  & 14.44$_{-1.39}^{+1.47}$  & 12.91$_{-1.84}^{+1.91}$  & 130.08$_{-12.32}^{+13.40}$  \\
\hline 
$\gamma_{HARPS}$ & [m\,s$^{-1}$] & \multicolumn{4}{c}{-16104.1107$_{-0.2826}^{+0.3698}$}\\
$\sigma_{(O-C)}$ & [m\,s$^{-1}$] & \multicolumn{4}{c}{1.65}\\
$\log{(\rm Post})$ &   & \multicolumn{4}{c}{-364.1867$_{-3.7161}^{+3.3177}$}\\
\hline
\end{tabular}
\end{center}
\end{table*}

\subsubsection{A 3rd planet with a period of 184 days?}
Although the signals at 11.9 and 33.9 days are clearly induced by planets orbiting around \object{HD\,20003}, the signal at 184 days is more difficult to interpret as it is very close to half a year, with strong aliases at 127 days and 359 days (close to 1/3 and 1 year).
As explained in Sec.~\ref{sec:analysis3}, signals at a year or harmonics of it can be induced by a discontinuity in the wavelength calibration introduced by tiny gaps between the different quadrants of the detector. However, the RV data that we analyze here have been corrected for this effect following the work of \citet{Dumusque-2015}. This correction has been already applied to the RVs of several stars, and has always been successful in removing the spurious signal. The fact that the RV residuals after removing the two planets, and the long-period effect of the magnetic cycle do not correlate with the barycentric Earth velocity (BERV, see right panel of Fig.\,\ref{udry:fig4.1}) disfavors the hypothesis that the 184-day signal is due to a discontinuity in the wavelength calibration introduced by tiny gaps between the different quadrants of the detector.
In conclusion, we would be inclined to claim that this signal at 184 days is a real planetary signal. However, although intensively tested, there is always the possibility that the data reduction system does not correct for all the instrumental effects in this particular case. We leave the signal as a potential one, and encourage other teams to confirm it using other pipelines or other facilities than {\footnotesize HARPS}.

\subsubsection{Discussion of the planetary system}
A surprising aspect of this system is the high eccentricity of the inner planet, $e$=0.38, while the planet at 33.9 days has an eccentricity close to 0. From a planet formation or a dynamical point of view, this is not straightforward to explain. There might be actually several possibilities to explain the observed configuration. It is out of the scope if this paper to check all of them in details, we just mention here the basic ideas. \\
\indent
1) The first and simplest possibility is that the high eccentricity of the planet at 11.9 days actually hides a planet at half the orbital period \citep{Anglada-Escude-2010}, in (or close to) a 2:1 resonance configuration. We therefore tried to fit a model with an extra planet with initial period at 5.95 days, setting the eccentricities of the 5.95 and 11.9-day planets to zero or letting them free to vary. Model comparison shows that, in both cases, the more complex solution is disfavoured with a $\Delta$BIC of 8.3 and 24.3, respectively. It seems therefore that the significant eccentricity of planet b is real. \\
\indent
2) The two planets are close to the 3:1 resonance. An increase of the eccentricity could have happened if, in the past, the two planets have been migrating trapped in the resonance. Then a specific event made the two planets go slightly out of resonance on the nowadays observed orbits. Possible scenarios for this event could be instability when the gas disappeared or the presence of an additional planet (e.g. resonant chain) that was ejected when the eccentricity increased (scattering).\\
\indent
3) Invoking additional bodies in the system raises the possibility of an external companion that could induce a Kozai-lidov effect on the inner planets in case of a large mutual inclination \citep[][]{Kozai-1962,Lidov-1961}. Such a companion could actually be hinted by our difficulty of modelling the long-period variation observed in the radial velocities. It is however not very clear to understand the differential effect on the two inner planets.\\
\indent
4) Finally, a very appealing potential explanation might rely on the spin-orbit evolution of planets feeling the tidal effect of the central star and the gravitational perturbation of another planetary companion. \citet{Correia-2012} have shown that "under some particular initial conditions, orbital and spin evolution cannot be dissociated and counter-intuitive behaviors can be observed, such as the secular increase of the eccentricity. This effect can last over long timescales and may explain the high eccentricities observed for moderate close-in planets".

%HD20781
\subsection{{\footnotesize HD}\,20781: A packed system with 2 super-Earths and 2 Neptune-mass planets}
\label{udry:hd20781}  

\object{HD\,20781} is part of a visual binary system including another star, \object{HD\,20782}, known to host a planet on a 595-day very eccentric orbit \citep{Jones-2006}. We take the opportunity of the discovery of a compact system of small planets around \object{HD\,20781} to provide here as well an updated solution for the planet around \object{HD\,20782} using {\footnotesize HARPS} data.

\subsubsection{Analysis of the {\footnotesize HD}\,20781 system}
\object{HD\,20781} was part of the original high-precision HARPS GTO survey and the star has been then followed for more than 11 years (4093 days). Over this time span, we gathered a total of 226 high signal-to-noise spectra ($<$S/N$>$ of 112 at 550 nm) corresponding in the end to 216 RV measurements binned over 1 hour. As reported in Table\,\ref{table:obsstat}, the typical precision of individual measurements is 0.76\,\ms\ including photon noise and calibration uncertainties. The raw RV rms is significantly higher, at 3.41\,\ms, pointing towards additional variations in the data of potentially stellar or planetary origin, assuming the instrumental effects are kept below the photon-noise level of the observations.

As a first approach we looked at the RV and activity index time series shown in Fig.\,\ref{udry:fig1}. No significant long-term variation is observed in \logrhk\ data and no long-term variation is visible in the GLS periodogram of the velocity time series. We conclude that there is no noticeable sign of a magnetic activity cycle for this star. The average value of \logrhk\ at $-5.03\pm0.01$ is also very low with a very small dispersion, similar to the Sun at minimum activity. 

\begin{figure*}[!h]
\center
  \includegraphics[angle=0,width=0.4\textwidth,origin=br]{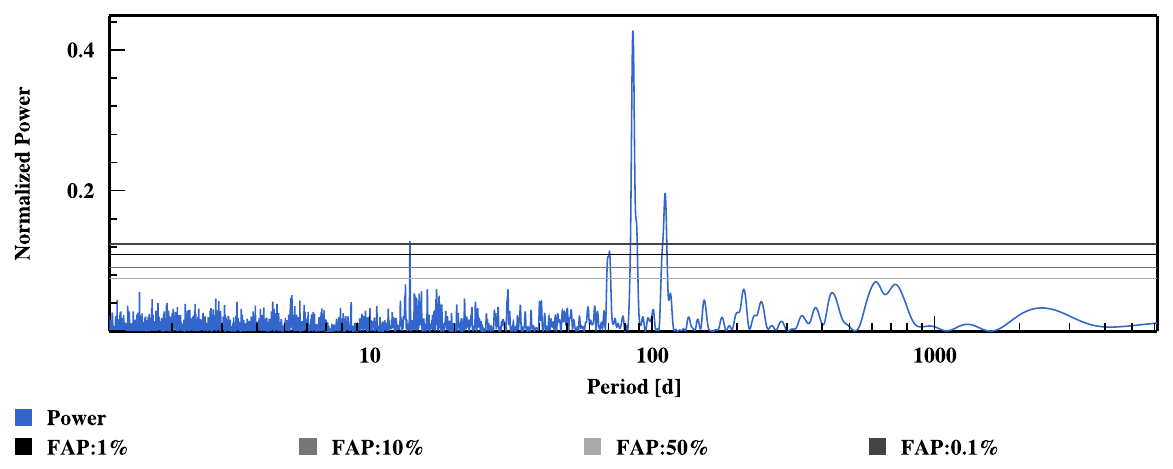} 
  \includegraphics[angle=0,width=0.4\textwidth,origin=br]{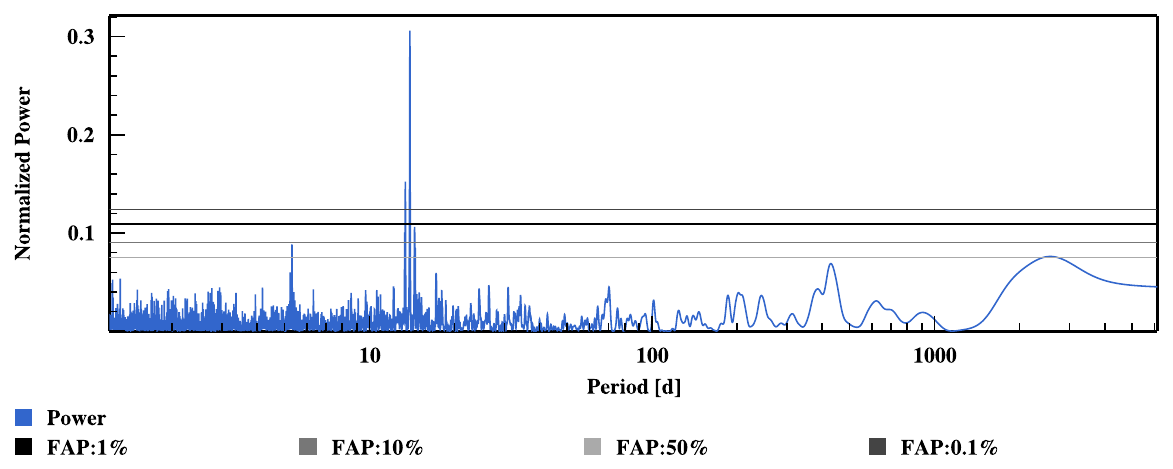} 
  \includegraphics[angle=0,width=0.4\textwidth,origin=br]{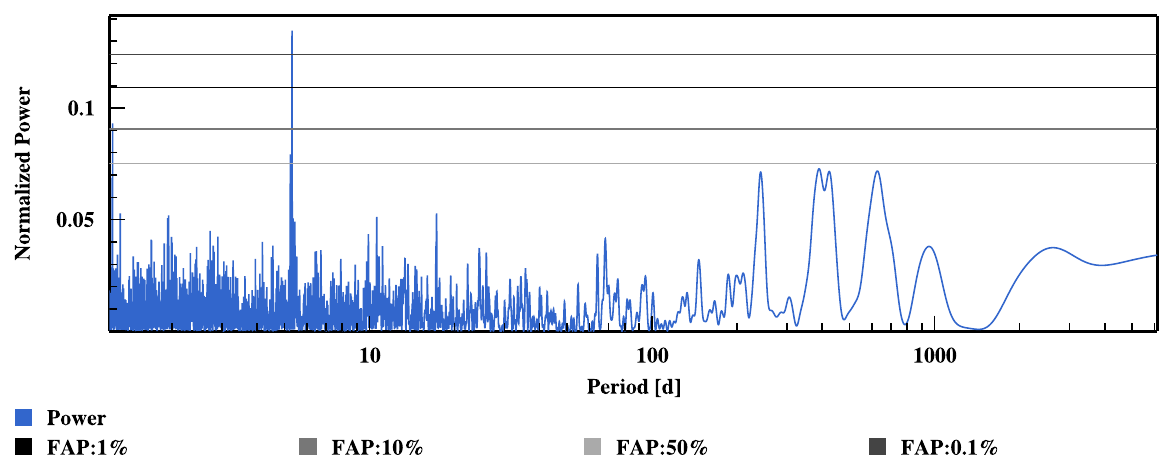} 
  \includegraphics[angle=0,width=0.4\textwidth,origin=br]{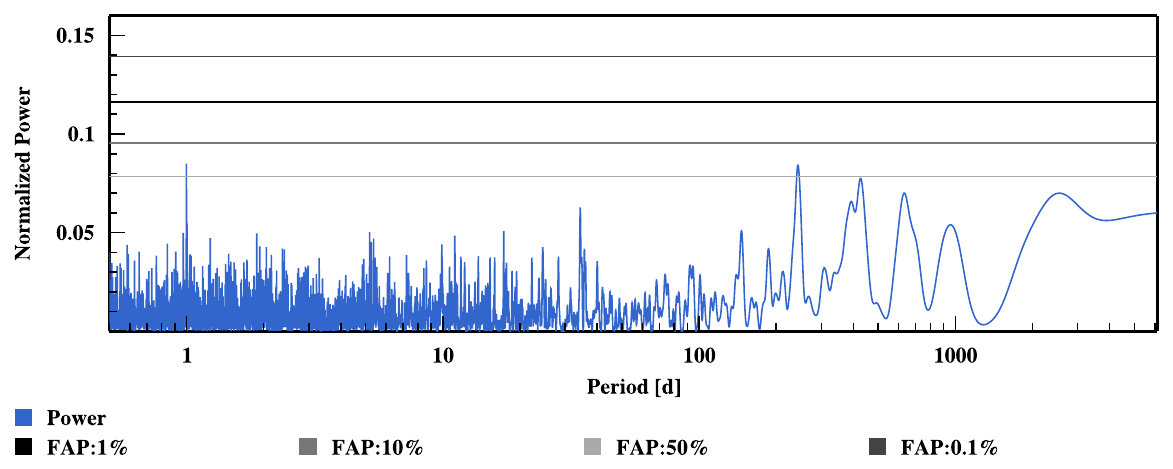} 
\caption[]{GLS periodogram of the residuals at each step, after removing one planet after the other in the analysis of {\footnotesize HD}\,20781 (from left to right and top to bottom). The GLS periodogram of the raw RVs is shown in Fig.~\ref{udry:fig1}.}
\label{udry:fig6}
\end{figure*}

Due to the small activity level and the large number of observations, the GLS periodogram of the velocity series is very clean, with significant peaks clearly coming out of the noise background. So, even if the final characterization of the planetary system parameters is performed through a Bayesian-based MCMC approach, a step-by-step analysis of the system, characterizing and then removing one planet after the other from the data, will provide an excellent illustration of the significance of the planet detection in this system. The most prominent peak in the GLS periodogram is at a period around 20 days. Deriving a Keplerian solution for this planet and then removing the corresponding signal from the raw RVs makes a clear signal at 86 days appear in the GLS periodogram of the residuals (Fig.\,\ref{udry:fig6}). Keeping on with the same approach, we can clearly identify, sequentially, significant signals first at 13.9 and then at 5.3 days. It has to be noted here that none of the significant periods in the data are close to the stellar rotational period, 46.8 days, estimated from the activity index (Table\,\ref{table:star}).

\begin{figure*}[!h]
\center
  \includegraphics[angle=0,width=0.4\textwidth,origin=br]{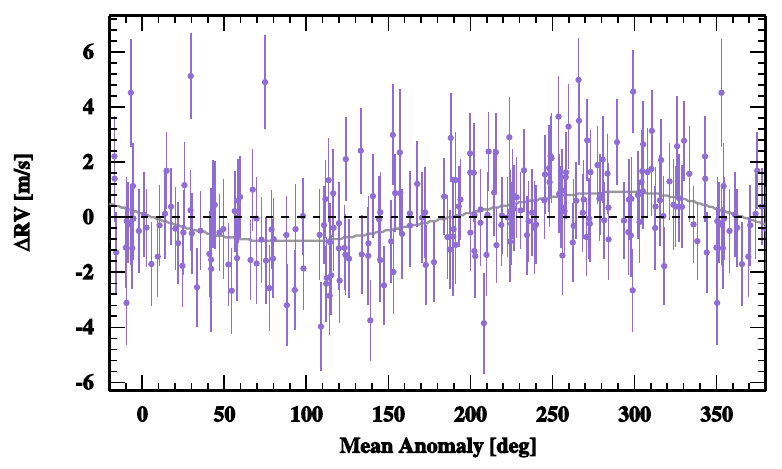} 
  \includegraphics[angle=0,width=0.4\textwidth,origin=br]{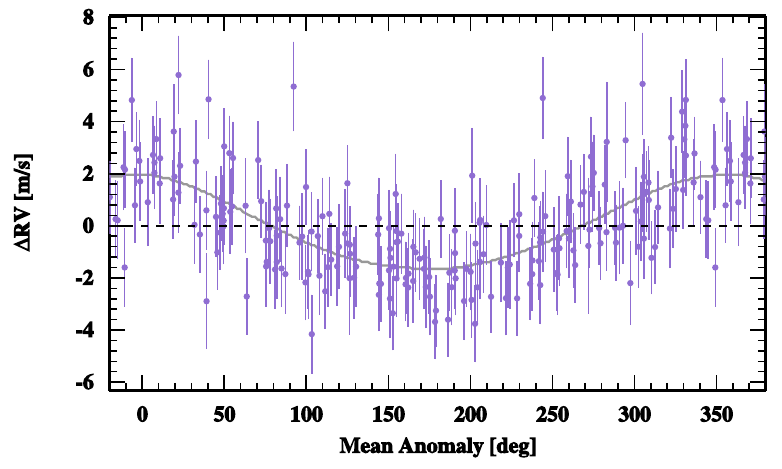} 
  \includegraphics[angle=0,width=0.4\textwidth,origin=br]{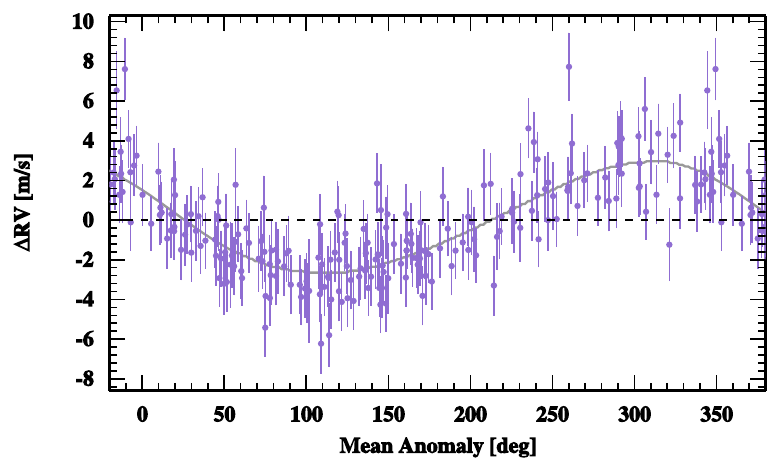} 
  \includegraphics[angle=0,width=0.4\textwidth,origin=br]{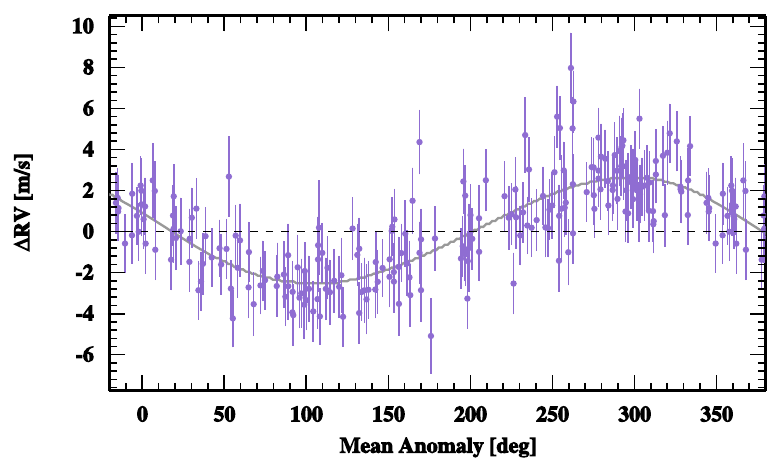} 
  \includegraphics[angle=0,width=0.4\textwidth,origin=br]{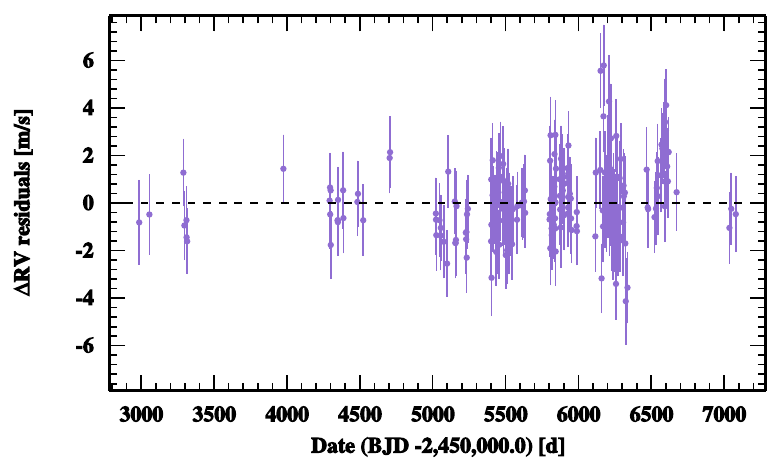} 
\caption[]{Phase-folded RV measurements of {\footnotesize HD}\,20781 with the best-fit solution represented as a black curve for each of the signal found in the data (from left to right and top to bottom: planet b, c, d and e). Error bars include photon and calibration noise, as well as instrumental + stellar jitters derived from the MCMC modeling. The residuals around the solution are displayed in the lower panel. Corresponding orbital elements are listed in Table\,\ref{HD20781_tab-mcmc-Summary_params}.}
\label{udry:fig7}
\end{figure*}

As described in Sec.\,\ref{sec:analysis}, the final determination of the planetary system parameters is performed using a MCMC sampler and a model composed of four Keplerians representing the planetary signals, and an extra white-noise jitter to consider potential stellar or instrumental noise not included in the RV error bars (the $\sigma_{JIT}$ parameter, see Table\,\ref{tab-mcmc-params}). Phase-folded planetary solutions, as well as the RV residuals around the best solution, are displayed in Fig.\,\ref{udry:fig7}. The best-fit parameters are reported in Table\,\ref{HD20781_tab-mcmc-Summary_params}. 

Looking at the periodograms of the \logrhk, the BIS SPAN and the FWHM in Fig.\,\ref{app:figHD20781}, to check if any announced planet matches any signal in the activity indicators, we see that the \logrhk\ times series presents signals at 115, 81 and 68 days, and the FWHM time series at 380 days. This later signal is probably due to interaction with the window functions, creating a signal near a year. The former signals are more difficult to interpret as they are not compatible with the estimated rotational period of the star, 46.8 days (Table\,\ref{table:star}). 
%Fitting the signal at 115 days removes the signal at 68 days, and the $p$-value of the signal at 81 days goes above 10\%.
In particular, we can note that the 86-day Neptune-mass planet has a period close to the detected peaks in the \logrhk\ time-series at 81 days, and furthermore is close to the 1-year alias of the 115-d signal (days). The amplitude of these signals however very low, at the level of $\sim$0.01\,dex, 20 times smaller than the solar magnetic cycle variation. Such signals can therefore probably not be responsible for the 2.6\,\ms\ periodic variation detected in the RVs at 86 days. 
%In addition, fitting the 115-day signal removes all power at 68 and 81 days, therefore we are confident that the 86-day signal detected in the RVs is associated to the presence of a Neptune-mass planet orbiting {\footnotesize HD}\,20781.

Our conclusion is thus that the system \object{HD\,20781} hosts two inner super-Earths with periods of 5.3 and 13.9 days and two outer Neptune-mass planets with periods of 29 and 86 days. With small eccentricities, stability is not a concern for this system of small-mass planets in a moderately compact configuration. 

% P0 = 29 jours
% P1 = 85
% P2 = 13
% P3 = 5.3

%HD20782
\subsubsection{{\footnotesize HD}\,20782: More planets in this visual binary system}

The star \object{HD\,20782} is the brightest companion of the HD\,20781-HD\,20782 binary system. It is known to harbor a 595-day very eccentric Jupiter planet \citep{Jones-2006}. A total of 71 high signal-to-noise spectra ($<$S/N$>$ of 181 at 550 nm) were obtained with {\footnotesize HARPS} on this target, which translates into 68 RV measurements after binning the data over 1 hour. Because of the very large semi-amplitude and eccentricity of the RV signal induced by the planet, we decided to include in addition to the {\footnotesize HARPS} very precise measurements the lower precision RV data obtained with UCLES \citep[published in][]{Jones-2006} and {\footnotesize CORALIE}.

\begin{figure}[t!]
\center
  \includegraphics[angle=0,width=0.4\textwidth,origin=br]{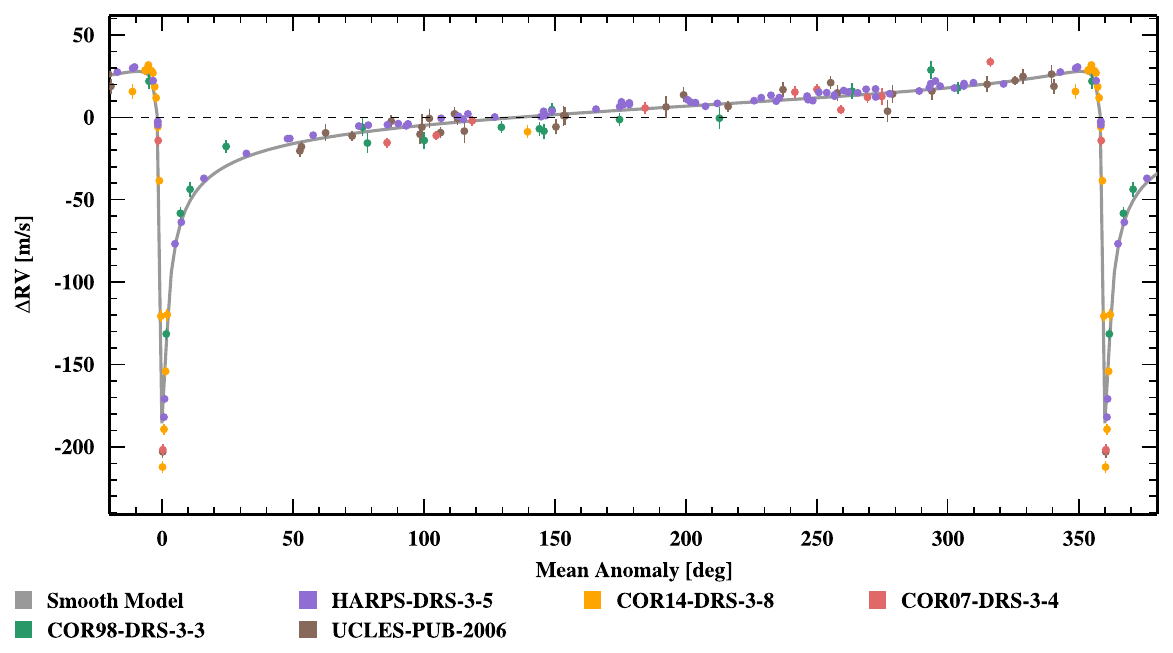} 
  \includegraphics[angle=0,width=0.4\textwidth,origin=br]{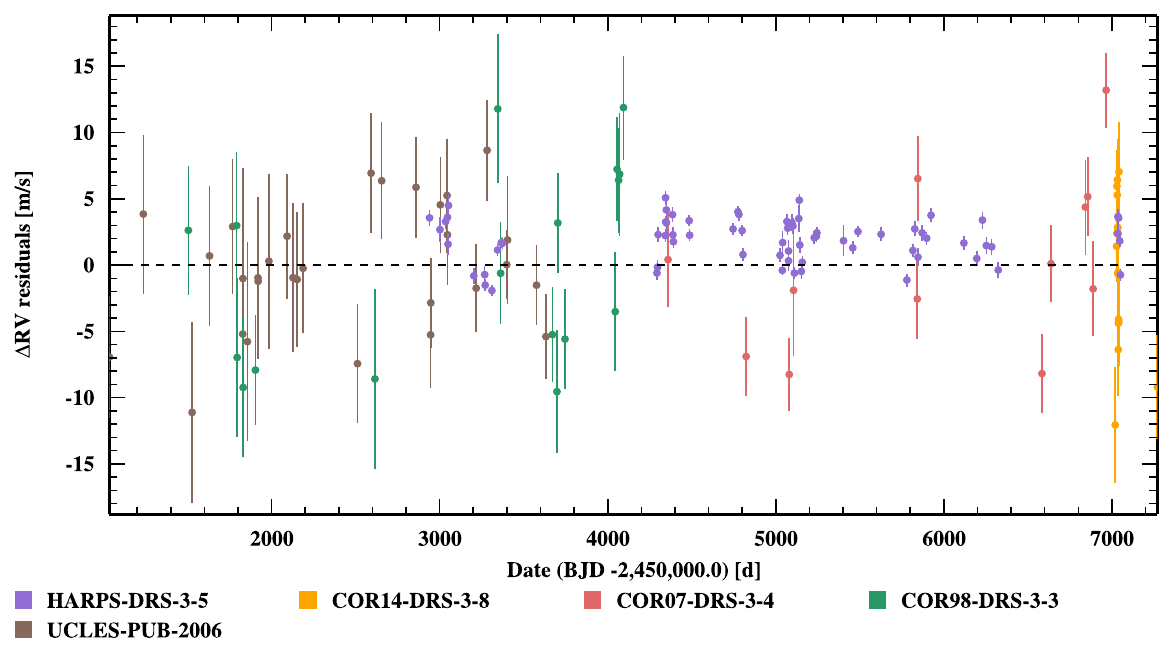} 
\caption[]{Best keplerian solution for the eccentric planet orbiting {\footnotesize HD}\,20782. \emph{Top:} Phase-folded RV measurements with the best solution represented as a black curve. \emph{Bottom:} RV residuals around the best Keplerian solution. 
%\emph{Bottom:} GLS periodogram of the RV residuals. Corresponding orbital elements are listed in Table \ref{HD20781_tab-mcmc-Summary_params}.
}
\label{udry:fig8}
\end{figure}

We searched for the best-fit parameters using a MCMC sampler. The solution converges to an extremely eccentric Jupiter-mass planet with a period of 597 days. The phase-folded planetary solution, as well as the RV residuals around the best solution and their GLS periodogram, are displayed in Fig.\,\ref{udry:fig8}. The best-fit parameters are reported in Table\,\ref{HD20781_tab-mcmc-Summary_params}. Although the eccentricity, the amplitude and the argument of periastron are compatible within one sigma with the values presented in \citet{Jones-2006}, the period and argument of periastron are not compatible, even if different by less than 2\%. This can be explained by the addition of the {\footnotesize CORALIE} and  {\footnotesize HARPS} data that allows to sample much better the periastron passage.

The high eccentricity of \object{HD\,20782}\,b has commonly been explained via the Kozai-Lidov mechanism \citep[][]{Kozai-1962,Lidov-1961} induced by \object{HD\,20781}.

%
% O-C = 1.45
\begin{table*}
\caption{Best-fitted solution for the planetary system orbiting {\footnotesize HD}\,20781 and {\footnotesize HD}\,20782. For each parameter, the median of the posterior distribution is considered, with error bars computed from the MCMC posteriors using a 68.3\,\% confidence interval. The value $\sigma_{(O-C)}$ corresponds to the weighted standard deviation of the residuals around the best solution. All the parameters probed by the MCMC can be found in Annex, in Tables\,\ref{HD20781_tab-mcmc-Probed_params} and \ref{HD20782_tab-mcmc-Probed_params}. See Table\,\ref{tab-mcmc-params} for definition of the parameters.}  
\label{HD20781_tab-mcmc-Summary_params}
\def\arraystretch{1.5}
\begin{center}
\begin{tabular}{lccccc|c}
\hline
\hline
Param. & Units & HD20781b & HD20781c & HD20781d & HD20781e & HD20782b \\
\hline
$P$ & [d] & 5.3135$_{-0.0010}^{+0.0010}$  & 13.8905$_{-0.0034}^{+0.0033}$  & 29.1580$_{-0.0100}^{+0.0102}$  & 85.5073$_{-0.0947}^{+0.0983}$  &  597.0643$_{-0.0256}^{+0.0256}$\\
$K$ & [m\,s$^{-1}$] & 0.91$_{-0.15}^{+0.15}$  & 1.81$_{-0.16}^{+0.16}$  & 2.82$_{-0.16}^{+0.17}$  & 2.60$_{-0.14}^{+0.14}$ & 118.43$_{-1.78}^{+1.78}$ \\
$e$ &   & 0.10$_{-0.07}^{+0.11}$  & 0.09$_{-0.06}^{+0.09}$  & 0.11$_{-0.06}^{+0.05}$  & 0.06$_{-0.04}^{+0.06}$ & 0.95$_{-0.001}^{+0.001}$  \\
$\omega$ & [deg] & 84.04$_{-108.41}^{+141.70}$  & 7.44$_{-70.21}^{+53.87}$  & 60.99$_{-30.03}^{+30.79}$  & 70.59$_{-67.58}^{+61.40}$ & 143.58$_{-0.66}^{+0.56}$ \\
$T_P$ & [d] & 55503.2027$_{-1.5934}^{+2.0979}$  & 55503.5204$_{-2.6815}^{+2.0425}$  & 55511.3258$_{-2.4382}^{+2.4394}$  & 55513.3912$_{-16.0090}^{+14.3623}$ & 55247.0150$_{-0.0837}^{+0.0770}$ \\
$T_C$ & [d] & 55503.2888$_{-0.2384}^{+0.2361}$  & 55506.4386$_{-0.4385}^{+0.3546}$  & 55513.2432$_{-0.4410}^{+0.4250}$  & 55517.4714$_{-1.4838}^{+1.3596}$ & 55246.1712$_{-0.0706}^{+0.0706}$ \\
\hline
$Ar$ & [AU] & 0.0529$_{-0.0027}^{+0.0024}$  & 0.1004$_{-0.0051}^{+0.0046}$  & 0.1647$_{-0.0083}^{+0.0076}$  & 0.3374$_{-0.0170}^{+0.0155}$  & 1.3649$_{-0.0495}^{+0.0466}$\\
M.$\sin{i}$ & [M$_{\rm Jup}$] & 0.0061$_{-0.0011}^{+0.0012}$  & 0.0168$_{-0.0021}^{+0.0022}$  & 0.0334$_{-0.0037}^{+0.0038}$  & 0.0442$_{-0.0049}^{+0.0049}$ & 1.4878$_{-0.1066}^{+0.1045}$ \\
M.$\sin{i}$ & [M$_{\rm Earth}$] & 1.93$_{-0.36}^{+0.39}$  & 5.33$_{-0.67}^{+0.70}$  & 10.61$_{-1.19}^{+1.20}$  & 14.03$_{-1.56}^{+1.56}$ & 472.83$_{-33.88}^{+33.22}$ \\
\hline 
$\mathrm{offset}_{UCLES}$ &  [m\,s$^{-1}$] &  & & & &\multicolumn{1}{c}{5.4431$_{-0.9370}^{+0.9302}$}\\
$\gamma_{COR98}$ &  [m\,s$^{-1}$] & & & & & \multicolumn{1}{c}{39928.1039$_{-1.9160}^{+1.8747}$}\\
$\gamma_{COR07}$ & [m\,s$^{-1}$] & & & & &  \multicolumn{1}{c}{39930.6146$_{-2.1696}^{+2.0828}$}\\
$\gamma_{COR14}$ & [m\,s$^{-1}$] & & & & & \multicolumn{1}{c}{39956.5569$_{-1.6638}^{+1.5850}$}\\
$\gamma_{HARPS}$ & [m\,s$^{-1}$] & \multicolumn{4}{c}{40369.2080$_{-0.1104}^{+0.1147}$} & 39964.8070$_{-0.2275}^{+0.2193}$ \\
$\sigma_{(O-C)}$ & [m\,s$^{-1}$] & \multicolumn{4}{c}{1.45} & 2.34\\
$\log{(\rm Post})$ &   & \multicolumn{4}{c}{-397.1395$_{-3.8525}^{+3.0943}$} & -379.6596$_{-3.2805}^{+2.6521}$\\
\hline
\end{tabular}
\end{center}
\end{table*}

%\subsubsection{Kozai-Lidov mechanism in the visual binary system}
%{\bf 
%The reciprocal effect is then probably expected as well but is however not observed for the compact system around \object{HD\,20781}. A simple way to explain this non-symmetry relies on the need for the perturbing companion to be significantly out of the orbital plane of the eccentric system for the Kozai-Lidov mechanism to be active, by more than ~40 degrees. So \object{HD\,20781} is probably more than 40 degrees out of the orbital plane of \object{HD\,20782}\,b, and this is not the case for \object{HD\,20782} with regard to the planes of the orbits of the planets around \object{HD\,20781}. We can then postulate that the mutual inclination between the orbital planes of the planets orbiting the two stars are smaller than 40 degrees.
%}

\begin{figure}[!h]
\center
  \includegraphics[angle=0,width=0.4\textwidth,origin=br]{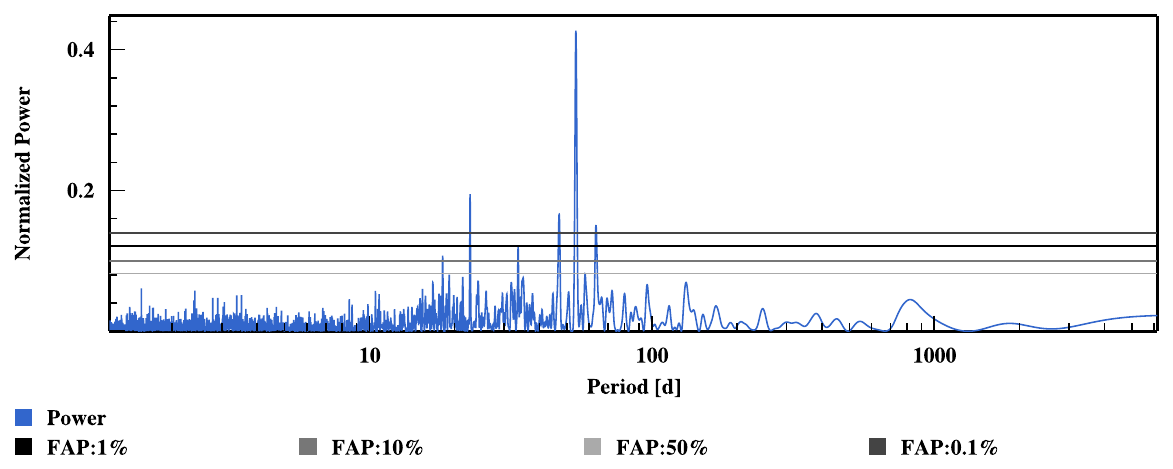} 
  \includegraphics[angle=0,width=0.4\textwidth,origin=br]{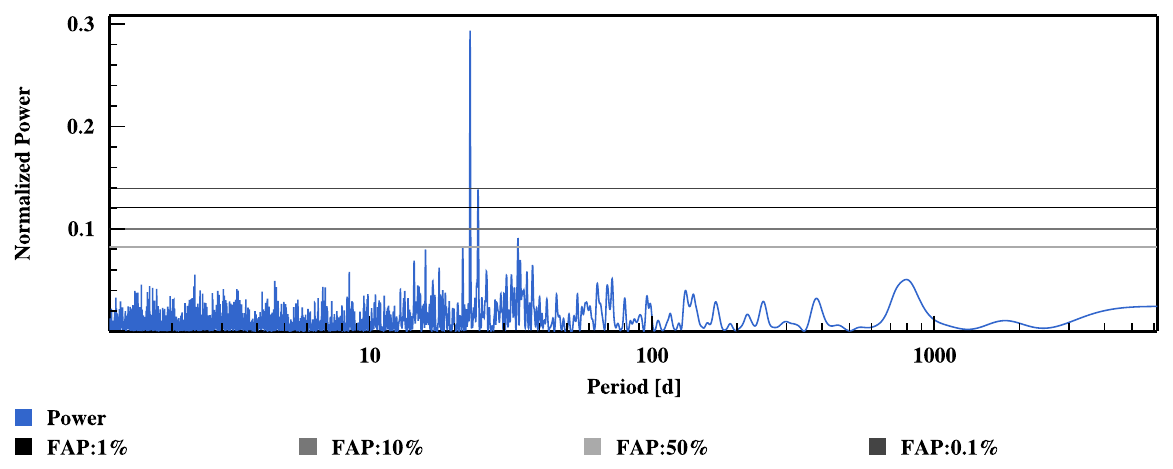} 
  \includegraphics[angle=0,width=0.4\textwidth,origin=br]{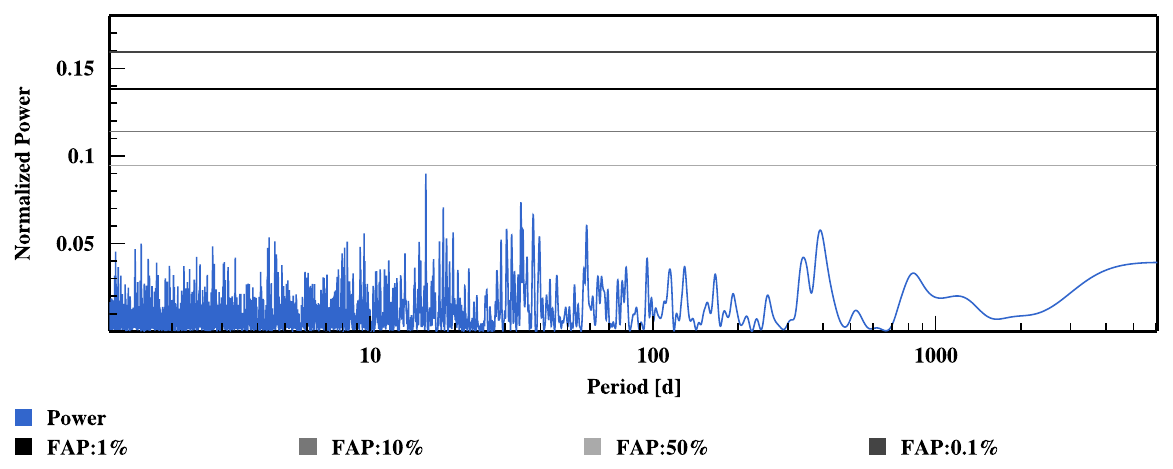} 
\caption[]{\emph{Top:} GLS periodogram of the residual RVs of {\footnotesize HD}\,21693 after removing the RV contribution of the magnetic cycle. \emph{Middle and bottom: } GLS periodogram of the residuals at each step, after removing one planet after the other in the analysis. The GLS periodogram of the raw RVs is shown in Fig.\,\ref{udry:fig1}.}
\label{udry:fig9}
\end{figure}

%HD21693
\subsection{{\footnotesize HD}\,21693: A system of 2 Neptune-mass planets close to a 5:2 resonance}
\label{udry:hd21693}  

Over a time span of 11 years (4106 days), 212 high signal-to-noise spectra ($<$S/N$>$ of 141 at 550\,nm) of \object{HD\,21693} were gathered, resulting in a total of 210 RV measurements when binning the data over 1 hour. The typical photon-noise and calibration uncertainty is 0.60\,\ms, which is significantly below the 4.72\,\ms\ observed dispersion of the RVs, pointing towards the existence of additional signals in the data. The raw RVs, their GLS periodogram and the calcium activity index of \object{HD\,21693} are shown in Fig.\,\ref{udry:fig1}. The variation of the \logrhk\ ranging from -5.02 to -4.83 highlights a significant magnetic cycle. This magnetic cycle is very similar in magnitude to the one of the Sun, but slightly shorter with a period of 10 years. When fitting a Keplerian signal to the \logrhk, we are left with significant signals in the residuals at 740 and 33.5 days (see Fig.\,\ref{app:figHD21693}). Those signals are also present in the FWHM and the bisector span of the CCF, although less significant. The signal at $\sim$740 days, close to two years is probably due to the sampling of the data, and the 33.5-day signal is likely the stellar rotation period. This value is compatible with the rotation period derived from the mean activity \logrhk\ level, i.e. 36 days (Table\,\ref{table:star}).

Looking at the raw RVs and their GLS periodogram in Fig.\,\ref{udry:fig1}, it is clear that the observed magnetic cycle has an impact on the measured RV measurements. To remove the RV contribution of the magnetic cycle, we remove from the RVs a Keplerian model that has the same parameters as the Keplerian fitted to \logrhk\ but with an amplitude free to vary. The GLS periodogram of the RV residuals after correcting for the magnetic cycle effect are displayed in the top panel of Fig.\,\ref{udry:fig9}. A highly significant signal at 54 days is present in the data. When removing this signal by fitting a Keplerian, an extra signal at 23 days is seen in the residuals (middle panel of Fig.\,\ref{udry:fig9}). No signal with $p$-value smaller than 1\,\% appears in the residuals of a two-Keplerian model; we therefore stop looking for extra signals. Note however that the most significant signal left in the GLS periodogram corresponds to a period of 16 days, likely the first harmonic of the stellar rotation period, which is expected from stellar activity \citep[][]{Boisse-2011}. 

\begin{figure}[!h]
\center
  \includegraphics[angle=0,width=0.4\textwidth,origin=br]{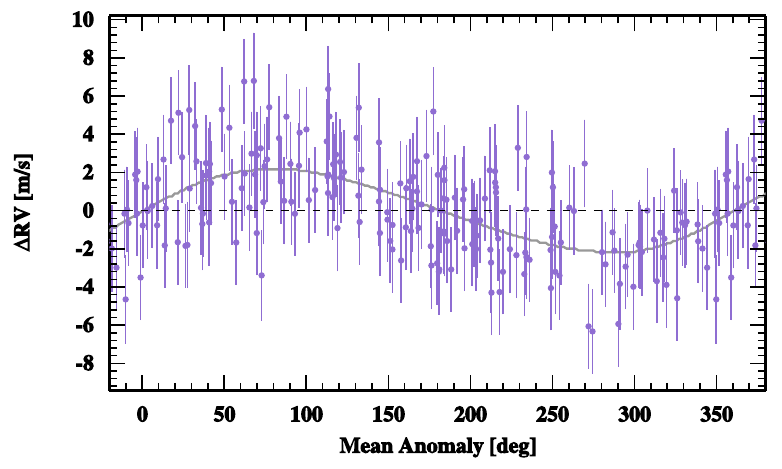} 
  \includegraphics[angle=0,width=0.4\textwidth,origin=br]{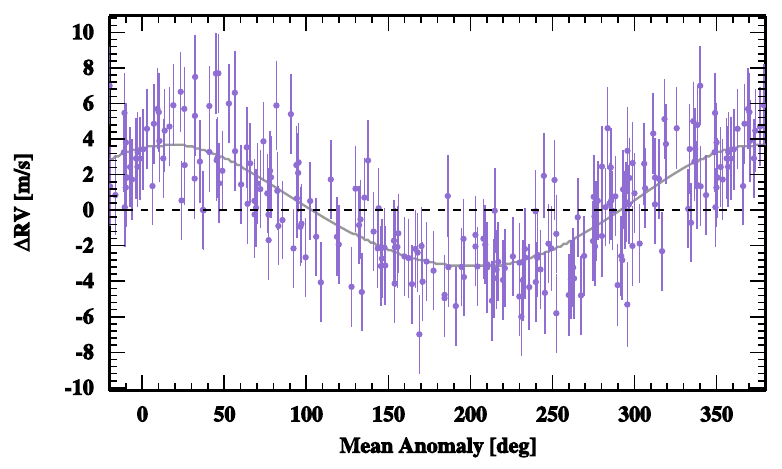} 
    \includegraphics[angle=0,width=0.4\textwidth,origin=br]{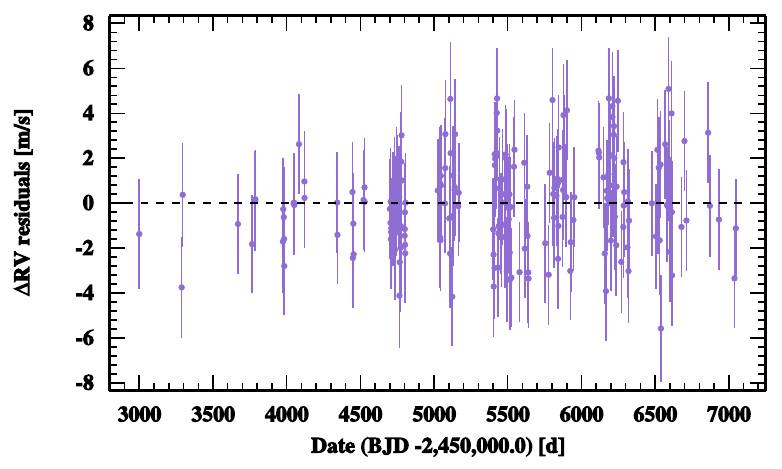}  
\caption[]{Phase-folded  RV measurements of {\footnotesize HD}\,21693 with the best planet solution represented as a black curve for each of the signal in the data (top to bottom: planet b and planet c). The residuals around the solution are displayed in the lower panel. Corresponding orbital elements are listed in Table\,\ref{HD21693_tab-mcmc-Summary_params}.}
\label{udry:fig10}
\end{figure}

After this preliminary phase looking for significant signals in the data, we search for the best-fit parameters with an MCMC, using a model composed of a linear correlation of the RVs with \logrhk\ to adjust the magnetic cycle effect, two Keplerian functions for the signals at 23 and 54 days, and two jitter values correspond to the instrumental plus stellar noise at the minimum and maximum of the magnetic cycle (see Sec.\,\ref{sec:analysis}). The best-fit for each planet and the RV residuals are displayed in Fig.\,\ref{udry:fig10}, and the best-fit parameters can be found in Table \ref{HD21693_tab-mcmc-Summary_params}. None of the two detected planet corresponds to signals found in the activity indicators (see Fig.\,\ref{app:figHD21693}).

When looking at the RV residuals, the scatter is still rather high even after removing all the significant signal detected in the data. Although in Hipparcos the star is catalogued as a G8 dwarf, the spectroscopic survey of nearby stars NSTAR finds that that \object{HD\,21693} is a G9IV-V therefore a slightly evolved star \citep[][]{Gray-2006}. Evolved stars presents higher photometric and RV jitter associated with more significant granulation, which might explain this significant residual jitter \citep[][]{Bastien-2014,Bastien-2013,Dumusque-2011:a}.

\begin{figure}[!h]
 \center
  \includegraphics[angle=0,width=0.4\textwidth,origin=br]{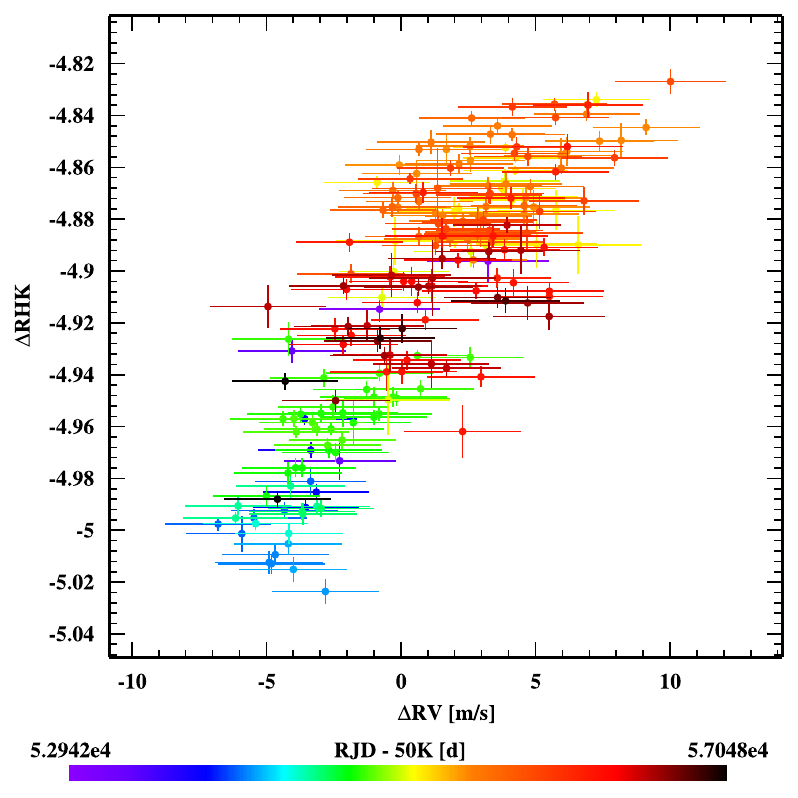} 
    \caption[]{RV residuals when removing all the detected signals except the magnetic cycle effect plotted as a function of the activity index \logrhk\ for {\footnotesize HD}\,21693. The observed correlation indicates that most of the RV residual variation is due to activity-related effects.}
\label{udry:fig10.1}
\end{figure}

The correlation between the activity index \logrhk\ and the RV residuals removing only the two-planet solution can be seen in Fig.\,\ref{udry:fig10.1}. Note that this correlation is considered in the MCMC model we used to fit the RV data.

Our analysis of the {\footnotesize HARPS} RV measurements of \object{HD\,21693} provides strong evidence that 2 Neptune-mass planets orbit the star, with periods of 22.7 and 53.7 days. With such periods, this planetary system is close to a 5:2 resonance but with a period ratio smaller than the 5:2 commensurability. If the system formed through a convergent migration scenario trapping the planets in the 5:2 resonance, some process has to be invoked to push the system out of the resonance configuration after the disappearance of the protoplanetary disk.

%P0 = 53 days
%P1 = 22 days

%O-C = 2.05
\begin{table}
\caption{Best-fitted solution for the planetary system orbiting {\footnotesize HD}\,21693. For each parameter, the median of the posterior is considered, with error bars computed from the MCMC chains using a 68.3\,\% confidence interval. $\sigma_{O-C}$ corresponds to the weighted standard deviation of the residuals around this best solutions. All the parameters probe by the MCMC can be found in Annex, in Table\,\ref{HD21693_tab-mcmc-Probed_params}.}  
\label{HD21693_tab-mcmc-Summary_params}
\def\arraystretch{1.5}
\begin{center}
\begin{tabular}{lccc}
\hline
\hline
Param. & Units & HD21693b & HD21693c \\
\hline
$P$ & [d] & 22.6786$_{-0.0087}^{+0.0085}$  & 53.7357$_{-0.0309}^{+0.0312}$  \\
$K$ & [m\,s$^{-1}$] & 2.20$_{-0.22}^{+0.22}$  & 3.44$_{-0.20}^{+0.20}$  \\
$e$ &   & 0.12$_{-0.08}^{+0.09}$  & 0.07$_{-0.05}^{+0.06}$  \\
$\omega$ & [deg] & -91.04$_{-50.35}^{+50.57}$  & -17.34$_{-55.35}^{+47.50}$  \\
$T_P$ & [d] & 55492.0549$_{-3.1310}^{+3.0968}$  & 55528.4064$_{-8.1820}^{+7.0762}$  \\
$T_C$ & [d] & 55480.7554$_{-0.6973}^{+0.7454}$  & 55543.5822$_{-1.2491}^{+1.0676}$  \\
\hline
$Ar$ & [AU] & 0.1455$_{-0.0063}^{+0.0058}$  & 0.2586$_{-0.0113}^{+0.0103}$  \\
M.$\sin{i}$ & [M$_{\rm Jup}$] & 0.0259$_{-0.0033}^{+0.0034}$  & 0.0547$_{-0.0056}^{+0.0056}$  \\
M.$\sin{i}$ & [M$_{\rm Earth}$] & 8.23$_{-1.05}^{+1.08}$  & 17.37$_{-1.79}^{+1.77}$  \\
\hline 
$\gamma_{HARPS}$ & [m\,s$^{-1}$] & \multicolumn{2}{c}{39768.8113$_{-0.1471}^{+0.1425}$}\\
$\sigma_{(O-C)}$ & [m\,s$^{-1}$] & \multicolumn{2}{c}{2.05}\\
$\log{(\rm Post})$ &   & \multicolumn{2}{c}{-440.9160$_{-3.1188}^{+2.2958}$}\\
\hline
\end{tabular}
\end{center}
\end{table}

%HD31527
\subsection{{\footnotesize HD}\,31527: A 3-Neptune system}
\label{udry:hd31527}  

In total, 257 high signal-to-noise spectra ($<$S/N$>$ of 180 at 550\,nm) of \object{HD\,31527} were gathered over a time span of 11 years (4135 days). This results in 245 observations of the star when data are binned over 1 hour. The 3.19 \ms observed dispersion of the RVs is much larger than the typical RV precision of 0.64\,\ms, pointing towards the existence of extra signals in the data. The raw RVs, their GLS periodogram and the calcium activity index of \object{HD\,31527} are shown in Fig.\,\ref{udry:fig1}. The calcium activity index \logrhk\ does not show any significant variation as a function of time, therefore we do not expect the RVs to be affected by long-period signals generally induced by magnetic cycles. The mean activity level of the star is equal to \logrhk$=-4.96$, close to solar minimum. The RVs should therefore be exempt as well of activity signal at the rotational period of the star and its harmonics due to active regions on the stellar surface \citep[][]{Boisse-2011}. This is confirmed when looking at the periodograms of the different activity indicators in Fig.\,\ref{app:figHD31527}. Only a signal at 400 days in the BIS SPAN is significant. This signal, not too far from a year, might be due the interplay between the time series and the window function.

\begin{figure}[!h]
\center
% already in Fig.2           \includegraphics[angle=0,width=0.4\textwidth,origin=br]{udry_fig11_s5per-step0.pdf} 
  \includegraphics[angle=0,width=0.4\textwidth,origin=br]{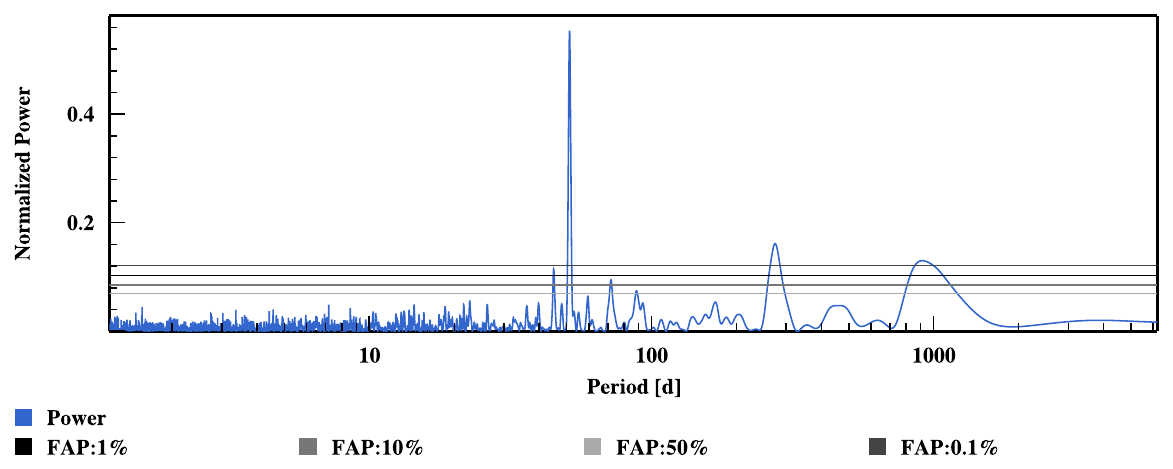} 
  \includegraphics[angle=0,width=0.4\textwidth,origin=br]{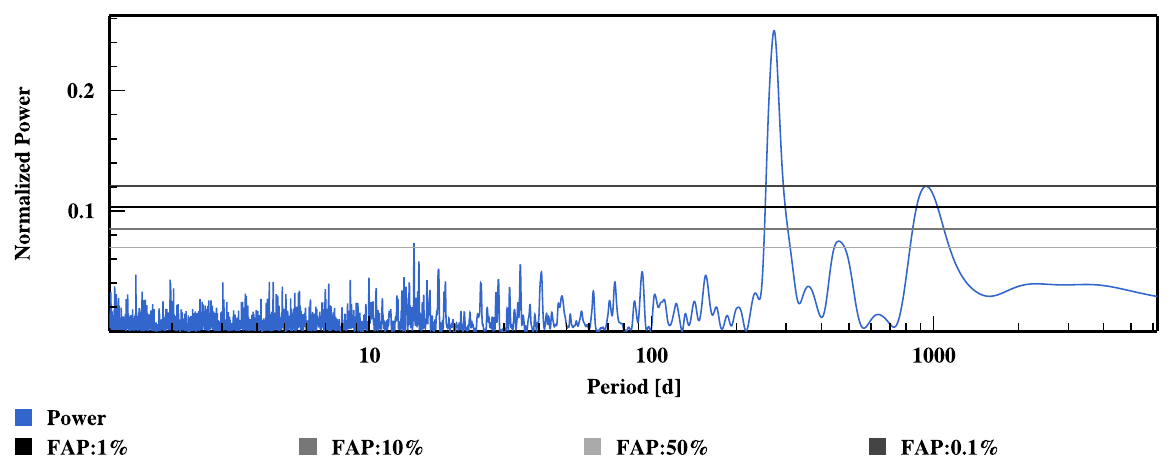} 
  \includegraphics[angle=0,width=0.4\textwidth,origin=br]{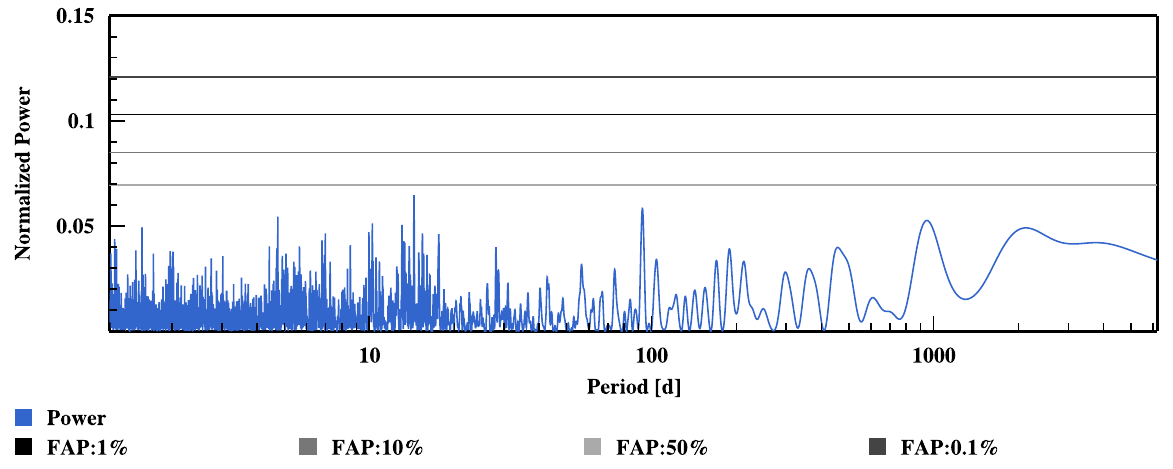} 
    \caption[]{\emph{Top:} GLS periodogram of the residual RVs of {\footnotesize HD}\,31527 after removing the best-fit Keplerian to account for the significant signal at 17 days seen in the raw RVs (bottom right panel of Fig.\,\ref{udry:fig1}). \emph{Middle and bottom: } GLS periodogram of the residuals at each step, after removing the second and third planet from the RVs.}
\label{udry:fig11}
\end{figure}

%P0 = 16 days
%P1 = 51 days
%P2 = 271 days
\begin{figure*}[!h]
\center
  \includegraphics[angle=0,width=0.4\textwidth,origin=br]{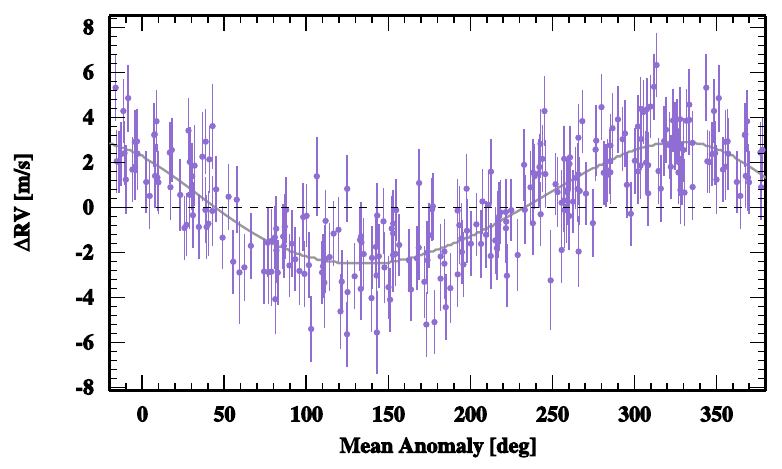} 
  \includegraphics[angle=0,width=0.4\textwidth,origin=br]{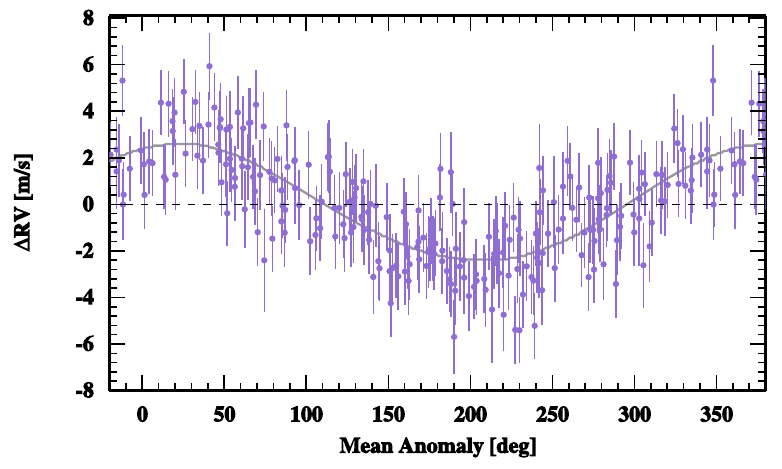} 
  \includegraphics[angle=0,width=0.4\textwidth,origin=br]{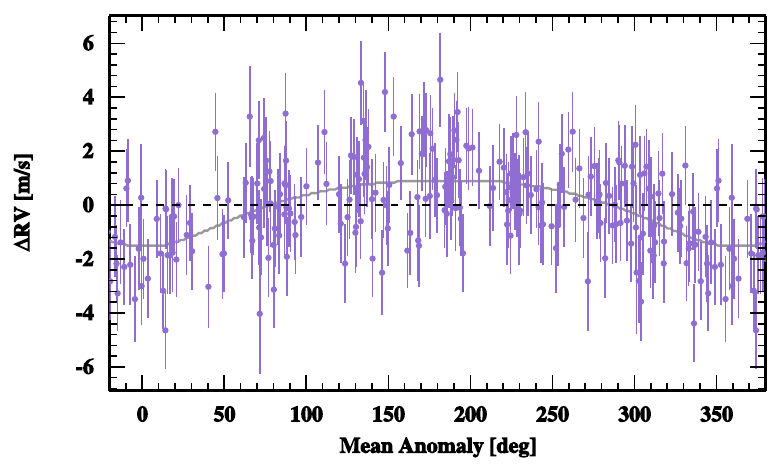} 
  \includegraphics[angle=0,width=0.4\textwidth,origin=br]{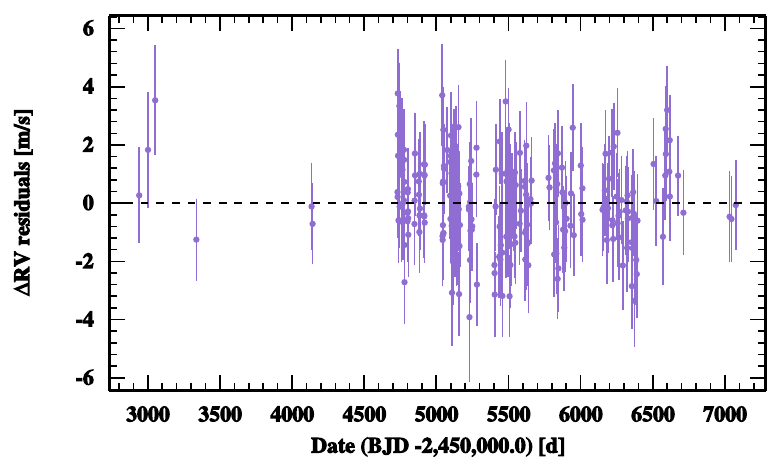} 
    \caption[]{Phase-folded  RV measurements of {\footnotesize HD}\,31527 with the best Keplerian solution for planet b, c, d and e represented as black curves (from left to right and top to bottom). Error bars include photon and calibration noise, as well as a jitter estimated from the MCMC analysis. The residuals around the solution are displayed in the bottom right panel. Corresponding orbital elements are listed in Table\,\ref{HD31527_tab-mcmc-Summary_params}.}
\label{udry:fig12}
\end{figure*}

%O-C = 1.41
\begin{table*}[!h]
\caption{Best-fitted solution for the planetary system orbiting {\footnotesize HD}\,31527. For each parameter, the median of the posterior distribution is considered, with error bars computed from the MCMC chains using a 68.3\,\% confidence interval. $\sigma_{O-C}$ corresponds to the weighted standard deviation of the residuals around the best solution. All the parameters probeb by the MCMC can be found in Annex, in Table\,\ref{HD31527_tab-mcmc-Probed_params}.}  
\label{HD31527_tab-mcmc-Summary_params}
\def\arraystretch{1.5}
\begin{center}
\begin{tabular}{lcccc}
\hline
\hline
Param. & Units & HD31527b & HD31527c & HD31527d \\
\hline
$P$ & [d] & 16.5535$_{-0.0035}^{+0.0034}$  & 51.2053$_{-0.0368}^{+0.0373}$  & 271.6737$_{-2.2471}^{+2.1135}$  \\
$K$ & [m\,s$^{-1}$] & 2.72$_{-0.13}^{+0.13}$  & 2.51$_{-0.14}^{+0.14}$  & 1.25$_{-0.16}^{+0.17}$  \\
$e$ &   & 0.10$_{-0.05}^{+0.05}$  & 0.04$_{-0.03}^{+0.05}$  & 0.24$_{-0.13}^{+0.13}$  \\
$\omega$ & [deg] & 41.12$_{-34.41}^{+29.46}$  & -23.25$_{-152.43}^{+68.94}$  & 179.00$_{-26.20}^{+31.11}$  \\
$T_P$ & [d] & 55499.5453$_{-1.5818}^{+1.3130}$  & 55526.3434$_{-21.5943}^{+9.7876}$  & 55718.8091$_{-17.3759}^{+21.3880}$  \\
$T_C$ & [d] & 55501.4585$_{-0.2667}^{+0.2640}$  & 55542.0635$_{-0.9373}^{+0.7471}$  & 55670.8324$_{-13.7509}^{+11.7715}$  \\
\hline
$Ar$ & [AU] & 0.1254$_{-0.0045}^{+0.0041}$  & 0.2663$_{-0.0095}^{+0.0088}$  & 0.8098$_{-0.0293}^{+0.0273}$  \\
M.$\sin{i}$ & [M$_{\rm Jup}$] & 0.0329$_{-0.0028}^{+0.0028}$  & 0.0445$_{-0.0039}^{+0.0040}$  & 0.0372$_{-0.0052}^{+0.0053}$  \\
M.$\sin{i}$ & [M$_{\rm Earth}$] & 10.47$_{-0.87}^{+0.89}$  & 14.16$_{-1.23}^{+1.28}$  & 11.82$_{-1.64}^{+1.70}$  \\
\hline 
$\gamma_{HARPS}$ & [m\,s$^{-1}$] & \multicolumn{3}{c}{25739.7025$_{-0.0971}^{+0.0952}$}\\
$\sigma_{(O-C)}$ & [m\,s$^{-1}$] & \multicolumn{3}{c}{1.41}\\
$\log{(\rm Post})$ &   & \multicolumn{3}{c}{-439.4138$_{-3.3929}^{+2.7350}$}\\
\hline
\end{tabular}
\end{center}
\end{table*}

As we can see in the GLS periodogram of the raw RVs (Fig.\,\ref{udry:fig1}), two extremely significant signals appear at 17 and 52 days. After fitting these signals with a two-Keplerian model, a third significant signal at 271 days can be seen in the RV residuals (middle panel of Fig.\,\ref{udry:fig11}). Finally, the residuals of a three-Keplerian model do not show any signal with $p$-value smaller than 1\,\% (bottom panel of Fig.\,\ref{udry:fig11}). The fact that no signal is present at the estimated rotation period of the star (19 days, Table \ref{table:star}) or its harmonics confirms that the RVs of \object{HD\,31527} are not affected by significant activity signal.

After this first search for significant signals in the data, we fitted, using a MCMC sampler, a three-Keplerian model to the data including a white-noise jitter component to account for stellar and instrumental uncertainties not included in the RV error bars. The best-fit for each planet, as well as the RV residuals, can be seen in Fig.\,\ref{udry:fig12}. We report in addition the best-fit parameters in Table\,\ref{HD31527_tab-mcmc-Summary_params}.

None of the signals announced here matches signals in the different activity indicators (see Fig.\,\ref{app:figHD31527}).We therefore conclude that \object{HD\,31527} hosts 3 Neptune-mass planets, with periods of 16.6, 51.2 and 272 days. The star is a G2 dwarf like the Sun, therefore the outer planet in this system lies on an orbit between the ones of Venus and the Earth, therefore in the habitable zone of its host star \citep[][]{Selsis-2007}. This planet, with a minimum mass of 13 Earth-masses, is however most likely composed of a large gas envelope \citep[][]{Rogers-2015,Wolfgang-2015,Weiss-2014}, except if it is similar to Kepler-10\,c in composition \citep[][]{Dumusque-2014}.

%HD45184
\subsection{{\footnotesize HD}\,45184: A system of two close-in Neptunes}
\label{udry:hd45184}  

\begin{figure}[!t]
\center
% already in Fig.2           \includegraphics[angle=0,width=0.45\textwidth,origin=br]{udry_fig13_s5per-step0.pdf} 
  \includegraphics[angle=0,width=0.4\textwidth,origin=br]{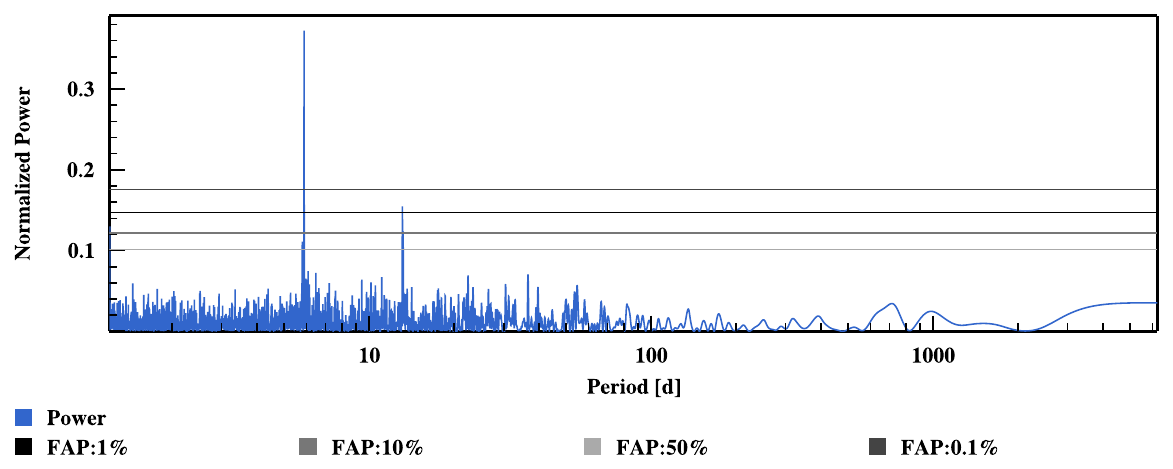} 
  \includegraphics[angle=0,width=0.4\textwidth,origin=br]{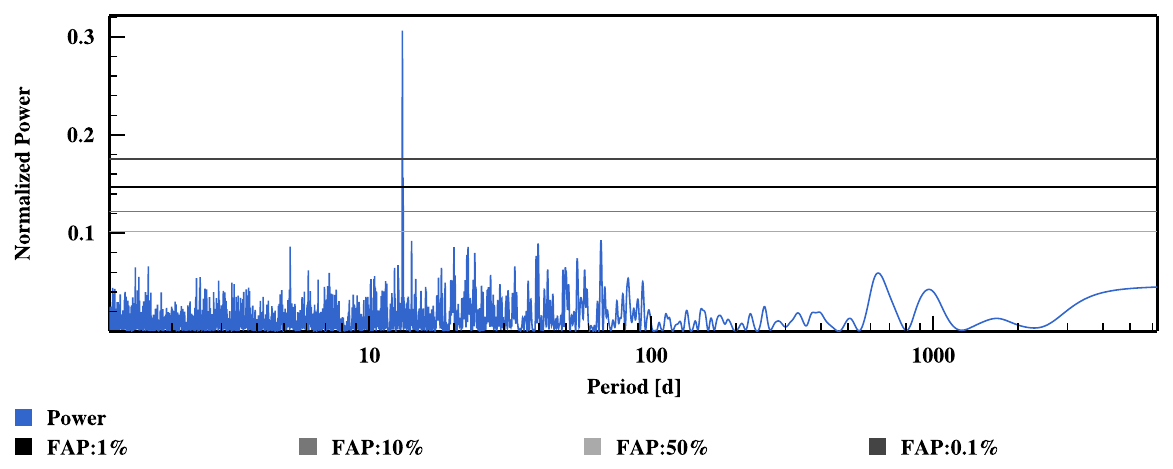} 
  \includegraphics[angle=0,width=0.4\textwidth,origin=br]{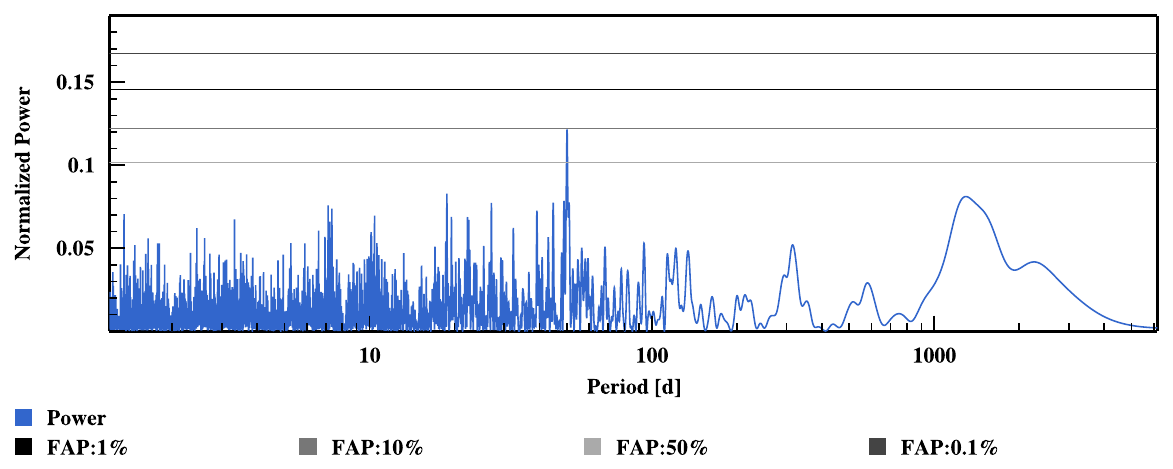} 
\caption[]{From top to bottom: GLS periodogram of the RVs after removing the effect induced by the stellar magnetic cycles and then one planet after the other in the analysis of {\footnotesize HD}\,45184. The GLS periodogram of the raw RVs is shown in Fig.\,\ref{udry:fig1}.}
\label{udry:fig13}
\end{figure}

We gathered a total of 309 high signal-to-noise spectra ($<$S/N$>$ of 221 at 550\,nm) of the G1.5 dwarf \object{HD\,45184} during a time span of 11 years (4160 days). This results in 178 RV measurements, when the data are binned over 1 hour. The average precision of the RVs is 0.41\,\ms\ considering only photon noise and calibration uncertainties. The raw RV rms is much higher, 4.72\,\ms, which implies that significant signals are present in the data. In Fig.\,\ref{udry:fig1}, we display the raw RVs and their GLS periodogram, and the \logrhk\ time series. We see that a significant magnetic cycle affects \logrhk, with values ranging from -5.00 to -4.86 with a periodicity of 5 years. This magnetic cycle is therefore smaller in amplitude and period than the solar cycle, despite the similarities of the derived stellar parameters with those of the Sun. To see if significant signals were present in the calcium activity index despite the long-period magnetic cycle, we fitted the \logrhk\ time series with a Keplerian. In the residuals, a strong signal at 20 days is present, likely corresponding to the stellar rotation period (see Fig.\,\ref{app:figHD45184}). This value is fully compatible with the rotation estimated using the \logrhk\ average level \citep[21.5~days, see Table\,\ref{table:star};][]{Mamajek-2008,Noyes-1984}.

Looking at the raw RVs of \object{HD\,45184} and their GLS periodogram in Fig.\,\ref{udry:fig1}, we see that the magnetic cycle observed in \logrhk\ has an influence on the RVs. To remove the RV contribution of the magnetic cycle, we fitted the \logrhk\ with a Keplerian, and removed the same Keplerian from the RVs leaving the amplitude as a free parameter. In the residuals, displayed in the top panel of Fig.\,\ref{udry:fig13}, we see a significant signal at 6 days. Once this signal is removed by fitting a Keplerian with the corresponding period, another signal at 13 days appears (see middle panel of Fig.\,\ref{udry:fig13}). Finally after removing a two-Keplerian model to the RVs corrected for the magnetic cycle effect, no signal with $p$-value smaller than 10\,\% remains. We therefore stop looking for extra signals in the data. Note that although not significant, the highest peak in the GLS periodogram of the RV residuals is at 18.6 days, likely the imprint of the stellar rotation period. The RVs are therefore slightly affected by stellar activity, however at a level that is not perturbing the detection and characterization of the two planets at 6 and 13 days.

%P0 = 5.9 days
%P1 = 13.1
\begin{figure}[!t]
\center
  \includegraphics[angle=0,width=0.4\textwidth,origin=br]{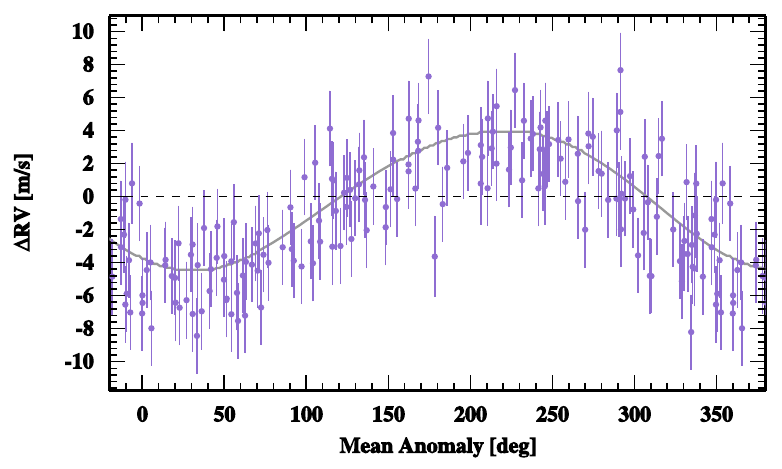} 
  \includegraphics[angle=0,width=0.4\textwidth,origin=br]{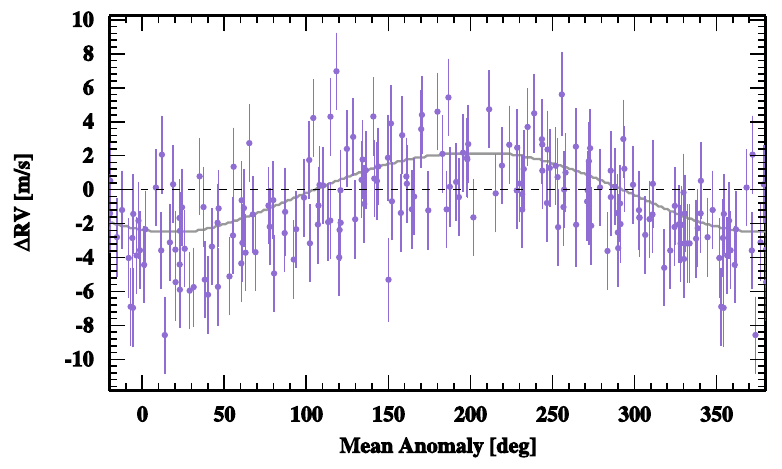} 
  \includegraphics[angle=0,width=0.4\textwidth,origin=br]{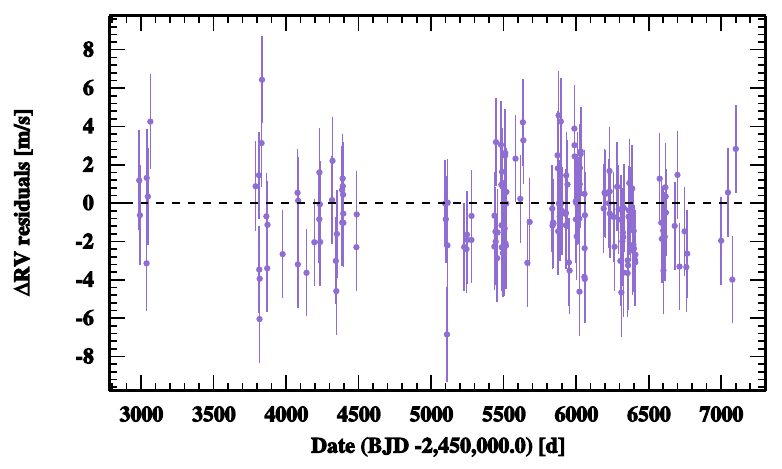} 
\caption[]{Phase-folded RV measurements of {\footnotesize HD}\,45184 with, from top to bottom, the best Keplerian solution for planets b and c, and the corresponding residuals. Error bars include photon and calibration noise, as well as a jitter effect (stellar + instrumental) determined in the MCMC analysis. Corresponding orbital elements are listed in Table\,\ref{HD45184_tab-mcmc-Summary_params}.}
\label{udry:fig14}
\end{figure}

After the preliminary phase of looking for significant signals, we fitted the RVs using a MCMC sampler and a model composed of a linear correlation of the RVs with \logrhk\ to adjust the magnetic cycle effect, two Keplerians to fit for the signals at 5.9 and 13.1 days, and two jitters that correspond to the instrumental plus stellar noise at the minimum and maximum of the magnetic cycle (see Sec.\,\ref{sec:analysis}). Each planet with its best fit can be seen in Fig.\,\ref{udry:fig14}, as well as the RV residuals after the best solution has been removed. The best-fit parameters are reported in Table\,\ref{HD45184_tab-mcmc-Summary_params}. None of the signals at 5.9 and 13.1 matched signals in the different activity indicators (see Fig.\,\ref{app:figHD45184}), and therefore those signal are associated with \emph{bona-fide} planets.

\begin{figure}[!h]
 \center
  \includegraphics[angle=0,width=0.4\textwidth,origin=br]{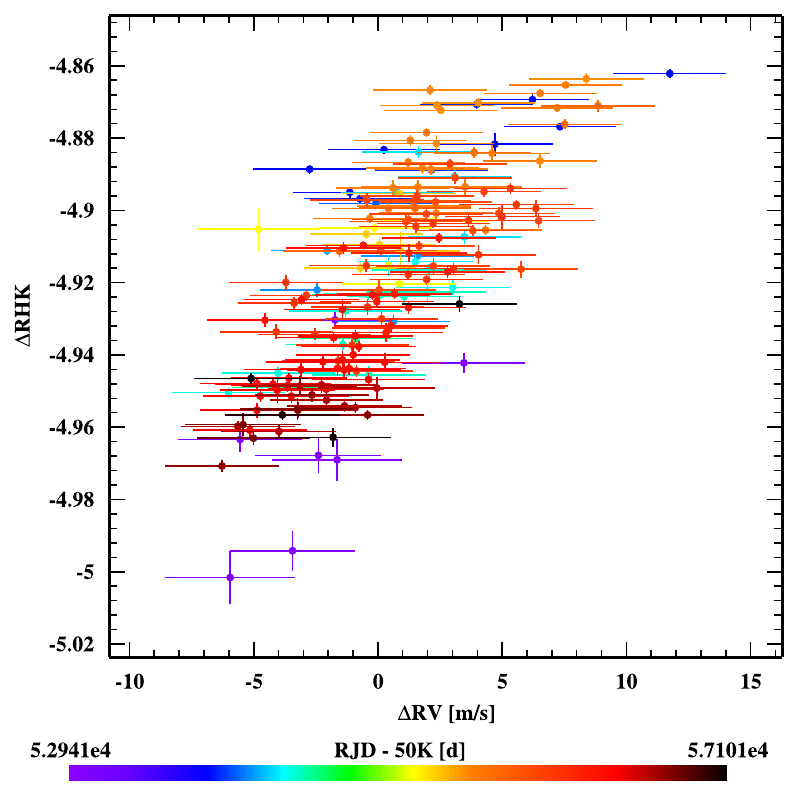} 
    \caption[]{RV residuals when removing all the detected signals except the magnetic cycle effect plotted as a function of the activity index \logrhk\, for {\footnotesize HD}\,45184. The observed correlation indicates that most of the RV residual variation is due to activity-related effects.}
\label{udry:fig15}
\end{figure}

In Fig.\,\ref{udry:fig15}, we show the RV residuals after removing the best-fit solution for planets b and c as a function of the \logrhk. The observed strong correlation indicates that most of the residuals are due to activity-related effects and motivates the use of our model that includes a correlation between \logrhk\ and RVs to mitigate the effect of long-term activity.

%O-C 2.15
\begin{table}
\caption{Best-fitted solution for the planetary system orbiting {\footnotesize HD}\,45184. For each parameter, the median of the posterior distribution is considered, with error bars computed from the MCMC chains using a 68.3\% confidence interval. $\sigma_{O-C}$ corresponds to the weighted standard deviation of the residuals around the best solution. All the parameters probed by the MCMC can be found in Annex, in Table\,\ref{HD45184_tab-mcmc-Probed_params}.}  
\label{HD45184_tab-mcmc-Summary_params}
\def\arraystretch{1.5}
\begin{center}
\begin{tabular}{lccc}
\hline
\hline
Param. & Units & HD45184b & HD45184c \\
\hline
$P$ & [d] & 5.8854$_{-0.0003}^{+0.0003}$  & 13.1354$_{-0.0025}^{+0.0026}$  \\
$K$ & [m\,s$^{-1}$] & 4.26$_{-0.23}^{+0.23}$  & 2.36$_{-0.23}^{+0.23}$  \\
$e$ &   & 0.07$_{-0.05}^{+0.05}$  & 0.07$_{-0.05}^{+0.07}$  \\
$\omega$ & [deg] & 145.80$_{-47.76}^{+49.43}$  & -197.97$_{-80.87}^{+119.85}$  \\
$T_P$ & [d] & 55500.2509$_{-0.7790}^{+0.7996}$  & 55497.4412$_{-2.9188}^{+4.4543}$  \\
$T_C$ & [d] & 55499.4150$_{-0.0903}^{+0.1050}$  & 55494.8821$_{-0.3065}^{+0.3364}$  \\
\hline
$Ar$ & [AU] & 0.0644$_{-0.0021}^{+0.0020}$  & 0.1100$_{-0.0036}^{+0.0034}$  \\
M.$\sin{i}$ & [M$_{\rm Jup}$] & 0.0384$_{-0.0032}^{+0.0033}$  & 0.0277$_{-0.0032}^{+0.0034}$  \\
M.$\sin{i}$ & [M$_{\rm Earth}$] & 12.19$_{-1.03}^{+1.06}$  & 8.81$_{-1.02}^{+1.09}$  \\
\hline 
$\gamma_{HARPS}$ & [m\,s$^{-1}$] & \multicolumn{2}{c}{-3757.6506$_{-0.1595}^{+0.1562}$}\\
$\sigma_{(O-C)}$ & [m\,s$^{-1}$] & \multicolumn{2}{c}{2.15}\\
$\log{(\rm Post})$ &   & \multicolumn{2}{c}{-382.5338$_{-2.9637}^{+2.3828}$}\\
\hline
\end{tabular}
\end{center}
\end{table}

%HD51608
\subsection{{\footnotesize HD}\,51608: 2 Neptune-mass planets}
\label{udry:hd51608}  

\begin{figure}[!t]
\center
% already in Fig.2 \includegraphics[angle=0,width=0.45\textwidth,origin=br]{udry_fig16_s6per-step0.pdf} 
  \includegraphics[angle=0,width=0.4\textwidth,origin=br]{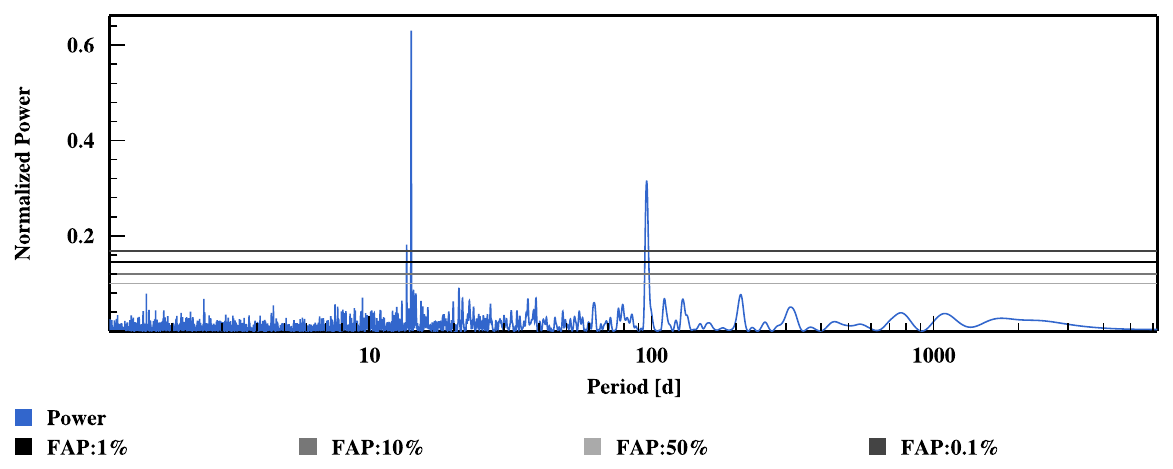} 
  \includegraphics[angle=0,width=0.4\textwidth,origin=br]{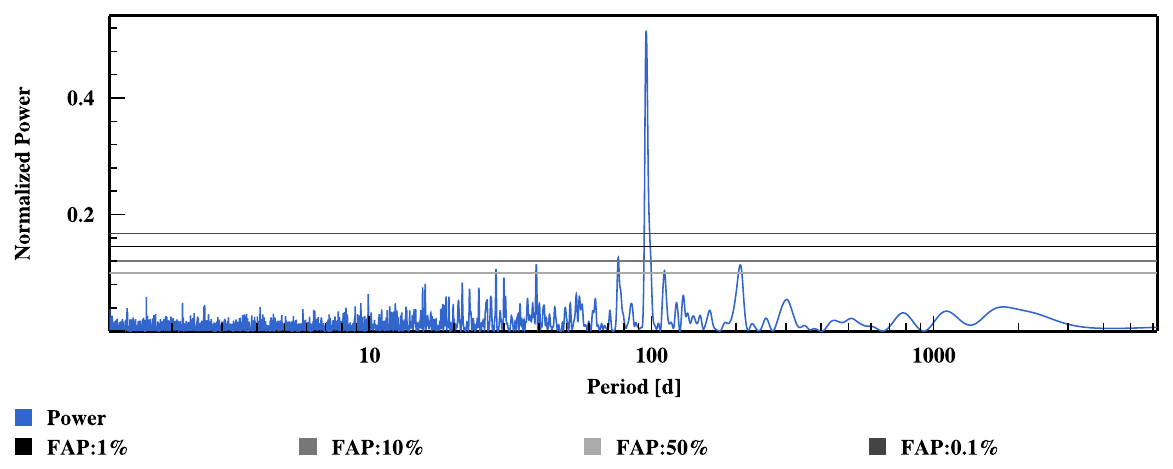} 
  \includegraphics[angle=0,width=0.4\textwidth,origin=br]{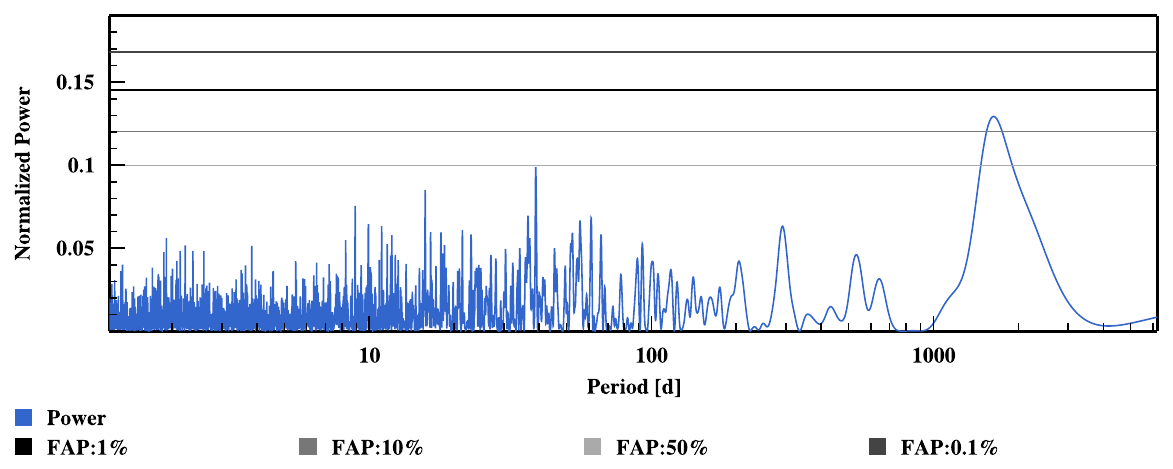} 
\caption[]{From top to bottom: GLS periodogram of the RV residuals of {\footnotesize HD}\,51608 after removing the effect of the magnetic cycle and, at each step, after removing one planet after the other. The GLS periodogram of the raw RVs is shown in Fig.\,\ref{udry:fig1}.}
\label{udry:fig16}
\end{figure}

%P0 = 14
%P1 = 95
\begin{figure}[!h]
\center
  \includegraphics[angle=0,width=0.4\textwidth,origin=br]{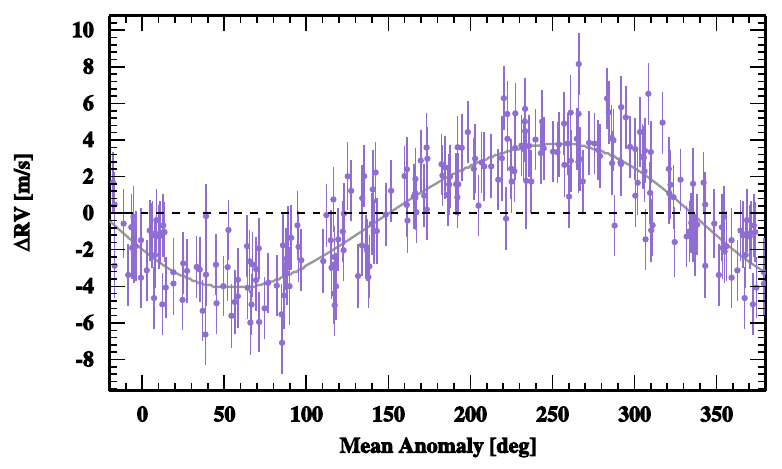} 
  \includegraphics[angle=0,width=0.4\textwidth,origin=br]{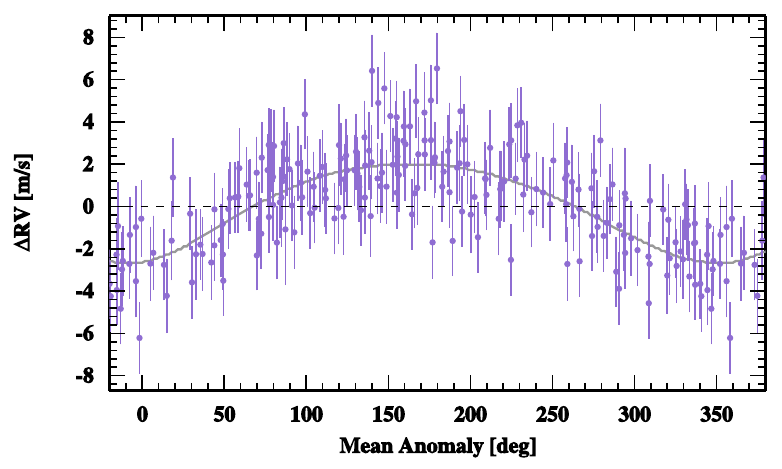} 
  \includegraphics[angle=0,width=0.4\textwidth,origin=br]{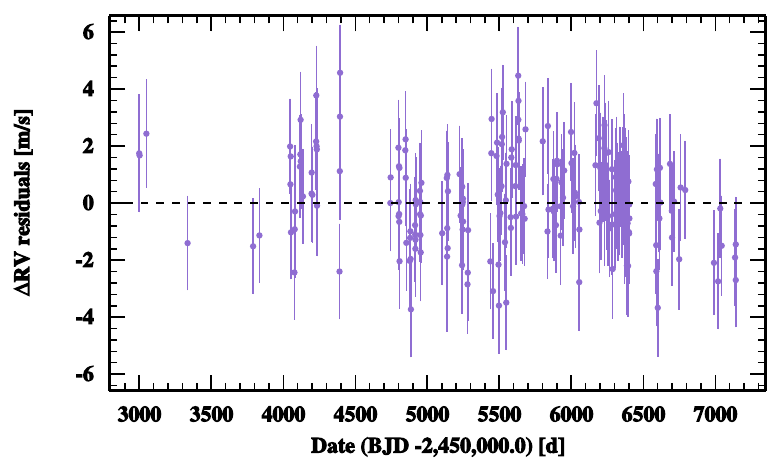}
\caption[]{Phase-folded RV measurements of {\footnotesize HD}\,51608 with, from top to bottom the best Keplerian solution for planet b, c and the residuals around the best-fitted solution. Corresponding orbital elements are listed in Table\,\ref{HD51608_tab-mcmc-Summary_params}.}
\label{udry:fig17}
\end{figure}

Over a time span of 11 years (4158 days), 218 high signal-to-noise spectra ($<$S/N$>$ of 133 at 550\,nm) of \object{HD\,51608} were gathered with {\footnotesize HARPS}, resulting in a total of 216 measurements binned over 1 hour with a typical photon-noise and calibration uncertainty of 0.62\,\ms. This value is significantly below the observed 4.07\,\ms\ dispersion of the RVs, pointing towards the existence of additional signals in the data. The raw RVs, their GLS periodogram and the calcium activity index of \object{HD\,51608} are shown in Fig.\,\ref{udry:fig1}. A small, albeit significant, long-term variation can be seen in \logrhk, with values in \logrhk\, ranging from -5.04 to -4.96 and with a period of 11 years. Although the period of this magnetic cycle is very similar to that of the Sun, its amplitude is much lower. After fitting this long-period signal in \logrhk, a signal with a $p$-value of $\sim$5\,\% and a period of 37 days is detected in the \logrhk\ residuals, and can also be seen in the BIS SPAN and the FWHM of the CCF (see Fig.\,\ref{app:figHD51608}). This is likely a signature of stellar activity as the mean \logrhk\ level gives an estimated rotation period of 40$\pm$4 days \citep[Table\,\ref{table:star};][]{Mamajek-2008,Noyes-1984}.

In the raw RVs, very strong signals at 14 and 96 days are present (see Fig.\,\ref{udry:fig1}). Once fitting a two Keplerian model to account for those signals, a long-period signal with a $p$-value smaller than 0.1\,\% appear in the GLS periodogram (see middle panel of Fig.\,\ref{udry:fig16}). This signal is induced by the stellar magnetic cycle, and we remove it as in the precedent cases by fitting a Keplerian to the \logrhk, and removing the same Keplerian from the RVs leaving the amplitude as a free parameter. After fitting the effect of the two planets plus the magnetic cycle, no significant signal with $p$-value smaller than 5\,\% is left in the residuals.

After this preliminary stage of checking significant signals in the data, we searched for the best-fit parameters using a MCMC sampler, and selecting a model composed of a linear correlation between the RVs and \logrhk\ two Keplerians to fit for the planetary signals at 14.1 and 96.0 days, and two jitters that correspond to the instrumental plus stellar noise at the minimum and maximum of the magnetic cycle (see Sec.\,\ref{sec:analysis}). The best-fit solution for the two planets are shown in Table\,\ref{HD51608_tab-mcmc-Summary_params} and illustrated in Fig.\,\ref{udry:fig17}, along with the RV residuals. The two signals detected in RVs are not matching any significant signal in the different activity indicators (see Fig.\,\ref{app:figHD51608}) and are therefore associated to \emph{bona-fide} planets.

\begin{figure}[!h]
\center
  \includegraphics[angle=0,width=0.4\textwidth,origin=br]{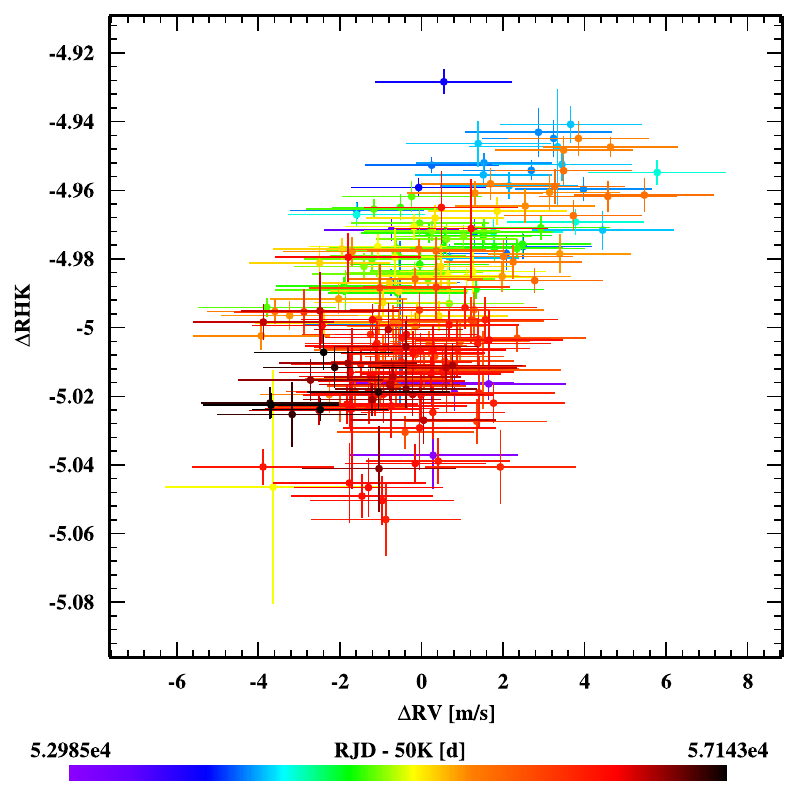} 
\caption[]{RV residuals around the best derived solution without considering the magnetic cycle effect plotted as a function of the activity index \logrhk\ for {\footnotesize HD}\,51608. The observed correlation indicates that most of the RV residual variation is due to activity-related effects.}
\label{udry:fig18}
\end{figure}

The correlation between the activity index \logrhk\ and the RV residuals removing only the two-planet solution can be clearly seen in Fig.\,\ref{udry:fig18}. Note that this correlation is considered in the model we used to fit the RV data.

With masses of 14.3 and 12.8 Earth-masses, the two planets orbiting \object{HD\,51608} are likely Neptune-type planets. There is however still the possibility to have (one of) them similar in composition to Kepler-10\,c \citep[][]{Dumusque-2014}.

%O-C = 1.6
\begin{table}
\caption{Best-fitted solution for the planetary system orbiting HD51608. For each parameter, the median of the posterior distribution is considered, with error bars computed from the MCMC chains using a 68.3\,\% confidence interval. $\sigma_{O-C}$ corresponds to the weighted standard deviation of the residuals around the best solution. All the parameters probed by the MCMC can be found in Annex, in Table\,\ref{HD51608_tab-mcmc-Probed_params}.}  
\label{HD51608_tab-mcmc-Summary_params}
\def\arraystretch{1.5}
\begin{center}
\begin{tabular}{lccc}
\hline
\hline
Param. & Units & HD51608b & HD51608c \\
\hline
$P$ & [d] & 14.0726$_{-0.0016}^{+0.0016}$  & 95.9446$_{-0.1366}^{+0.1555}$  \\
$K$ & [m\,s$^{-1}$] & 3.95$_{-0.16}^{+0.16}$  & 2.36$_{-0.17}^{+0.17}$  \\
$e$ &   & 0.09$_{-0.04}^{+0.04}$  & 0.14$_{-0.07}^{+0.07}$  \\
$\omega$ & [deg] & 117.45$_{-28.91}^{+30.27}$  & -165.07$_{-33.93}^{+29.34}$  \\
$T_P$ & [d] & 55494.5239$_{-1.1073}^{+1.1795}$  & 55498.3725$_{-9.0495}^{+7.6262}$  \\
$T_C$ & [d] & 55493.6039$_{-0.1646}^{+0.1684}$  & 55474.0824$_{-2.5650}^{+2.5542}$  \\
\hline
$Ar$ & [AU] & 0.1059$_{-0.0046}^{+0.0043}$  & 0.3809$_{-0.0164}^{+0.0153}$  \\
M.$\sin{i}$ & [M$_{\rm Jup}$] & 0.0402$_{-0.0037}^{+0.0038}$  & 0.0450$_{-0.0048}^{+0.0051}$  \\
M.$\sin{i}$ & [M$_{\rm Earth}$] & 12.77$_{-1.19}^{+1.20}$  & 14.31$_{-1.53}^{+1.63}$  \\
\hline 
$\gamma_{HARPS}$ & [m\,s$^{-1}$] & \multicolumn{2}{c}{39977.2351$_{-0.1147}^{+0.1159}$}\\
$\sigma_{(O-C)}$ & [m\,s$^{-1}$] & \multicolumn{2}{c}{1.60}\\
$\log{(\rm Post})$ &   & \multicolumn{2}{c}{-409.1398$_{-3.1572}^{+2.4423}$}\\
\hline
\end{tabular}
\end{center}
\end{table}

%HD134060
\subsection{{\footnotesize HD}\,134060: A short-period Neptune on an eccentric orbit with a long-period more massive companion}
\label{udry:hd134060}  

\begin{figure}[!t]
\center
% already in Fig.2           \includegraphics[angle=0,width=0.45\textwidth,origin=br]{udry_fig19_s7per-step0.pdf} 
  \includegraphics[angle=0,width=0.4\textwidth,origin=br]{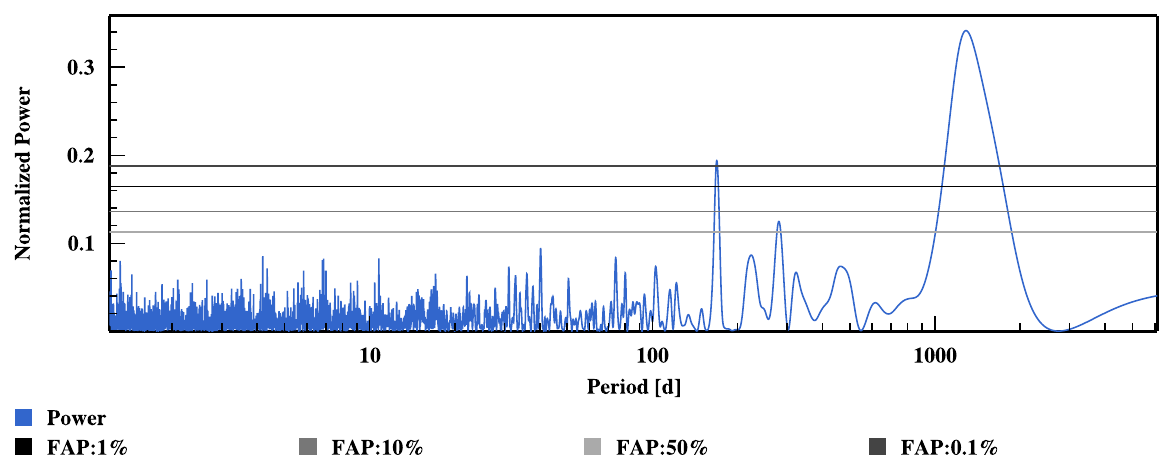} 
  \includegraphics[angle=0,width=0.4\textwidth,origin=br]{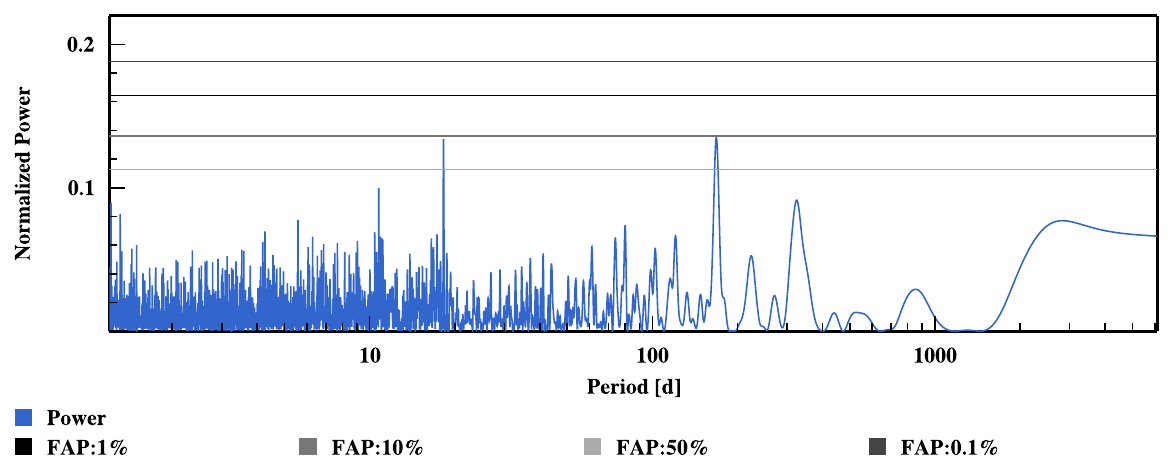} 
\caption[]{From top to bottom: GLS periodogram of the residuals at each step, after removing one planet after the other in the analysis of {\footnotesize HD}\,134060. The GLS periodogram of the raw RVs is shown in Fig.\,\ref{udry:fig1}.}
\label{udry:fig19}
\end{figure}

\subsubsection{Data analysis}
A total of 335 high signal-to-noise spectra ($<$S/N$>$ of 199 at 550\,nm) of \object{HD\,134060} have been gathered over a time span of 11 years (4083 days). When binning the measurements over one hour, we are left with 155 RV measurements, with a typical photon-noise plus calibration uncertainty of 0.40\,\ms. This is an order of magnitude below the observed dispersion of the RVs, 3.68\,\ms. The raw RV with the corresponding GLS periodogram and the \logrhk\ time series are displayed in Fig.\,\ref{udry:fig1}. The \logrhk\ time series presents only a very tiny long-term trend, with very low-level variations between -5.05 and -5. We therefore do not expect strong signals induced by stellar activity. This is confirmed by the fact that no significant signals appears in the periodogram of the different activity indicators in Fig.\,\ref{app:figHD134060}.

\begin{figure}[!t]
\center
  \includegraphics[angle=0,width=0.4\textwidth,origin=br]{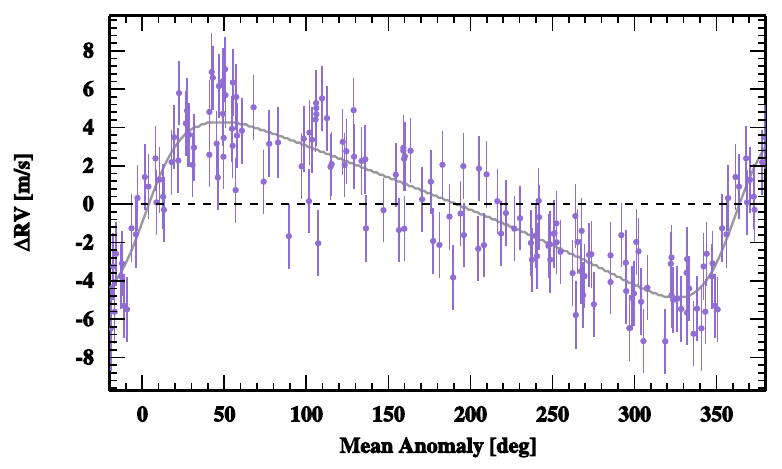} 
  \includegraphics[angle=0,width=0.4\textwidth,origin=br]{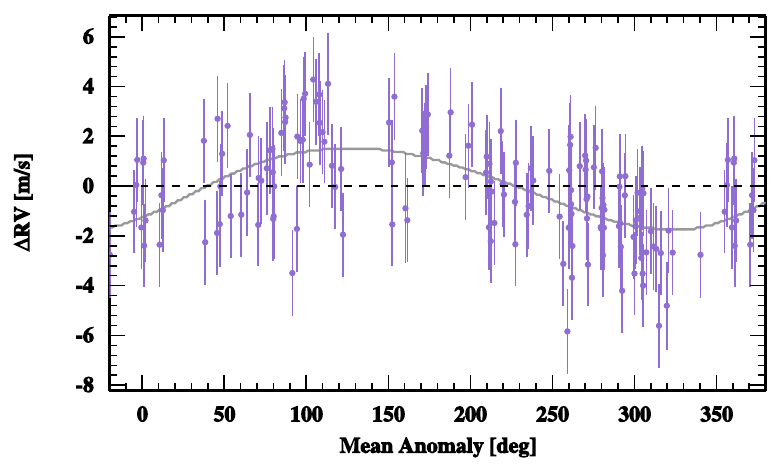} 
  \includegraphics[angle=0,width=0.4\textwidth,origin=br]{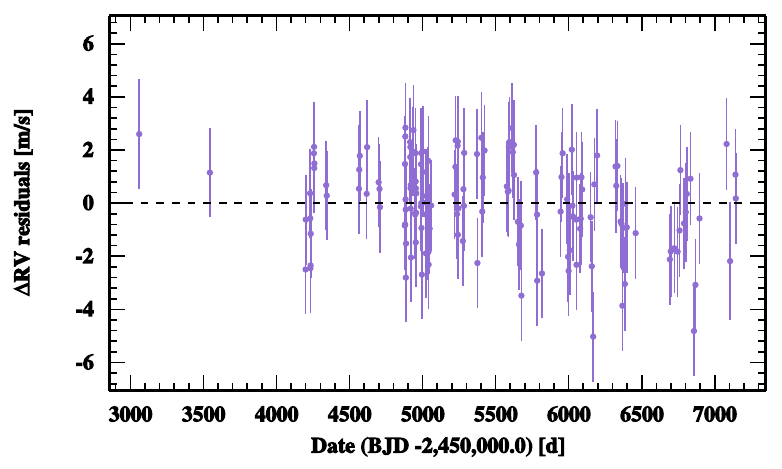} 
\caption[]{Phase-folded RV measurements of {\footnotesize HD}\,134060 with the best fitted solution for planet b and c represented as a black curve. The RV residuals around the best-fit solution are displayed in the lower panel. Corresponding orbital elements are listed in Table\,\ref{HD134060_tab-mcmc-Summary_params}. }
\label{udry:fig20}
\end{figure}

One very significant signal at 3.3 days can be seen in the GLS periodogram of the raw RVs (see Fig.\,\ref{udry:fig1}). Once this signal is fitted with a Keplerian, another significant peak appears at 1292 days, as can be seen in Fig.\,\ref{udry:fig19}. After fitting simultaneously those two signals, nothing is left in the RV residuals with $p$-values smaller than 10\,\%.

\begin{table}
\caption{Best-fitted solution for the planetary system orbiting {\footnotesize HD}\,134060. For each parameter, the median of the posterior distribution is considered, with error bars computed from the MCMC posteriors using a 68.3\,\% confidence interval. The value $\sigma_{(O-C)}$ corresponds to the weighted standard deviation of the residuals around the best solution. All the parameters probed by the MCMC can be found in Annex, in Table\,\ref{HD134060_tab-mcmc-Probed_params}. See Table \ref{tab-mcmc-params} for definition of the parameters.}  
\label{HD134060_tab-mcmc-Summary_params}
\def\arraystretch{1.5}
\footnotesize
\begin{center}
\begin{tabular}{lccc}
\hline \hline
Param. & Units & HD134060b & HD134060c \\
\hline
$P$ & [d] & 3.2696$_{-0.0001}^{+0.0001}$  & 1291.5646$_{-44.2197}^{+48.0333}$  \\
$K$ & [m\,s$^{-1}$] & 4.61$_{-0.22}^{+0.22}$  & 1.65$_{-0.23}^{+0.24}$  \\
$e$ &   & 0.45$_{-0.04}^{+0.04}$  & 0.11$_{-0.07}^{+0.13}$  \\
$\omega$ & [deg] & -98.23$_{-6.75}^{+6.61}$  & -132.73$_{-55.51}^{+121.28}$  \\
$T_P$ & [d] & 55499.6542$_{-0.0407}^{+0.0420}$  & 55232.8160$_{-186.5492}^{+444.3209}$  \\
$T_C$ & [d] & 55498.1943$_{-0.1037}^{+0.1065}$  & 56057.2704$_{-48.5676}^{+61.2663}$  \\
\hline
$Ar$ & [AU] & 0.0444$_{-0.0014}^{+0.0013}$  & 2.3928$_{-0.0951}^{+0.0929}$  \\
M.$\sin{i}$ & [M$_{\rm Jup}$] & 0.0318$_{-0.0024}^{+0.0025}$  & 0.0922$_{-0.0133}^{+0.0139}$  \\
M.$\sin{i}$ & [M$_{\rm Earth}$] & 10.10$_{-0.75}^{+0.79}$  & 29.29$_{-4.24}^{+4.43}$  \\
\hline 
$\gamma_{HARPS}$ & [m\,s$^{-1}$] & \multicolumn{2}{c}{37987.9484$_{-0.1489}^{+0.1512}$}\\
$\sigma_{(O-C)}$ & [m\,s$^{-1}$] & \multicolumn{2}{c}{1.64}\\
$\log{(\rm Post})$ &   & \multicolumn{2}{c}{-304.2166$_{-2.9181}^{+2.1293}$}\\
\hline 
\end{tabular}
\end{center}
\end{table}

To get the best possible orbital parameters for those two planets with reliable error bars, we perform an MCMC analysis with a model composed of two Keplerians plus a white-noise jitter to account for stellar and instrumental uncertainties not included in the RV error bars. The best-fit parameters can be found in Table\,\ref{HD134060_tab-mcmc-Summary_params}. The planetary signals, folded in phase, can be seen in Fig.\,\ref{udry:fig20} along with the RV residuals shown in the bottom plot.

\subsubsection{Origin of the high eccentricity of the inner planet?}

The MCMC converges to a solution with an inner planet of minimum mass 10.1\,M$_{\oplus}$ and period 3.27 days, on a relatively high eccentricity orbit with $e=0.45$. An inner eccentric planets in a system with an additional outer planet is reminiscent of the system \object{HD\,20003} above. The main differences are that in \object{HD\,134060} the outer companion is farther out and more massive, and the inner planet is experiencing stronger tides from the star making the circularization timescale shorter. Nevertheless some of the potential mechanisms proposed for explaining the high eccentricity might be considered as well.\\
\indent
1) The high eccentricity can hide a planetary system in 2:1 resonance, thus the existence of another planet at half its orbital period \citep{Anglada-Escude-2010}. We therefore tried to fit a model with an extra planet at 1.65 days, fixing the eccentricities at zero or leaving them free to vary. In both cases, the more complex solution is disfavoured by Bayesian model comparison with a $\Delta$BIC of 3.1 and 23.5, respectively. We therefore keep the simplest solution with the relatively high eccentricity of the inner planet. \\
\indent
2) This planet has a long-period companion that has a minimum mass three times larger. In the case of the two planetary orbits being strongly non-coplanar it is likely that the long-period planet is perturbing its inner companion through a Kozai-Lidov mechanism causing libration of its orbit \citep[][]{Kozai-1962,Lidov-1961}. During this process, the eccentricity of the inner planet can reach very high values. The inner planet starts then to interact with its host star during close approaches, inducing a circularization of the inner orbit. Because of the conservation of the total angular momentum, eccentricity can increase only if the inclination of the orbit changes. Inner planets under the influence of a Kozai-Lidov mechanism are therefore likely to be on inclined orbits relative to the stellar rotational plane. This can be measured using the Rossiter-McLaughlin effect if by chance the planet transits its host star. This has been checked for \object{HD\,134060} by \citet{Gillon-2017} using the Spitzer space telescope with unfortunately a null result. Another difficulty here is that for a planet at 3.3 days, the circularization timescale is normally very short for typical values of the tidal dissipation parameters, therefore a priori preventing us from observing the system in such a configuration. \\
\indent
3) Scattering is always a valid possibility when high eccentricities are involved. As for the previous point, preventing the system to then circularize is also the challenge. \\
\indent
4) The explanation invoking spin-orbit coupling during the evolution of the planet orbit under the influence of tides from the central star and gravitational perturbation from the outer companion \citep{Correia-2012} might finally be the most appealing process. It has however still to be demonstrated that the process works for the present set of parameters. As often, the challenge is in the balance of timescales. To help, \object{HD\,134060} seems to be a relatively young sub-giant with an age estimate of 1.75\,Gyr \citep{Delgado-Mena2015}.

In any case, with the high eccentricity of its inner planet \object{HD\,134060} provides a very interesting system to study planetary formation and evolution. 
%

%HD136352
\subsection{{\footnotesize HD}\,136352: A 3-planet system}
\label{udry:hd136352}  

\begin{figure}[!t]
\center
% already in Fig.2           \includegraphics[angle=0,width=0.45\textwidth,origin=br]{udry_fig21_s8per-step0.pdf} 
  \includegraphics[angle=0,width=0.4\textwidth,origin=br]{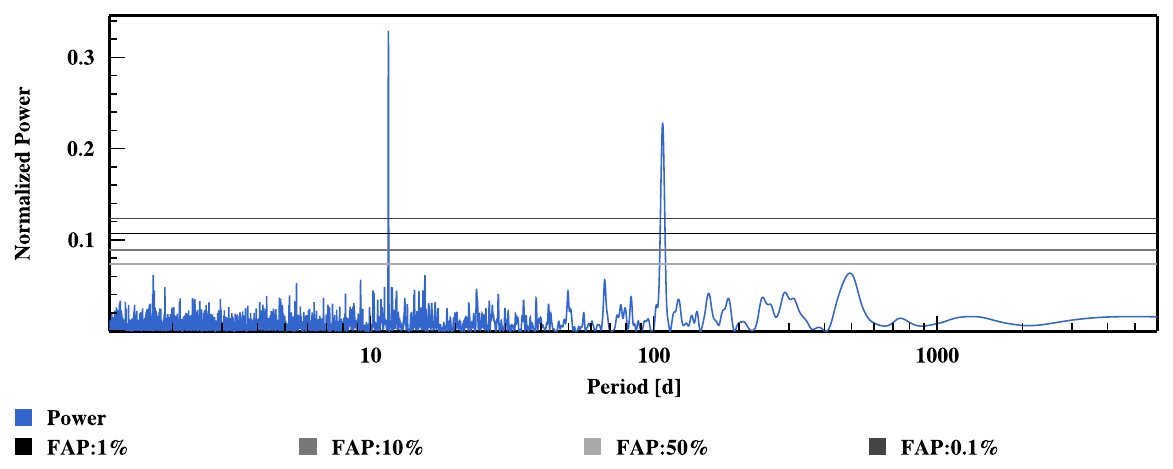} 
  \includegraphics[angle=0,width=0.4\textwidth,origin=br]{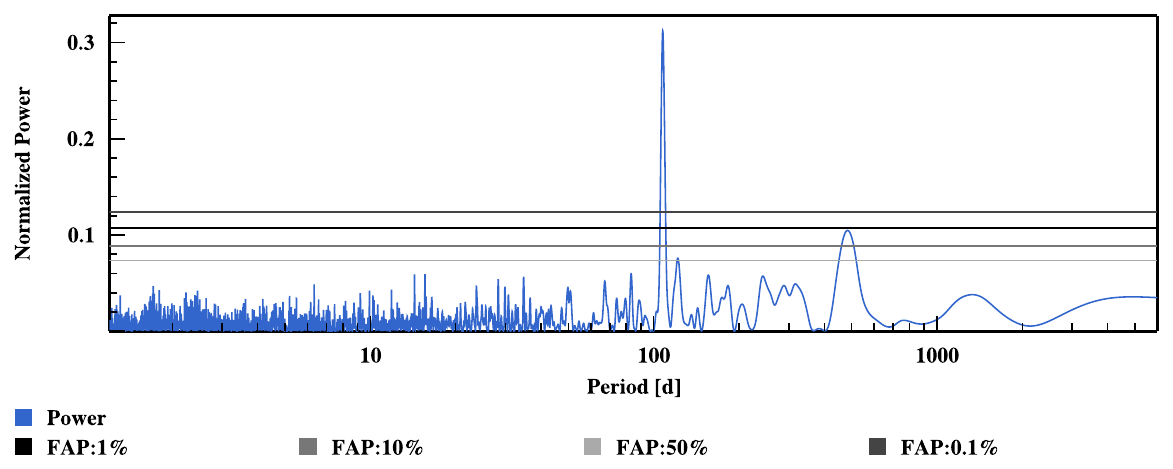} 
  \includegraphics[angle=0,width=0.4\textwidth,origin=br]{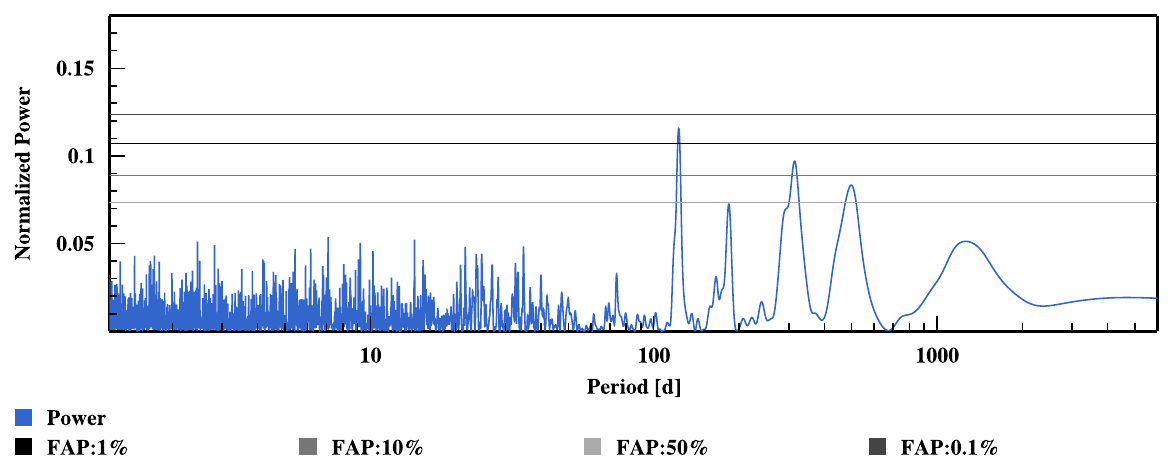} 
\caption[]{From top to bottom: GLS periodogram of the residuals at each step, after removing one planet after the other in the analysis of {\footnotesize HD}\,136352. The GLS periodogram of the raw RVs is shown in Fig.\,\ref{udry:fig1}.}
\label{udry:fig21}
\end{figure}

\object{HD\,136352} was part of the original high-precision HARPS GTO survey and the star has been followed for nearly 11 years (3993 days). Over this time span, we gathered a total of 649 high signal-to-noise spectra ($<$S/N$>$ of 231 at 550\,nm) corresponding in the end to 240 RV measurements binned over 1 hour. As reported in Table\,\ref{table:obsstat}, the typical precision of individual measurements is 0.33\,\ms including photon noise and calibration uncertainties, an order of magnitude smaller than the observed raw RV rms of 2.74\,\ms. This suggests that significant signals, of stellar or planetary origin, are present in the data. As a first approach we looked at the \logrhk\ activity index time series in Fig.\,\ref{udry:fig1}. No significant long-term variation is observed in the \logrhk\ data and no long-term variation is visible in the GLS periodogram of the velocity time series. We conclude that there is no noticeable sign of a magnetic activity cycle for this star. The average value of \logrhk\ at $-4.95$ is low with a small dispersion of $\sim0.01$, close to the Sun at minimum activity. No significant effect of stellar activity in the RV measurements is thus expected for this star. This is confirmed by the fact that no significant signals appears in the periodogram of the different activity indicators in Fig.\,\ref{app:figHD136352}.

Due to the small activity level and the large number of observations, the GLS periodogram of the velocity series is actually very clean, with peaks at 27.6, 11.6 and 108 days, in order of decreasing significance (see Fig.~\ref{udry:fig21}). After fitting those three signals with Keplerians, a study of the GLS periodogram of the RV residuals shows a peaks at 123 days with a $p$-value between 1 and 0.1\,\%, thus an interesting signal that we will consider in the MCMC analysis.

\begin{figure*}[!h]
\center
  \includegraphics[angle=0,width=0.4\textwidth,origin=br]{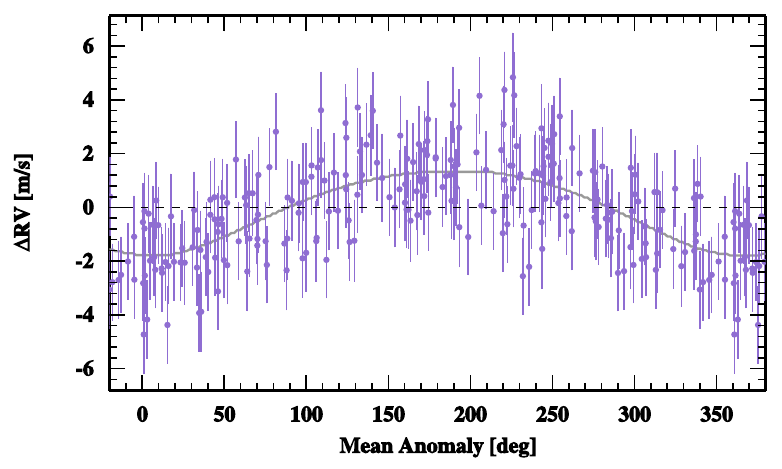} 
  \includegraphics[angle=0,width=0.4\textwidth,origin=br]{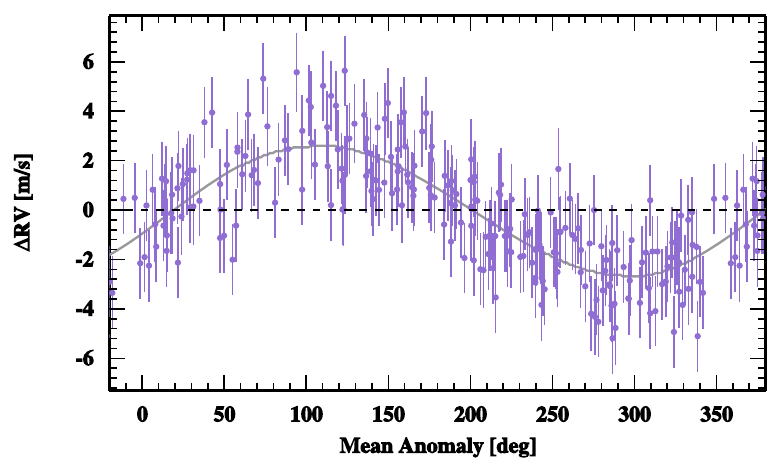} 
  \includegraphics[angle=0,width=0.4\textwidth,origin=br]{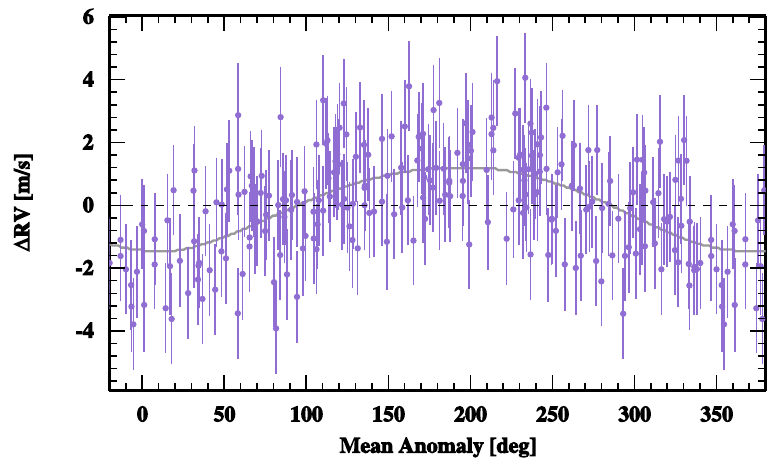}
  \includegraphics[angle=0,width=0.4\textwidth,origin=br]{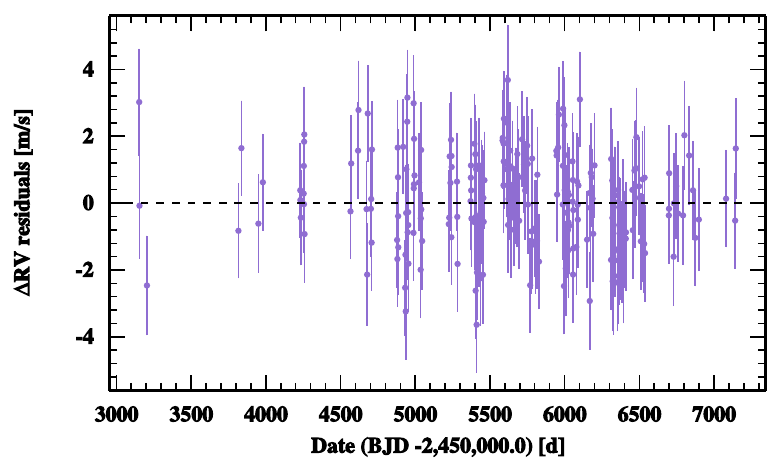}
\caption[]{Phase-folded  RV measurements of {\footnotesize HD}\,136352 with, from left to right and top to bottom the best fitted solution for planet b, c and d. The residuals around the solution are displayed in the lower right panel. Corresponding orbital elements are listed in Table\,\ref{HD136352_tab-mcmc-Summary_params}. }
\label{udry:fig22}
\end{figure*}

The best orbital parameters for the three planets orbiting \object{HD\,136352} are searched for using a MCMC sampler using a model composed of three Keplerians and an extra white-noise jitter to account for instrumental and stellar uncertainties not included in the RV error bars. Phase-folded planetary solutions are displayed in Fig.\,\ref{udry:fig22}, as well as the RV residuals around the best solution. The best-fit parameters are reported in Table\,\ref{HD136352_tab-mcmc-Summary_params}. 

%P0 = 27 days
%P1 = 11 days
%P2 = 107 days

Our analysis of {\footnotesize HD}\,136352 converges to the detection of three planet orbiting this G4V star. With minimum masses of 4.8, 8.6 and 10.8\,$M_{\oplus}$, {\footnotesize HD}\,136352 host three super-Earth on orbits ranging from 11 to 108 days. The two inner planets, with periods of 11.6 and 27.6 days, are close to a 5:2 resonance and, contrary to the two planets in \object{HD\,21693}, with a period ratio larger than 2.5.
%
%comparison k3 27.6, 11.6 and 108 days with k4 27.6, 11.6, 108 and 213 days
%Delta BIC 39.6 in favor of k3

%comparison k3 27.6, 11.6 and 108 days with k3 0.50, 11.6 and 27.6 days
%Delta BIC 16.85 in favor of k3 27.6, 11.6 and 108 days

\begin{table*}[!h]
\caption{Best-fitted solution for the planetary system orbiting {\footnotesize HD}\,136352. For each parameter, the median of the posterior distribution is considered, with error bars computed from the MCMC posteriors using a 68.3\,\% confidence interval. The value $\sigma_{(O-C)}$ corresponds to the weighted standard deviation of the residuals around the best solution. All the parameters probed by the MCMC can be found in Annex, in Table\,\ref{HD136352_tab-mcmc-Probed_params}. See Table\,\ref{tab-mcmc-params} for definition of the parameters.}  \label{HD136352_tab-mcmc-Summary_params}
\def\arraystretch{1.5}
\footnotesize
\begin{center}
\begin{tabular}{lcccc}
\hline \hline
Param. & Units & HD136352b & HD136352c & HD136352d \\
\hline
$P$ & [d] & 11.5824$_{-0.0025}^{+0.0024}$  & 27.5821$_{-0.0086}^{+0.0089}$  & 107.5983$_{-0.2669}^{+0.2796}$  \\
$K$ & [m\,s$^{-1}$] & 1.59$_{-0.13}^{+0.13}$  & 2.65$_{-0.14}^{+0.14}$  & 1.35$_{-0.15}^{+0.15}$  \\
$e$ &   & 0.14$_{-0.08}^{+0.08}$  & 0.04$_{-0.03}^{+0.05}$  & 0.09$_{-0.07}^{+0.10}$  \\
$\omega$ & [deg] & -185.64$_{-37.67}^{+35.97}$  & -110.22$_{-75.76}^{+74.58}$  & 170.89$_{-76.56}^{+63.23}$  \\
$T_P$ & [d] & 55496.7009$_{-1.2084}^{+1.1393}$  & 55490.0572$_{-5.8036}^{+5.7741}$  & 55494.9775$_{-22.9117}^{+18.6996}$  \\
$T_C$ & [d] & 55494.3913$_{-0.3259}^{+0.3534}$  & 55477.8619$_{-0.3730}^{+0.4105}$  & 55472.4695$_{-3.2834}^{+4.6515}$  \\
\hline
$Ar$ & [AU] & 0.0934$_{-0.0040}^{+0.0037}$  & 0.1666$_{-0.0072}^{+0.0066}$  & 0.4128$_{-0.0175}^{+0.0163}$  \\
M.$\sin{i}$ & [M$_{\rm Jup}$] & 0.0151$_{-0.0018}^{+0.0018}$  & 0.0340$_{-0.0033}^{+0.0034}$  & 0.0270$_{-0.0036}^{+0.0037}$  \\
M.$\sin{i}$ & [M$_{\rm Earth}$] & 4.81$_{-0.56}^{+0.59}$  & 10.80$_{-1.05}^{+1.08}$  & 8.58$_{-1.15}^{+1.18}$  \\
\hline 
$\gamma_{HARPS}$ & [m\,s$^{-1}$] & \multicolumn{3}{c}{-68709.0314$_{-0.0964}^{+0.0937}$}\\
$\sigma_{(O-C)}$ & [m\,s$^{-1}$] & \multicolumn{3}{c}{1.35}\\
$\log{(\rm Post})$ &   & \multicolumn{3}{c}{-420.5739$_{-3.1788}^{+2.5027}$}\\
\hline 
\end{tabular}
\end{center}
\end{table*}

As mentioned above, an interesting signal is present in the residuals at 123 days. Despite the period being suspicious (1/3 of a year), we ran another MCMC trial including this fourth signal as an additional planet (Keplerian). The fit converged towards a non-eccentric signal with an amplitude of 0.65\,\ms\ and a period of 122.6~days. However, when comparing the three- and four-planet model solutions, the case with three planets is strongly favored, with a $\Delta$BIC of 39.6. We also tried to add to the three-planet model a polynomial of the first or the second order to check if this 123-day signal could be due to a very-long period companion, whose orbit is not covered by the data, creating an alias at a period of one year or one of its harmonics. This approach, however, did not reduce the signal at 123 days and was disfavoured by a model comparison using the BIC.
We also looked for other possibilities of the three planet scenario, fitting the aliases of the 11 and 27.6-day signals, 0.91 and 0.96 day, respectively. These other possibilities are also ruled out with difference in BIC$>$17. Finally we tested the sensitivity of the GLS periodogram to outliers. We found that by simply removing two of them, for example JD 2455411 and 24556168, the amplitude of the peak found at 123 days goes above a  $p$-value of 1\,\%. Therefore, several arguments points in the direction of this interesting, albeit not conclusively significant, signal at 123 days being more likely an artefact induced by noise in the data or interaction with the window function rather than a \emph{bona-fide} planet. We therefore keep for {\footnotesize HD}\,136352 the three planet solution, with periods of 11.6, 27.6 and 108 days.
 
\section{Conclusion}
\label{sec:concl}

\begin{figure*}[!h]
\center
  \includegraphics[angle=0,width=1\textwidth,origin=br]{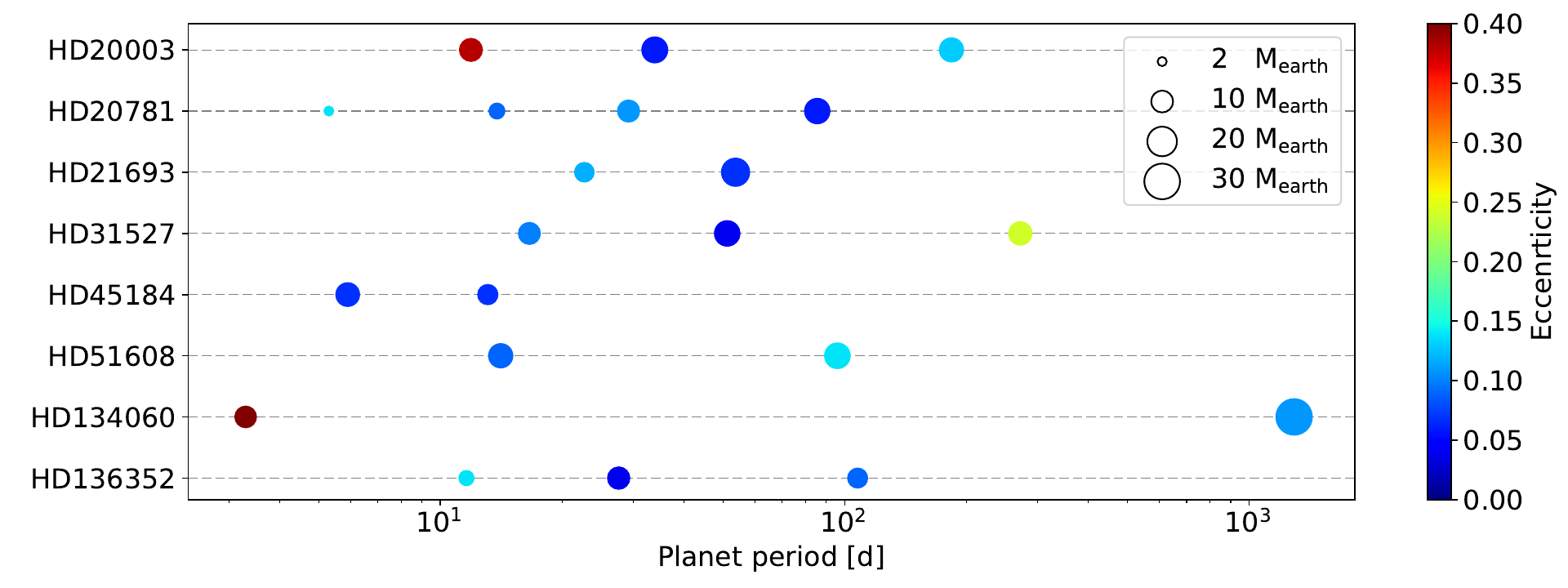} 
\caption[]{Summary of the confirmed planetary systems with indication of their periods (x-axis), masses (symbol size) and eccentricities (colour code). }
\label{udry:fig23}
\end{figure*}

We have reported the discovery and detailed characterization of eight planetary systems hosting twenty low-mass planets, plus a possible additional planet on a 6-month orbit, discovered with the  {\footnotesize HARPS}  Echelle spectrograph mounted on the 3.6-m ESO telescope located at La Silla Observatory. We also improved the characterization of the extremely eccentric Jupiter orbiting \object{HD20782}. Figure.\,\ref{udry:fig23} illustrates these detections in a diagram showing the planet distribution in periods, with indication of their masses and eccentricities. The planets in our sample can be divided in three categories:
\begin{itemize}
\item Very small-mass planets with minimum masses below 6\,M$_{\oplus}$, that are found orbiting \object{HD\,20781} and \object{HD\,136352}, on short periods, less than 15 days.
\item More massive planets in the super-Earth to Neptune transition regime, spanning a range in minimum masses from 8 to 17\,M$_{\oplus}$, and a range in periods from a few days to nearly a year. They represent the bulk of our detections. 
\item Massive planets found on long-period orbits.
\end{itemize}

The RV technique sensitivity goes down when moving towards small-mass and long-period planets. It is therefore not surprising that most of our detections are in the second mass category. The lack of massive objects on short-period orbits is already well established, and the detection of extremely small-mass planets, although numerous from \emph{Kepler} statistics \citep[e.g.][]{Coughlin:2016aa} and previous RV surveys \citep[e.g.][]{Butler-2017,Mayor-2011}, is challenging for RV surveys because of the perturbing effect of additional stellar signals \citep[][]{Dumusque-2016b} and the preponderance of appearance of these planets in multiple systems.

The detection of these planets had been announced in \citet{Mayor-2011} studying statistical properties of the systems discovered with {\footnotesize HARPS}. However, with more data in hand, it was possible to discover new \emph{bona-fide} planets. This is the case for the two inner super-Earth orbiting \object{HD20781} and the 13.1-day period super-Earth orbiting \object{HD45184}. There is also the 184-day signal found in the timeseries of \object{HD20003} even if, as explained in Sec.\,\ref{udry:hd20003}, we cannot exclude an instrumental origin to this signal. These new detections show that gathering more data helps in detecting small-mass planets on short-period orbits as well as long-period signals. Although expensive, characterizing those signals is therefore a necessary task to get an as complete as possible view of planetary systems.

The most important targets in today's exoplanet studies are the transiting small-mass planets orbiting bright stars. They allow for exquisite determination of their orbital and physical properties, bringing priceless constrains to our understanding of planet formation, of the establishment of planetary system architecture, and of the diversity of planet internal compositions. Those candidates would also be excellent candidates for further atmospheric characterization. A dedicated \emph{Spitzer} survey looking for potential transit events induced by the shortest-period planets in our sample was conducted by \citet{Gillon-2017}. Unfortunately, after searching for transits of the innermost planets orbiting \object{HD\,20003}, \object{HD\,20781}, \object{HD\,31527}, \object{HD\,45184}, \object{HD\,51608} and \object{HD\,134060}, no detection was reported in this study. The search will continue, in particular with the Swiss-ESA led CHEOPS satellite to be launched end of 2018 \citep{Fortier-2014}.

Some systems presented here are interesting in terms of formation process and architecture. \object{HD\,20003} hosts two Neptune-mass planets with periods of 11.9 and 33.9 days, thus close to a 3:1 commensurability. This configuration might explain the relatively high eccentricity measured on the innermost planet in this system, especially if some resonance is involved in the process. We note also that the two Neptune-mass planets orbiting \object{HD\,21693} and the two innermost super-Earth found around \object{HD\,136352} are close to a 5:2 resonance. Finally, \object{HD\,134060} is also an interesting dynamical system as it harbors a small-mass planet on a fairly eccentric 3-day orbit, accompanied by a more massive long-period planet at $\sim$1300 days.  
Several possibilities for the formation of such configurations have been mentioned in the text. The list is probably not exhaustive and further theoretical work is need to analyse these systems in a deeper way. Here we just checked that the stability of the proposed solutions is verified over a few tens of thousands of years, avoiding the systems to quickly fall apart.

\begin{acknowledgements}
We thank the Swiss National Science Foundation (SNSF) and the Geneva University for their continuous support to our planet search programs.
This work has been in particular  carried out in the frame of the National Centre for Competence in Research `PlanetS' supported by SNSF.
XD is grateful to the Society in Science--The Branco Weiss Fellowship for its financial support.
N.C.S. was supported by Funda\c{c}\~ao para a Ci\^encia e a Tecnologia (FCT, Portugal) through the research grant through national funds and by FEDER through COMPETE2020 by grants UID/FIS/04434/2013 \& POCI-01-0145-FEDER-007672 and PTDC/FIS-AST/1526/2014 \& POCI-01-0145-FEDER-016886, as well as through Investigador FCT contract nr. IF/00169/2012/CP0150/CT0002 funded by FCT and POPH/FSE (EC). 
PF acknowledges support by Funda\c{c}\~ao para a Ci\^encia e a Tecnologia (FCT) through Investigador FCT contracts of reference IF/00169/2012/CP0150/CT0002, respectively, and POPH/FSE (EC) by FEDER funding through the program ``Programa Operacional de Factores de Competitividade - COMPETE''. PF further acknowledges support from Funda\c{c}\~ao para a Ci\^encia e a Tecnologia (FCT) in the form of an exploratory project of reference IF/01037/2013/CP1191/CT0001.
ACC acknowledges support from STFC consolidated grant ST/M001296/1.
This research has made use of the SIMBAD database and of the VizieR catalogue access tool operated at CDS, France, and used the DACE platform developed in the frame of PlanetS (\url{https://dace.unige.ch}).  \end{acknowledgements}

\bibliographystyle{aa}
\bibliography{Udry2015}

%\clearpage

\begin{appendix}

\section{Periodogram of the activity and CCF indicators}

To check if any signal detected in RVs corresponds to signals measured in activity indicator, we show here the Lomb-Scargle periodograms of the \logrhk, the BIS SPAN, and the FWHM.
Those indicators have been shown to be sensitive to activity, and therefore any signal appearing both in the RVs and at least one of those indicators might be induced by stellar activity. Because magnetic cycles will be seen as a long-period significant signals in all those indicators, we removed any long-term signal either by fitting a Keplerian to adjust at best the observed magnetic cycle, as in Fig.\,\ref{udry:fig1}, or by adjusting a second order polynomial.

\begin{figure}[!h]
\center
 \includegraphics[angle=90,width=0.42\textwidth,origin=br]{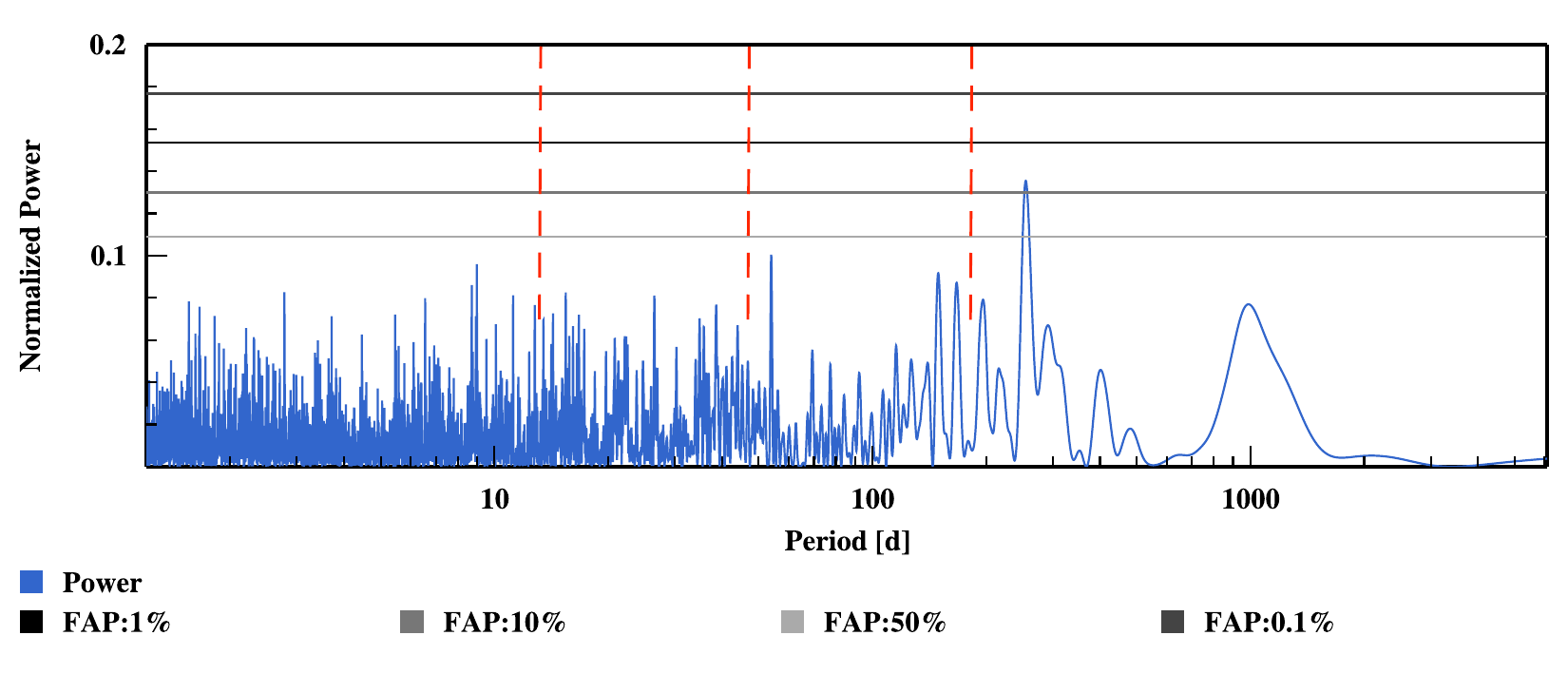}
  \includegraphics[angle=90,width=0.42\textwidth,origin=br]{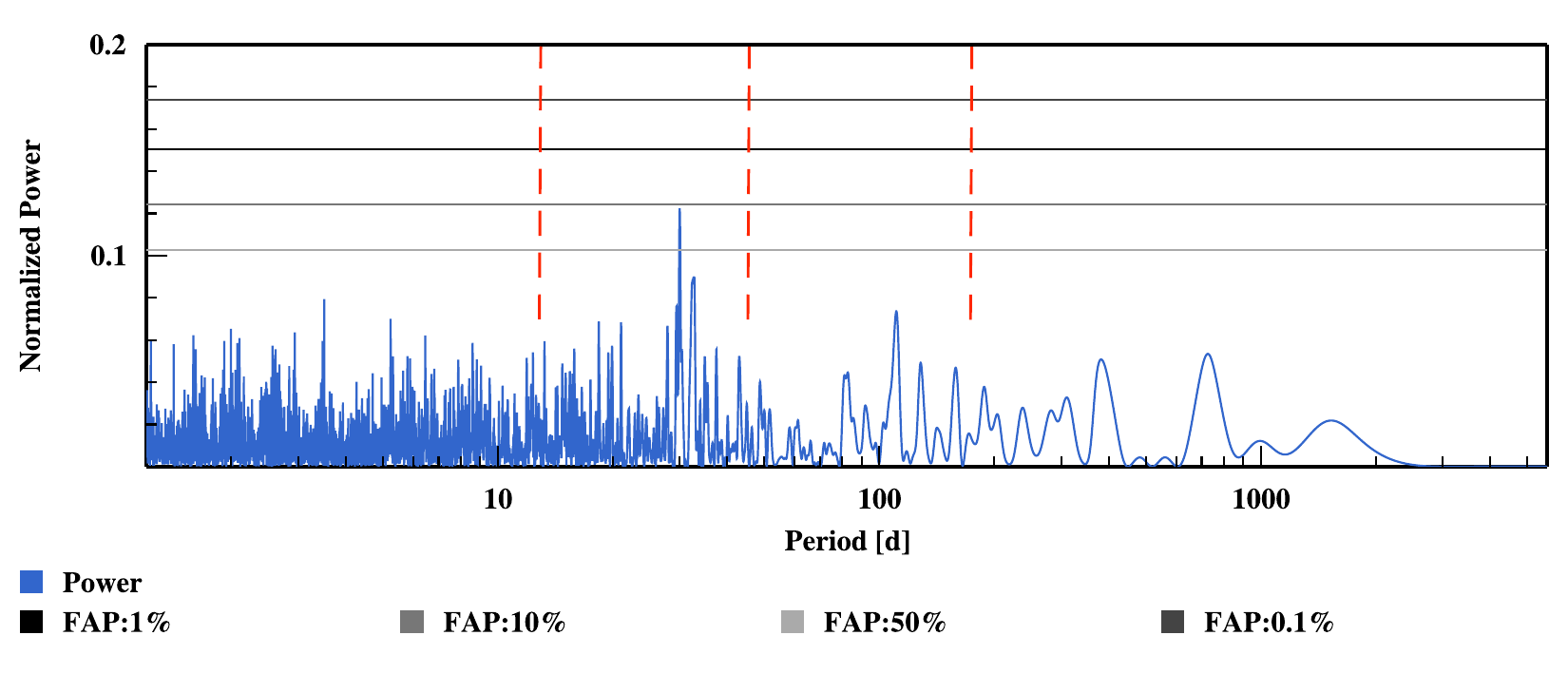}
   \includegraphics[angle=90,width=0.42\textwidth,origin=br]{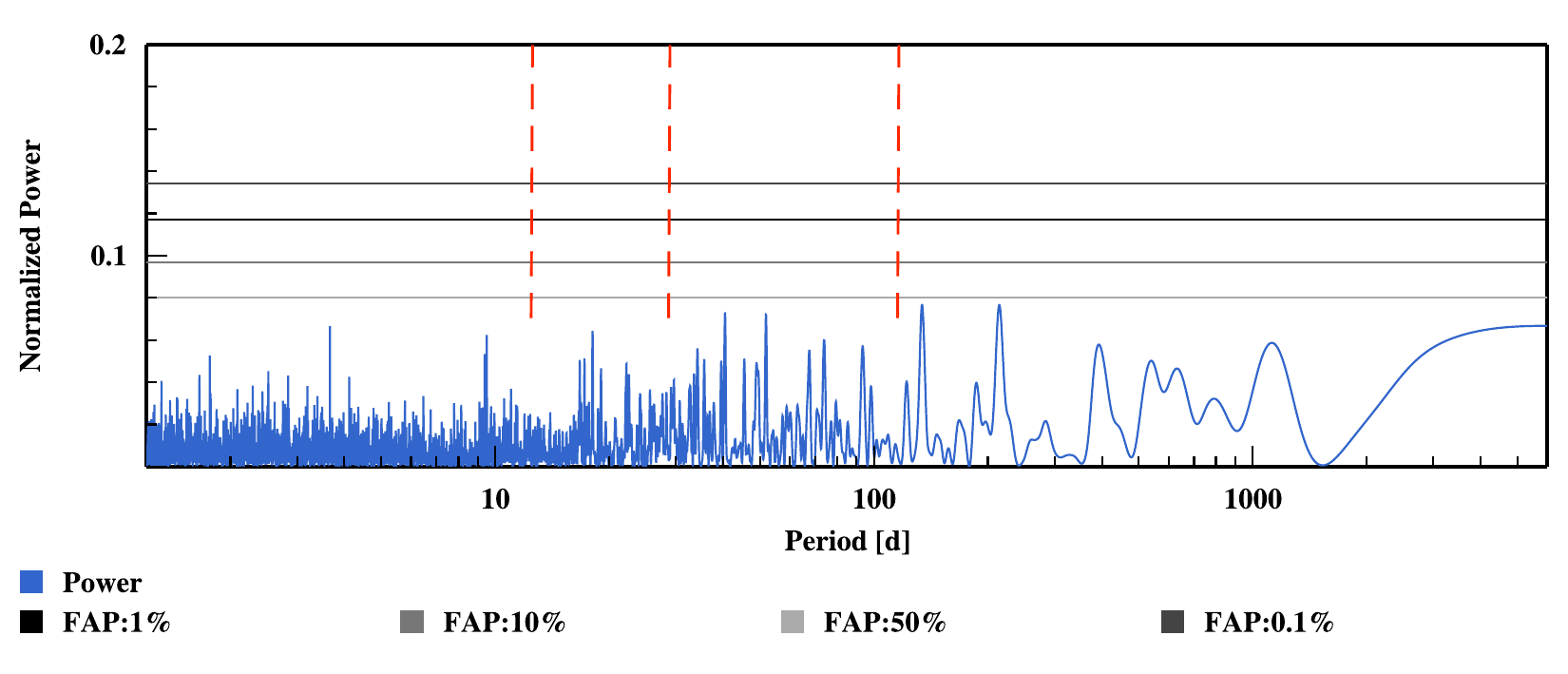}
   \caption[]{From top to bottom, periodograms of the \logrhk, BIS SPAN, and FWHM residuals of \object{HD20003} after fitting either a Keplerian to adjust at best the observed magnetic cycle or a second order polynomial to take into account any drift that could be instrumental. Planetary signals announced in Sec.\,\ref{sec:systems} are represented by dashed vertical red lines.}
\label{app:figHD20003}
\end{figure}

\begin{figure}[!h]
\center
 \includegraphics[angle=90,width=0.42\textwidth,origin=br]{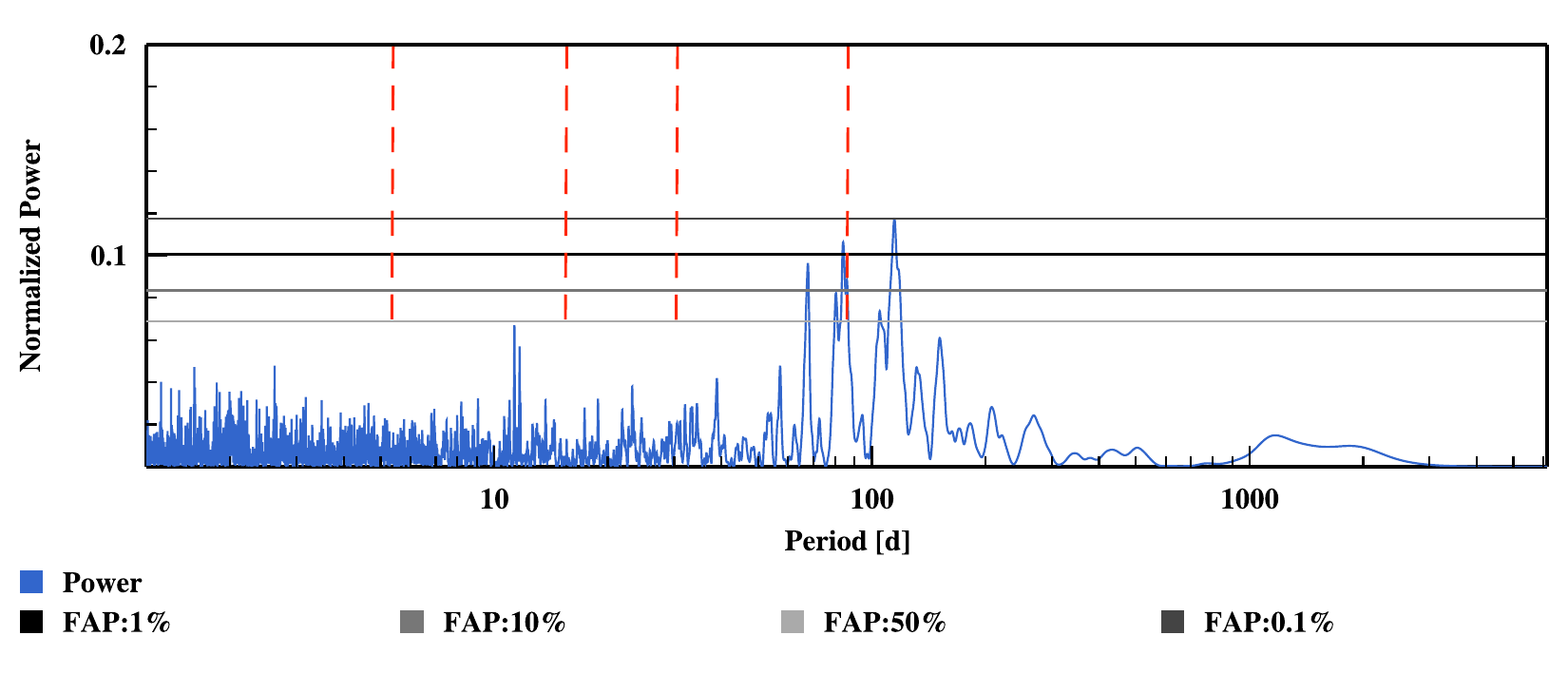}
  \includegraphics[angle=90,width=0.42\textwidth,origin=br]{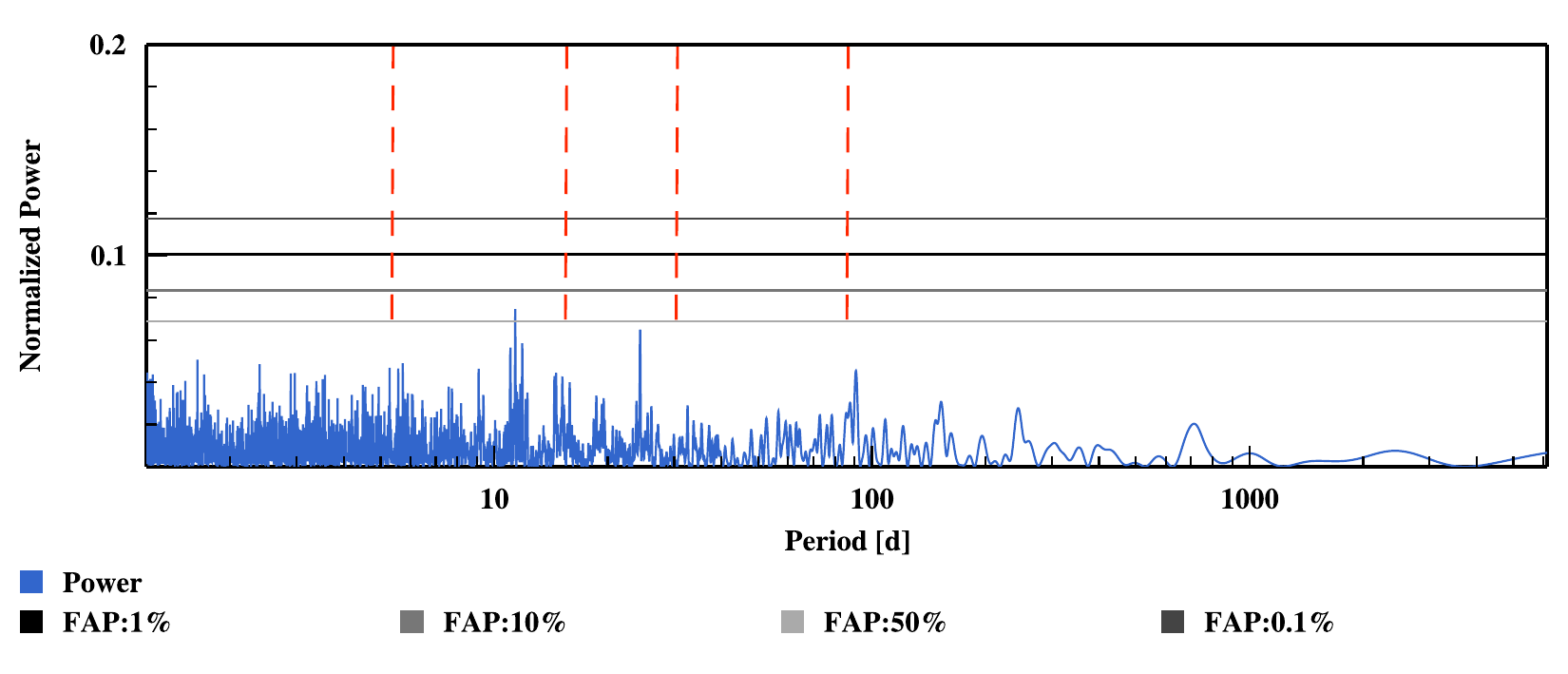}
   \includegraphics[angle=90,width=0.42\textwidth,origin=br]{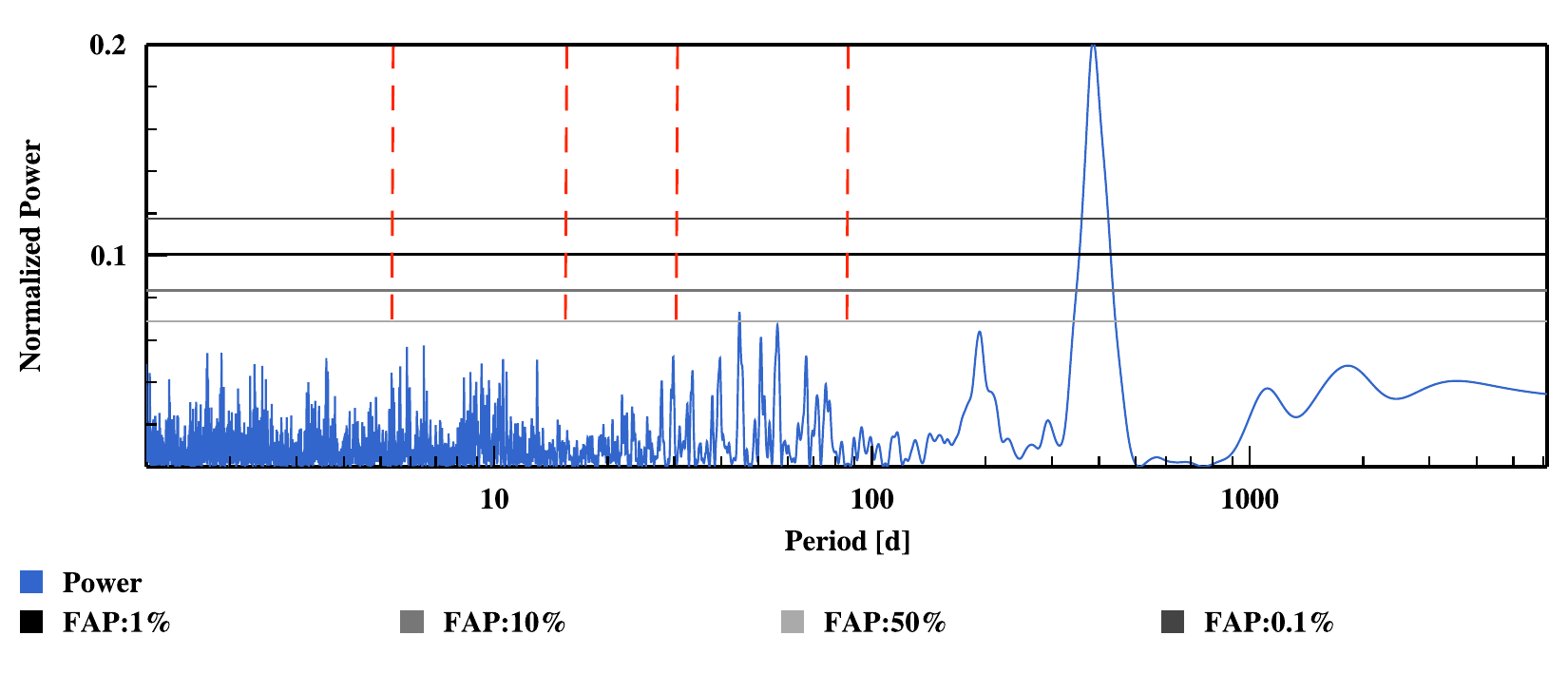}
   \caption[]{Same as Fig.\,\ref{app:figHD20003} but for \object{HD20781}}
\label{app:figHD20781}
\end{figure}

\begin{figure}[!h]
\center
 \includegraphics[angle=90,width=0.42\textwidth,origin=br]{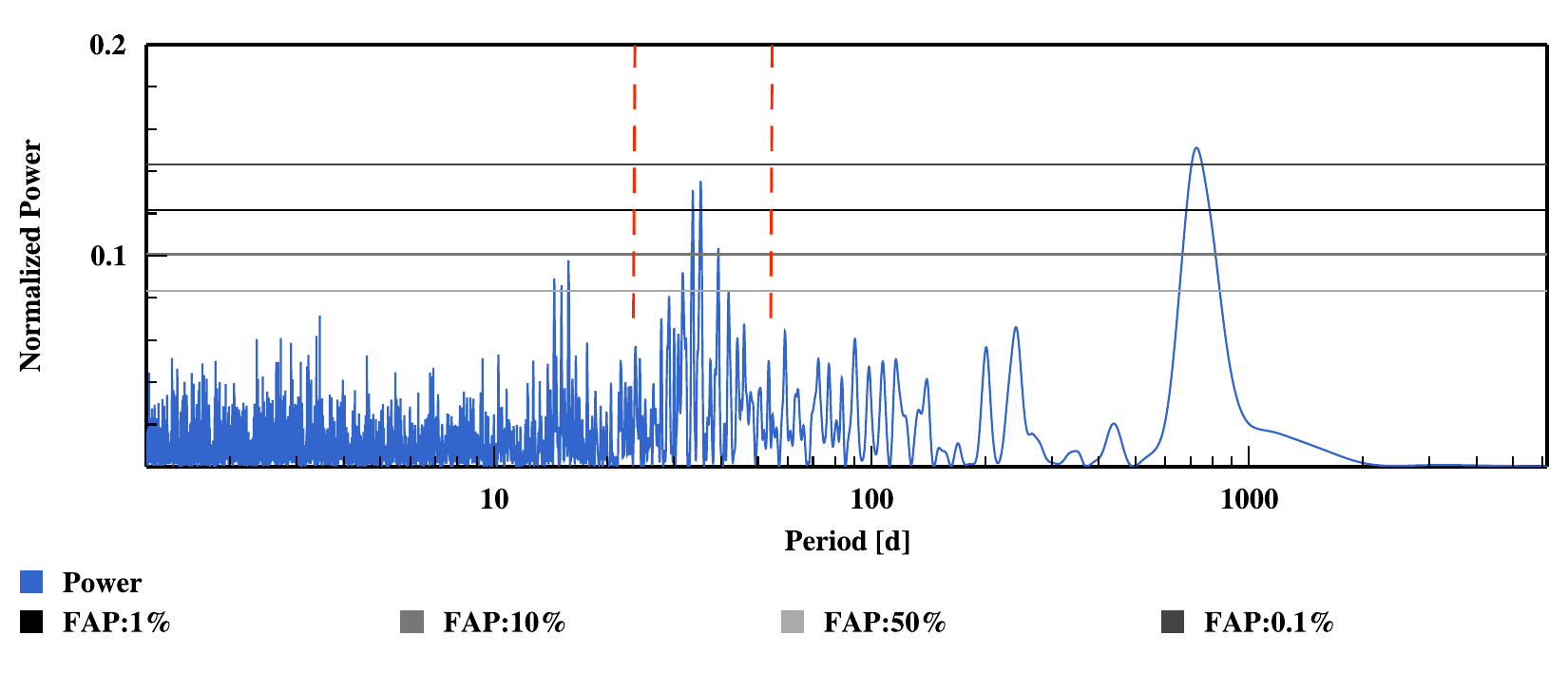}
  \includegraphics[angle=90,width=0.42\textwidth,origin=br]{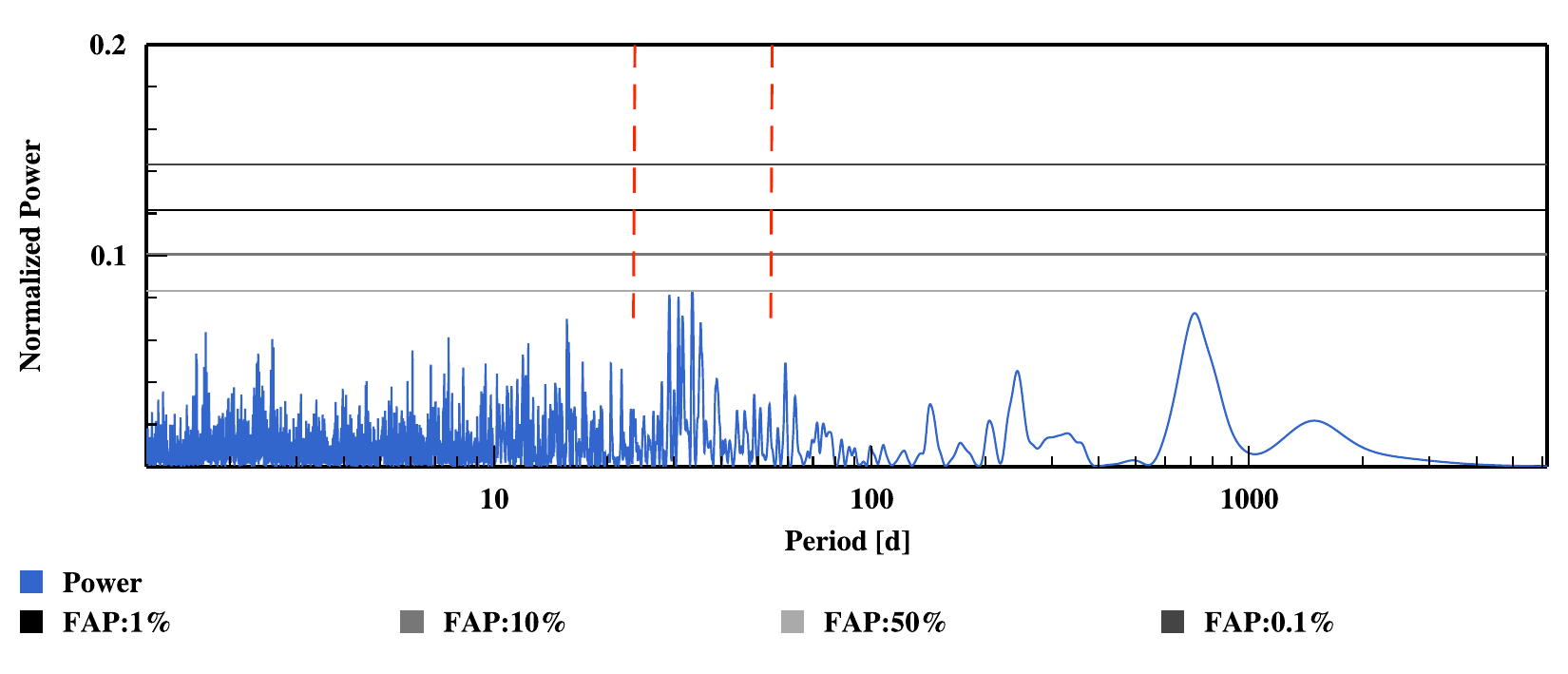}
   \includegraphics[angle=90,width=0.42\textwidth,origin=br]{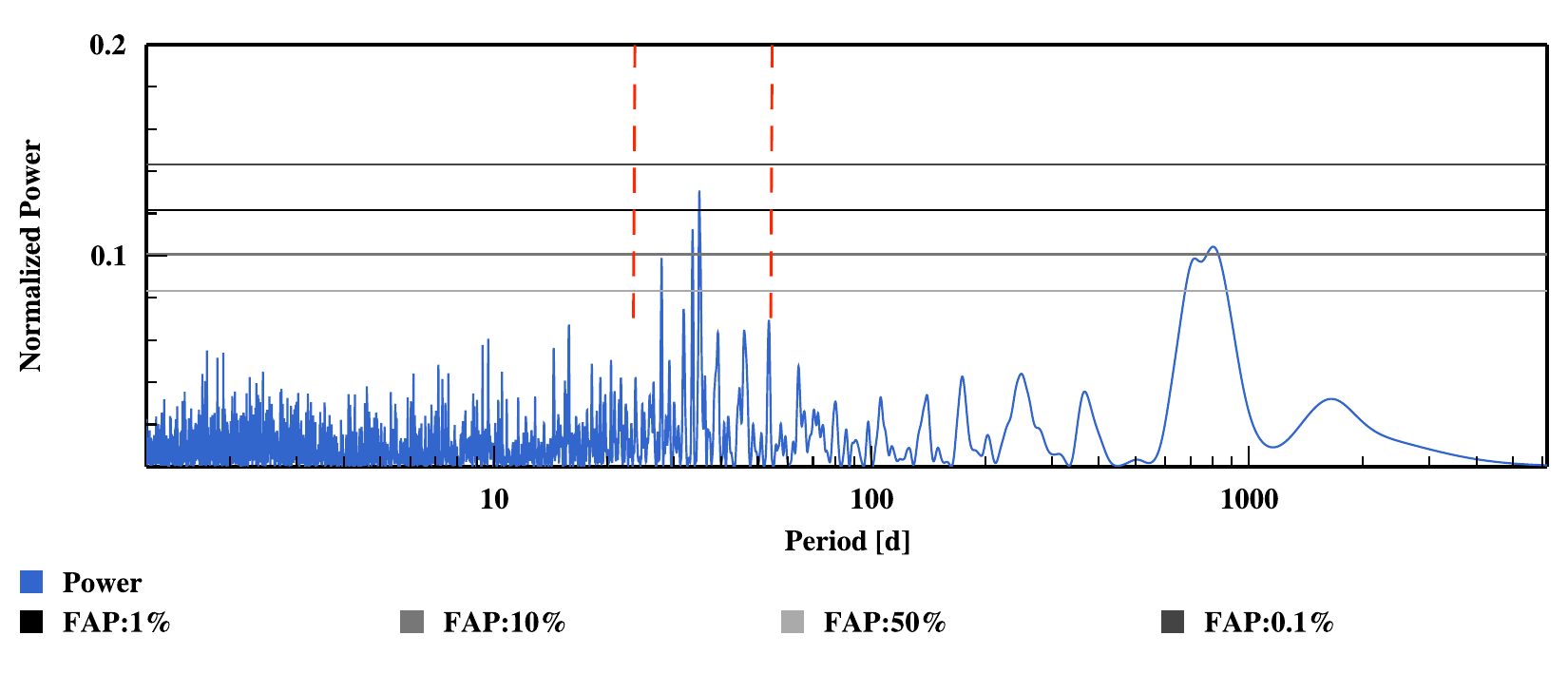}
   \caption[]{Same as Fig.\,\ref{app:figHD20003} but for \object{HD21693}}
\label{app:figHD21693}
\end{figure}

\begin{figure}[!h]
\center
 \includegraphics[angle=90,width=0.42\textwidth,origin=br]{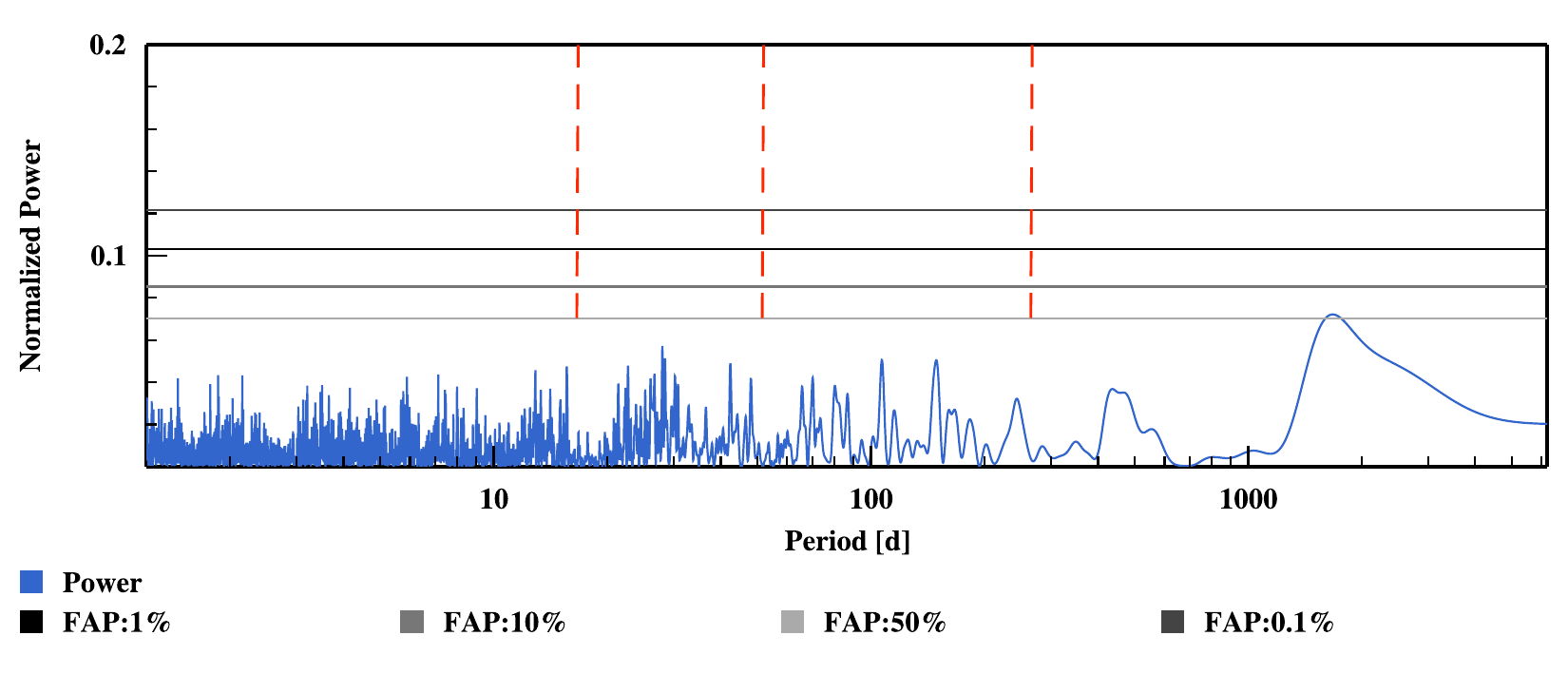}
  \includegraphics[angle=90,width=0.42\textwidth,origin=br]{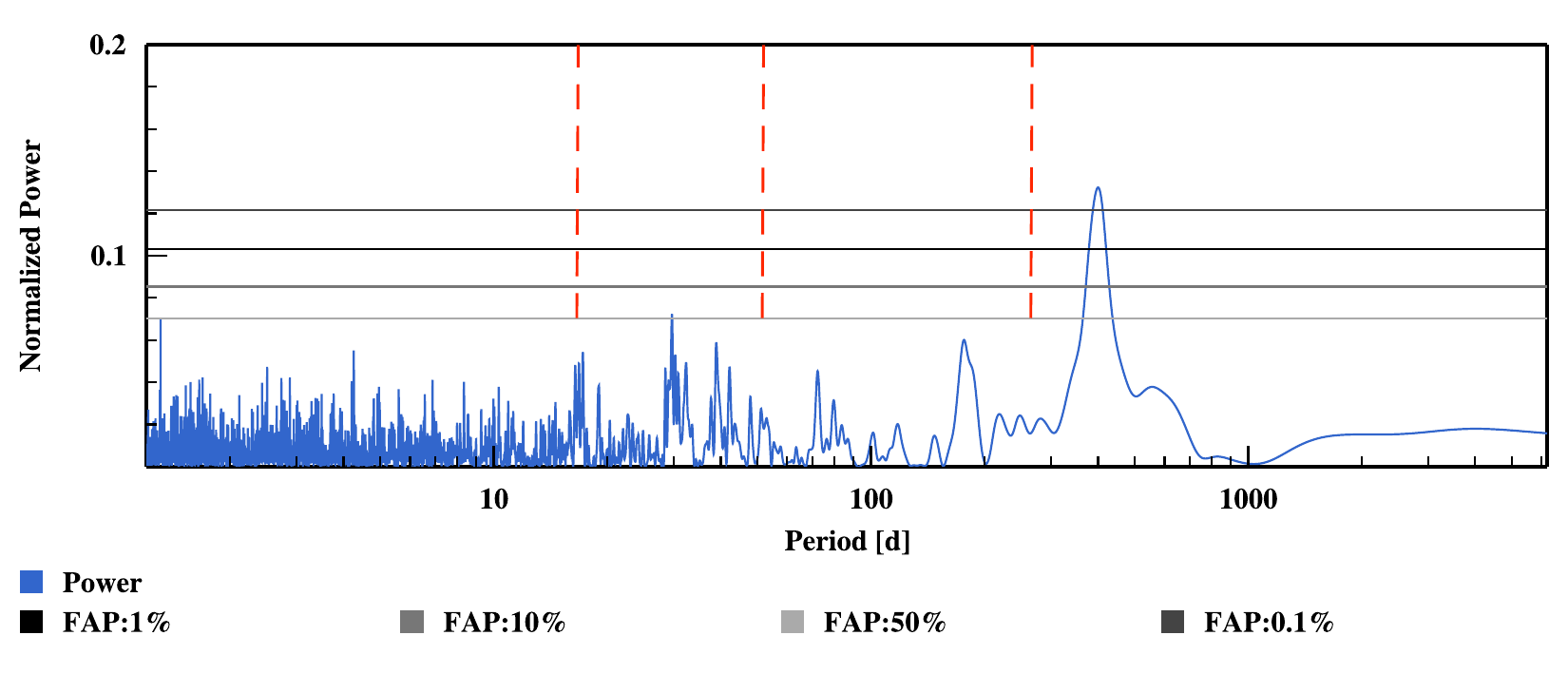}
   \includegraphics[angle=90,width=0.42\textwidth,origin=br]{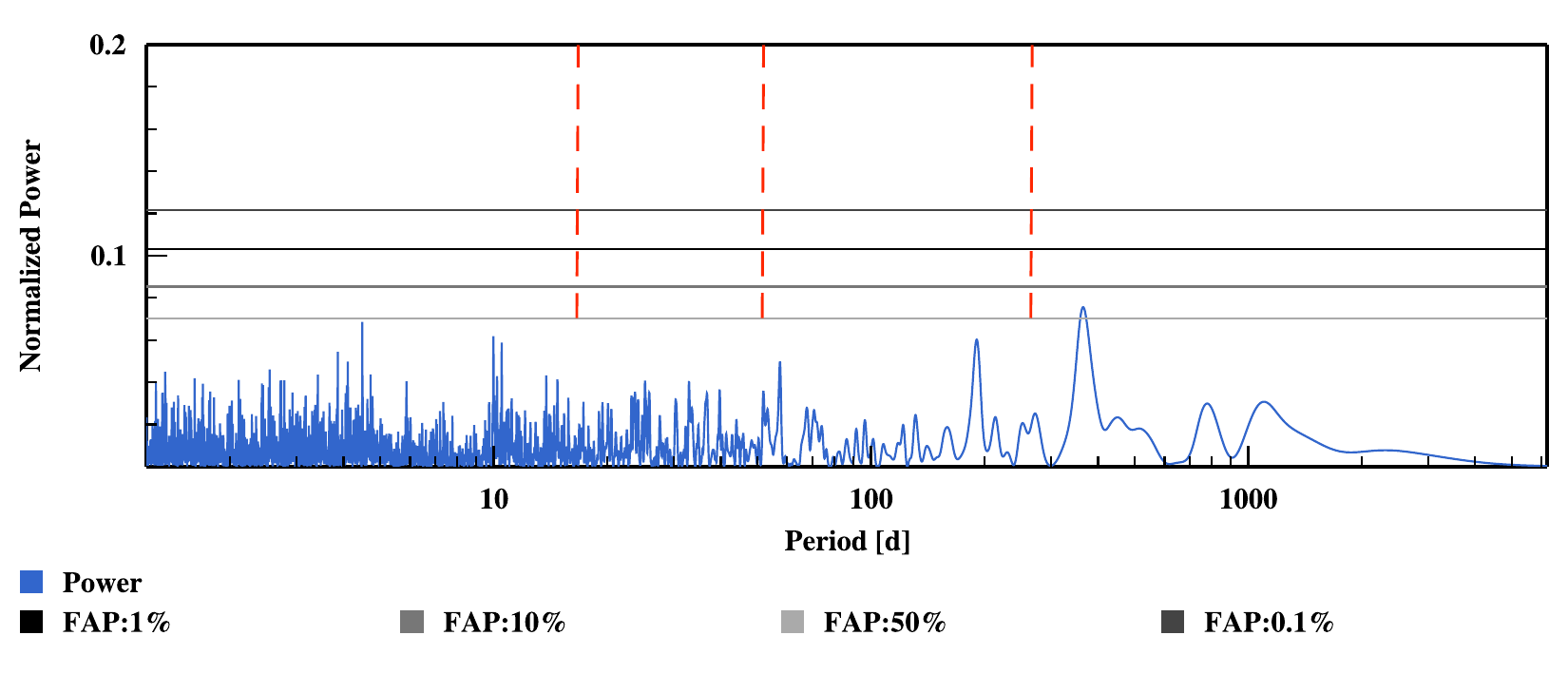}
   \caption[]{Same as Fig.\,\ref{app:figHD20003} but for \object{HD31527}}
\label{app:figHD31527}
\end{figure}

\begin{figure}[!h]
\center
 \includegraphics[angle=90,width=0.42\textwidth,origin=br]{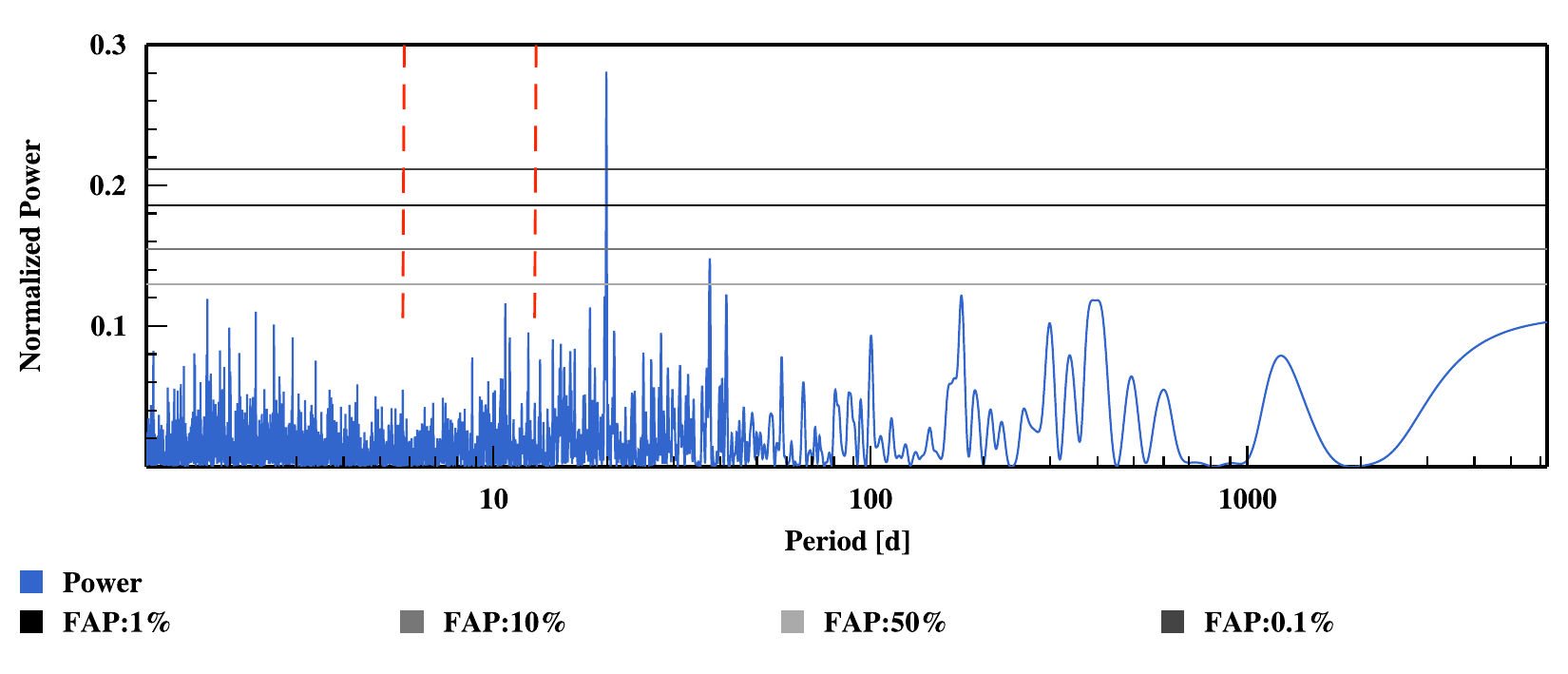}
  \includegraphics[angle=90,width=0.42\textwidth,origin=br]{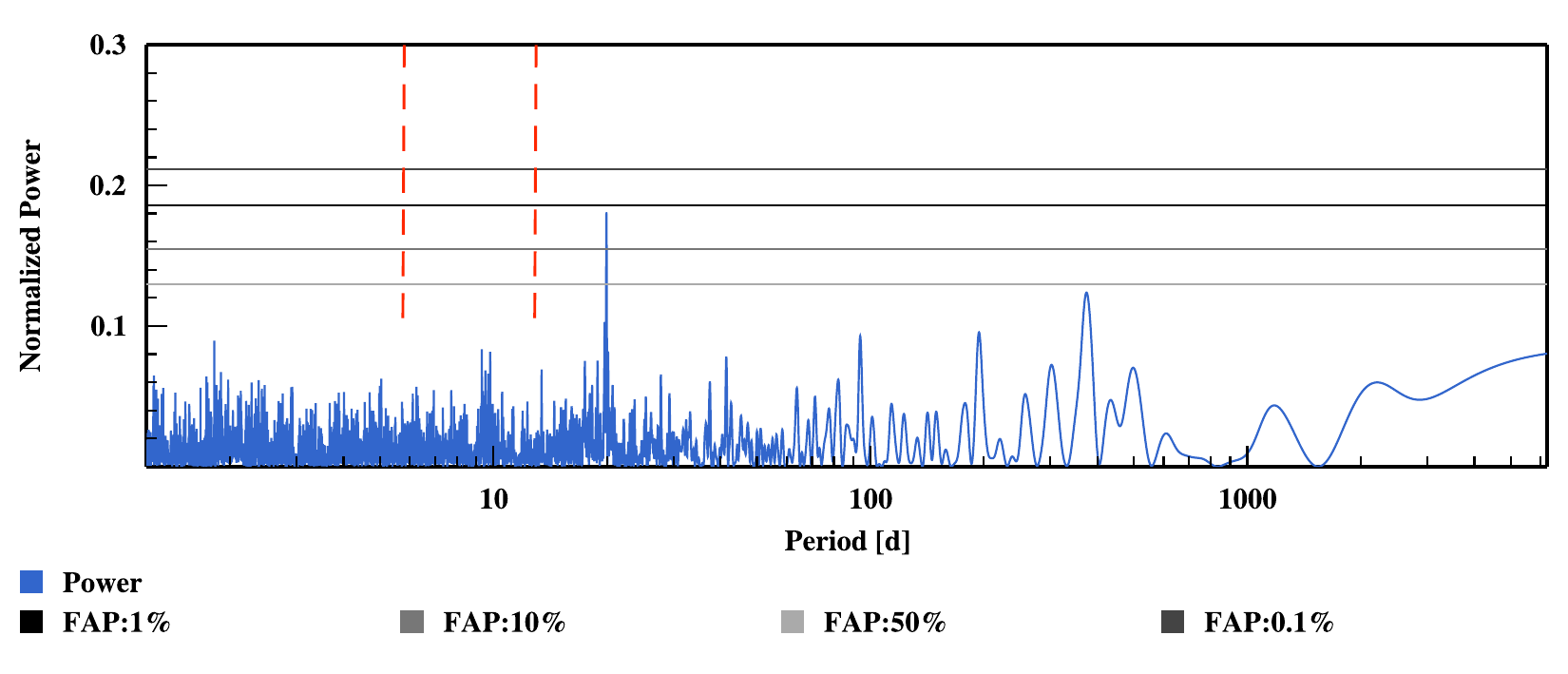}
   \includegraphics[angle=90,width=0.42\textwidth,origin=br]{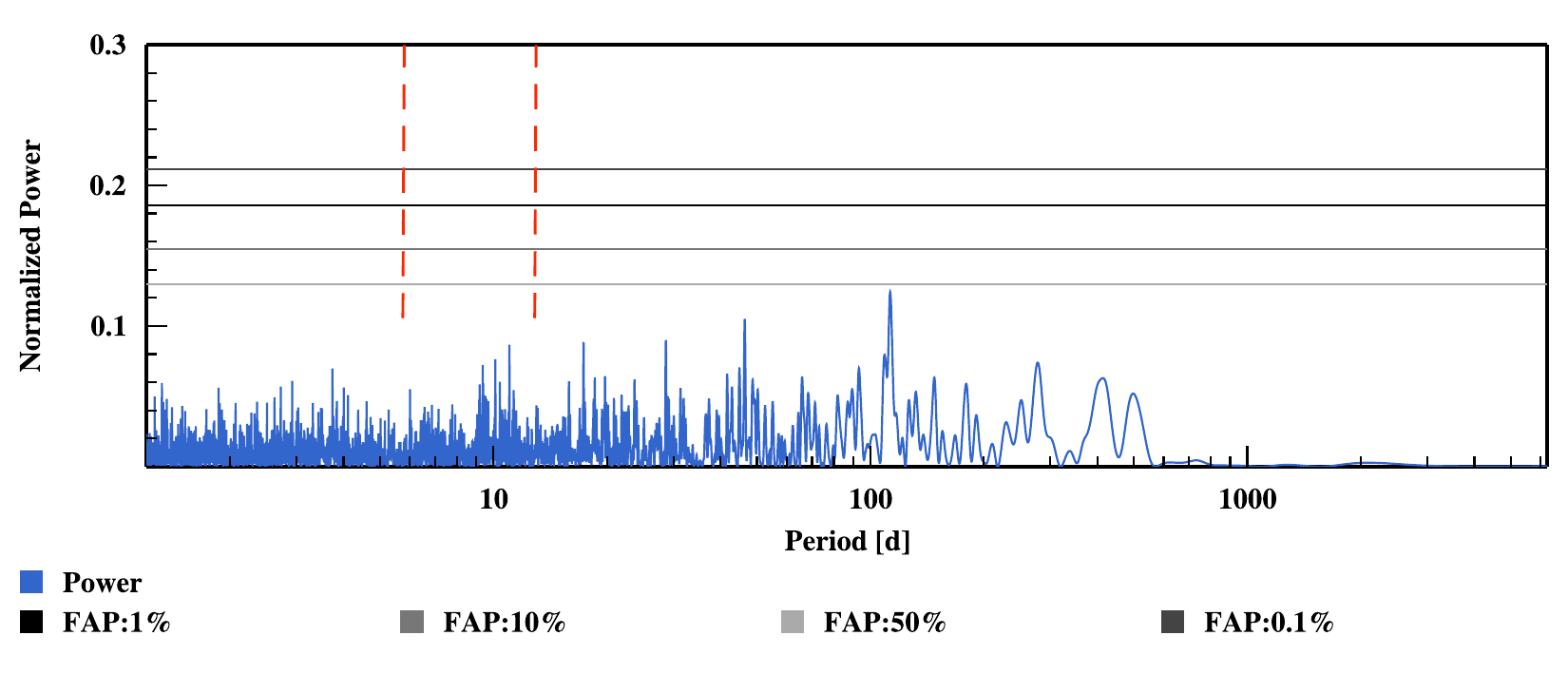}
   \caption[]{Same as Fig.\,\ref{app:figHD20003} but for \object{HD45184}}
\label{app:figHD45184}
\end{figure}

\begin{figure}[!h]
\center
 \includegraphics[angle=90,width=0.42\textwidth,origin=br]{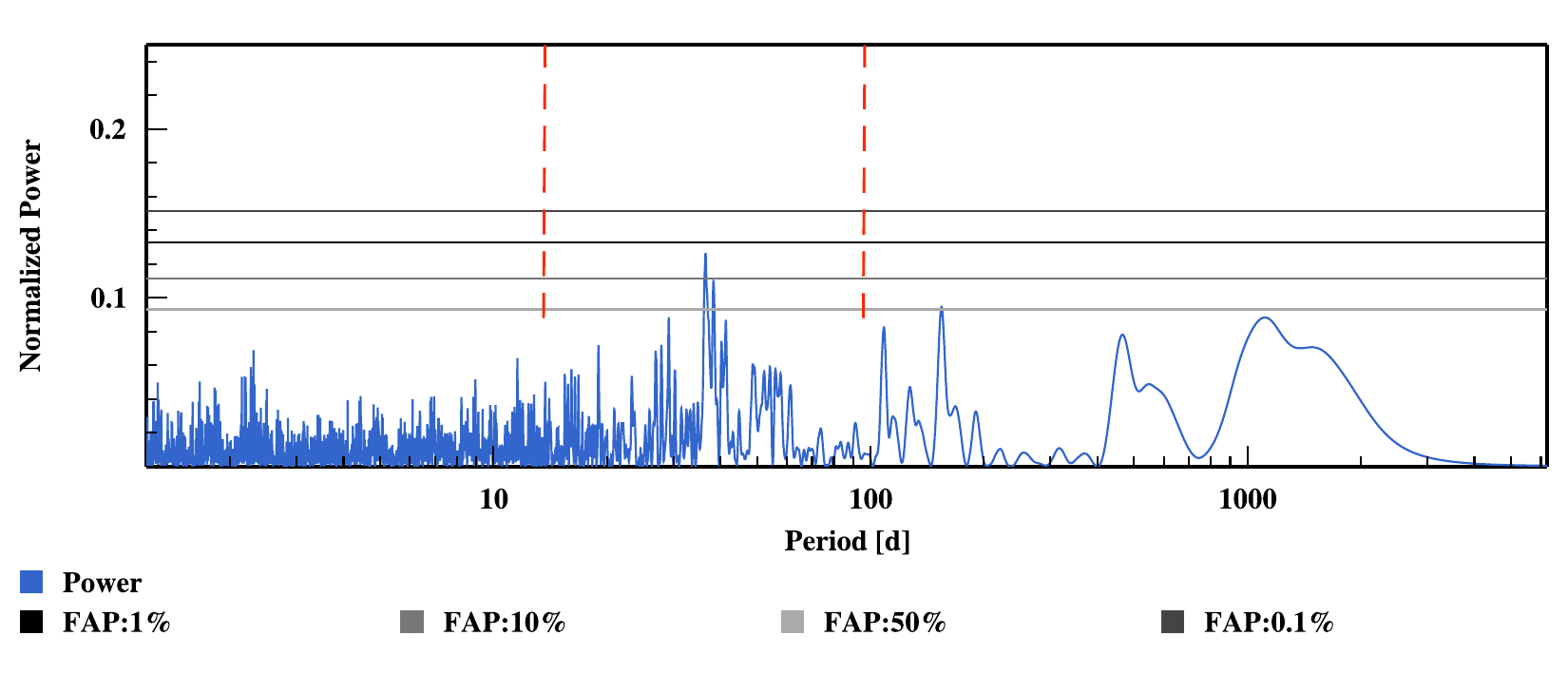}
  \includegraphics[angle=90,width=0.42\textwidth,origin=br]{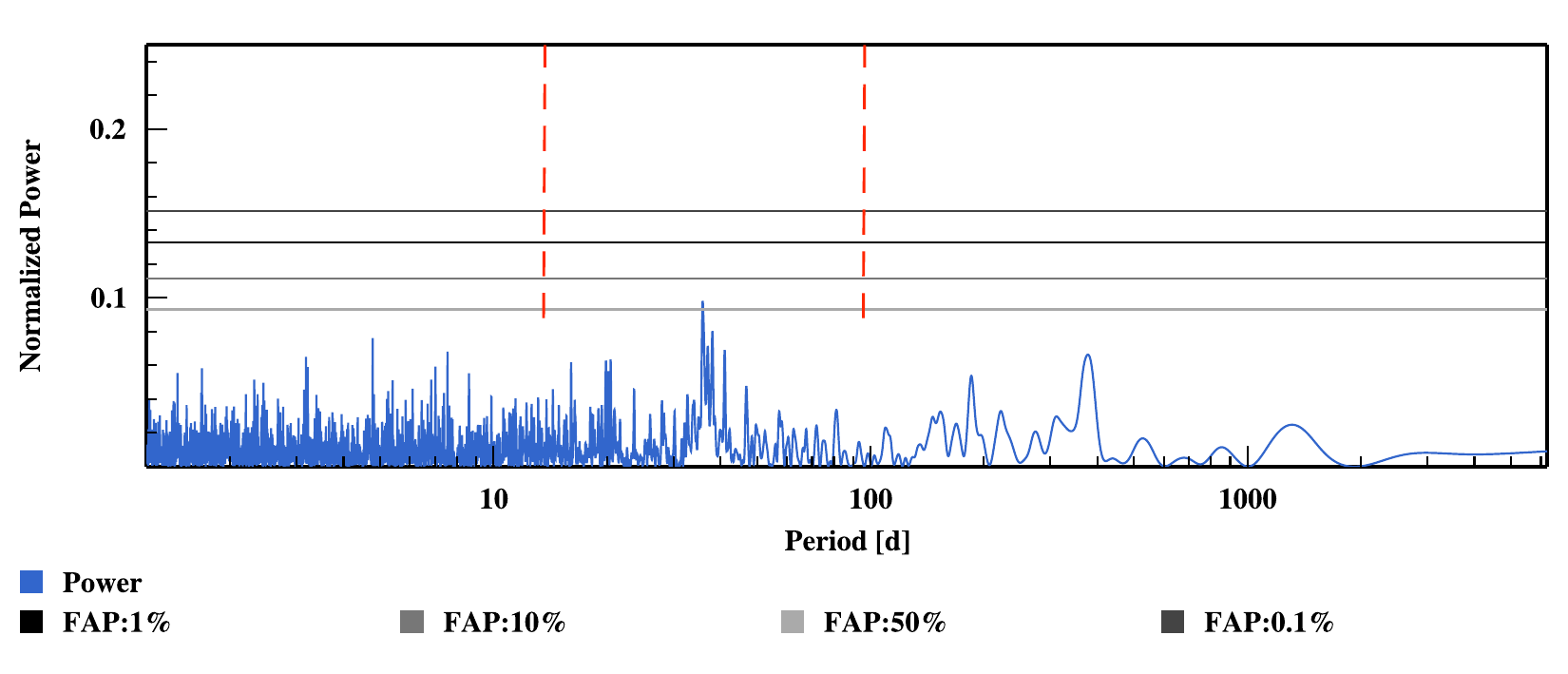}
   \includegraphics[angle=90,width=0.42\textwidth,origin=br]{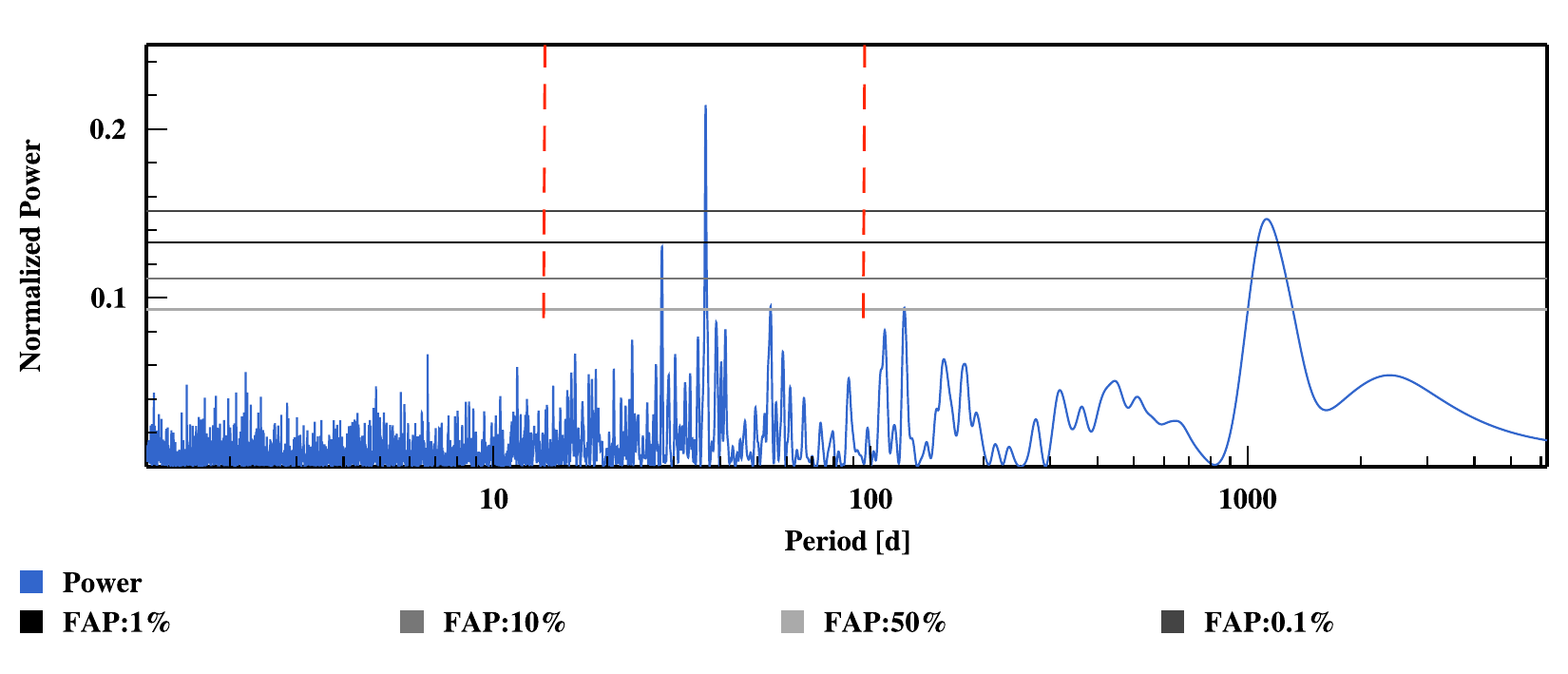}
   \caption[]{Same as Fig.\,\ref{app:figHD20003} but for \object{HD51608}}
\label{app:figHD51608}
\end{figure}

\begin{figure}[!h]
\center
 \includegraphics[angle=90,width=0.42\textwidth,origin=br]{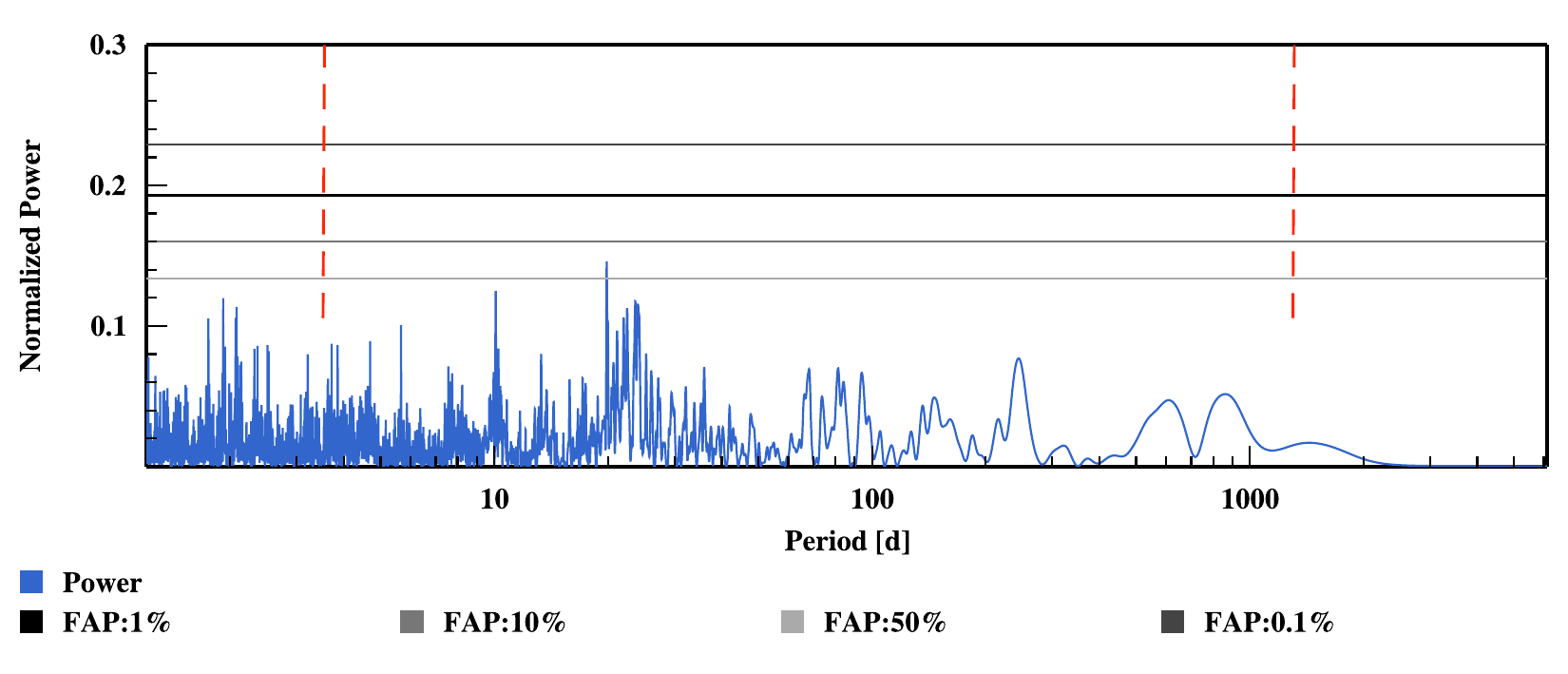}
  \includegraphics[angle=90,width=0.42\textwidth,origin=br]{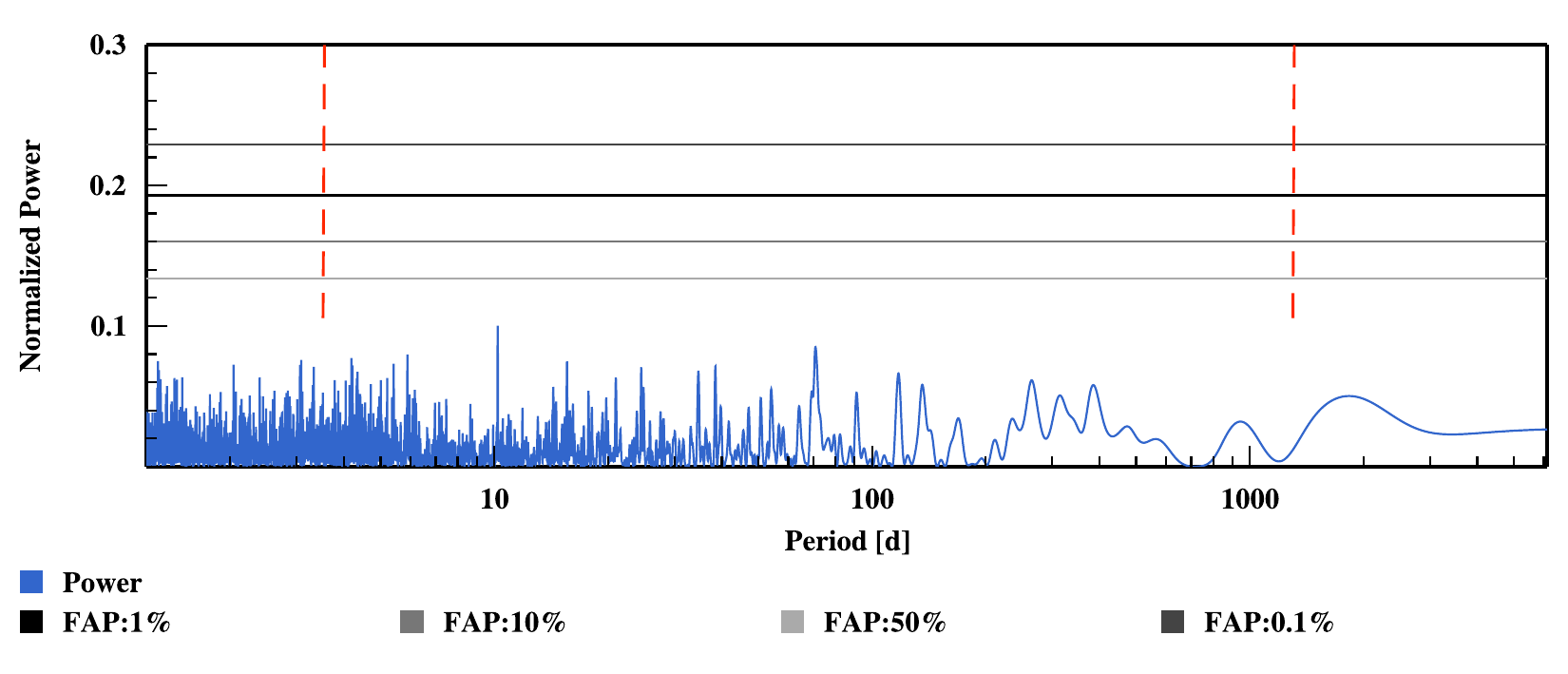}
   \includegraphics[angle=90,width=0.42\textwidth,origin=br]{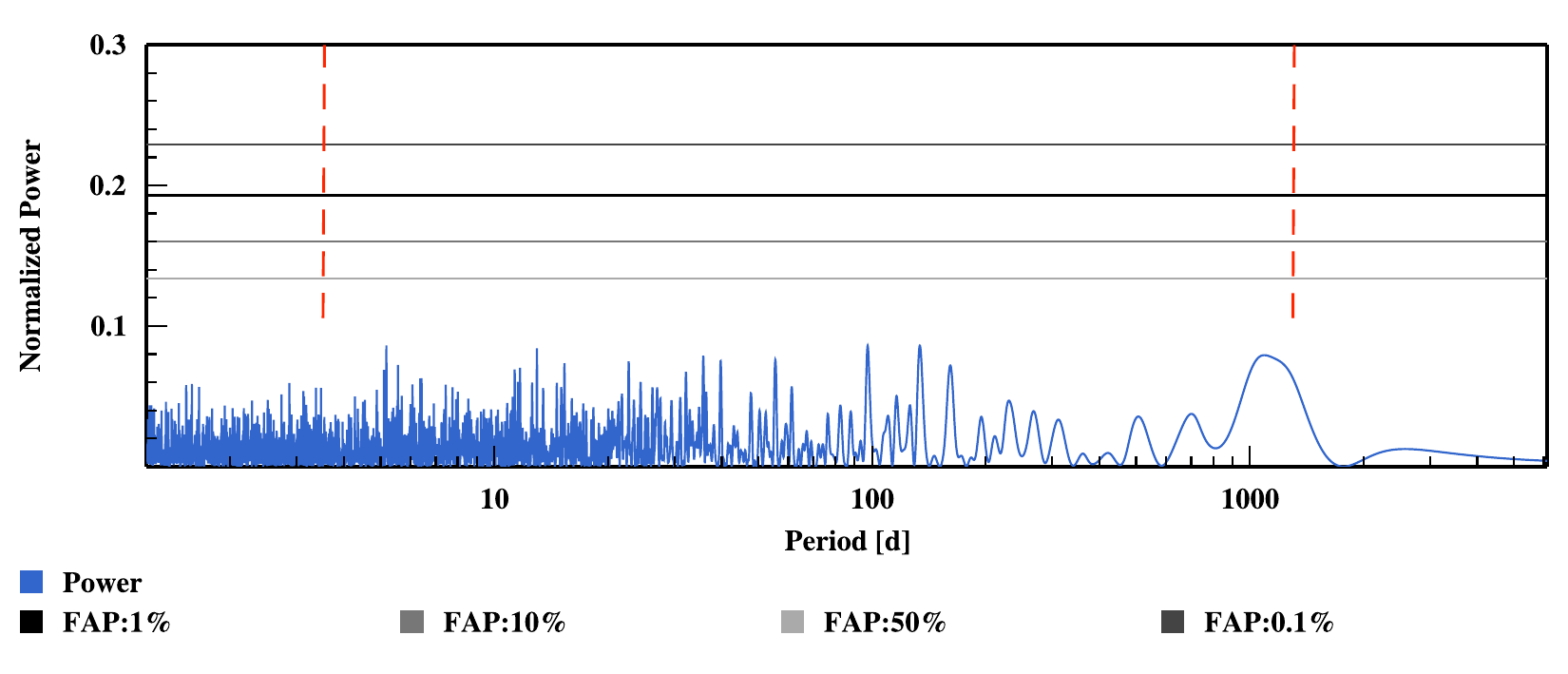}
   \caption[]{Same as Fig.\,\ref{app:figHD20003} but for \object{HD134060}}
\label{app:figHD134060}
\end{figure}

\begin{figure}[!h]
\center
 \includegraphics[angle=90,width=0.42\textwidth,origin=br]{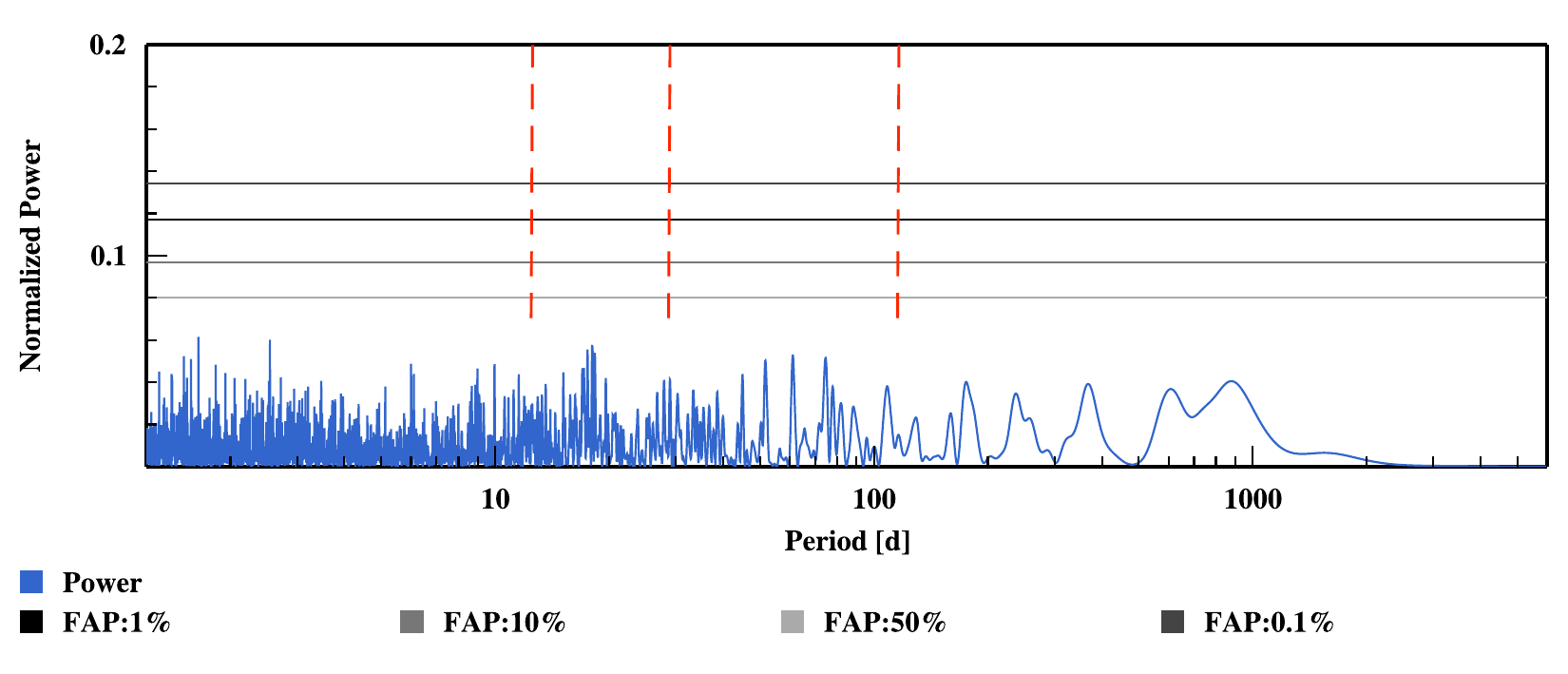}
  \includegraphics[angle=90,width=0.42\textwidth,origin=br]{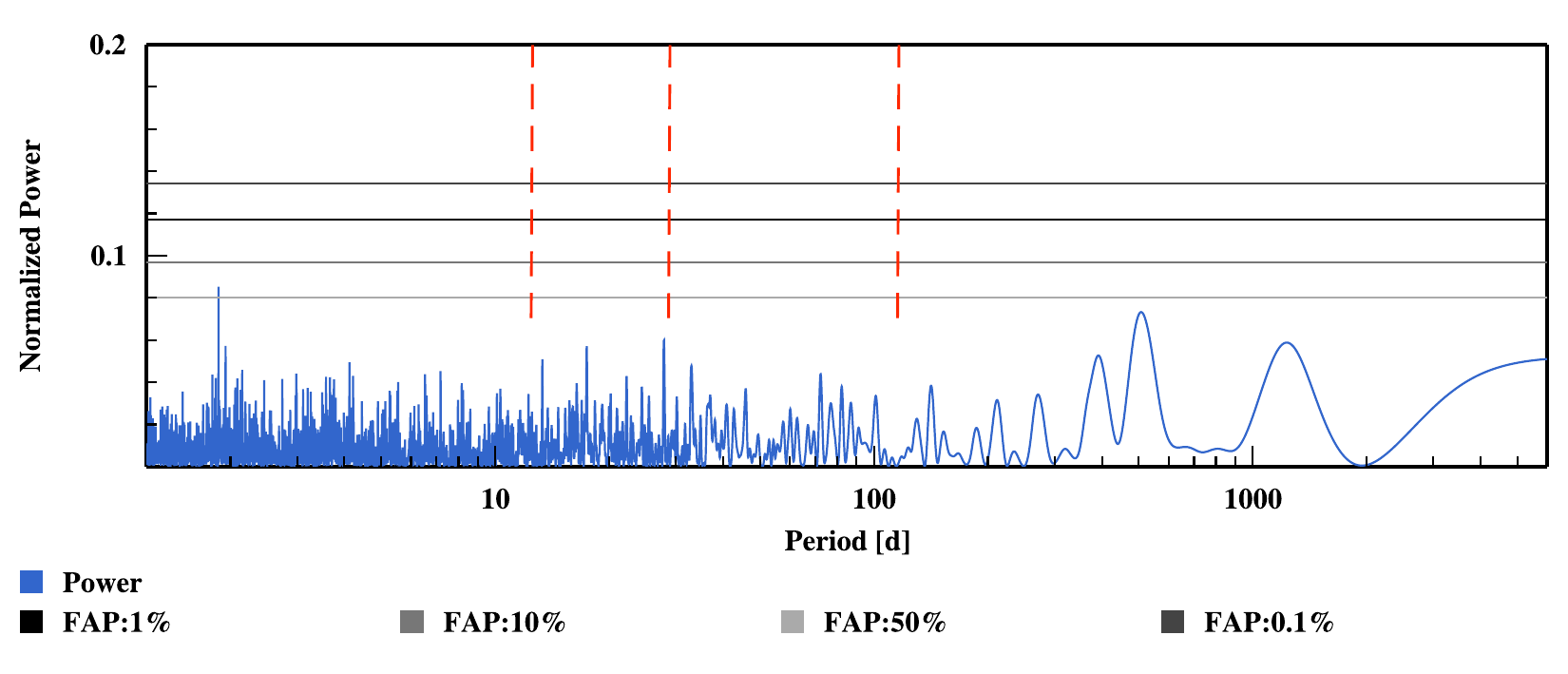}
   \includegraphics[angle=90,width=0.42\textwidth,origin=br]{HD136352_FWHM_res_d1_final}
   \caption[]{Same as Fig.\,\ref{app:figHD20003} but for \object{HD136352}}
\label{app:figHD136352}
\end{figure}

\clearpage
%\newpage

\section{Parameters probed by MCMC}

We present here the parameter estimates issued from the MCMC modeling of the planetary systems.

\begin{table*}
\scriptsize
\caption{Parameters probed by the MCMC used to fit the RV measurements of HD20003. The maximum likelihood solution (Max(Like)), the median (Med), mode (Mod) and standard deviation (Std) of the posterior distribution for each parameter is provided, as well as the 68.3\,\% (CI(15.85),CI(84.15)) and 95.45\,\% (CI(2.275),CI(97.725)) confidence intervals. The prior for each parameter can be of type: $\mathcal{U}$: uniform, $\mathcal{N}$: normal, $\mathcal{SN}$:split normal, $\mathcal{TN}$:truncated normal}.  
\label{HD20003_tab-mcmc-Probed_params}
\def\arraystretch{1.5}
\begin{center}
\begin{tabular}{lcclcccccccc}
\hline
\hline
Param. & Units & Max(Like) & Med & Mod &Std & CI(15.85) & CI(84.15) &CI(2.275) & CI(97.725) & Prior\\
\hline
\multicolumn{11}{c}{ \bf Likelihood}\\
\hline
$\log{(\rm Post})$&               &  -355.440573&  -364.186739&  -365.130519&     3.064118&  -367.902792&  -360.869032&  -372.251620&  -358.366379&               \\ 
$\log{(\rm Like)}$&               &  -354.388255&  -363.200339&  -364.552124&     3.075716&  -366.886986&  -359.786081&  -371.228449&  -357.313993&               \\ 
$\log{(\rm Prior)}$&               &    -1.052318&    -1.000952&    -1.023733&     0.258084&    -1.317939&    -0.740211&    -1.736133&    -0.528578&               \\ 
\hline 
M$_{\star}$    &[M$_{\odot}$]  &     0.947719&     0.877722&     0.888341&     0.086110&     0.777570&     0.972946&     0.677797&     1.077890&$\mathcal{N}(0.875,0.1)$\\ 
\hline 
$\sigma_{JIT}$ &[m\,s$^{-1}$]  &     1.45&     1.59&     1.63&     0.34&     1.17&     1.24&     0.77&     0.93&$\mathcal{U}$  \\ 
\hline 
$\gamma_{HARPS}$&[m\,s$^{-1}$]  &-16104.48&-16104.11&-16104.25&     0.31&-16104.39&-16103.74&-16104.65&-16103.07&$\mathcal{U}$  \\ 
\hline 
\hline 
$\log{(P)}$    &[d]            &     1.073639&     1.073654&     1.073651&     0.000052&     1.073593&     1.073713&     1.073536&     1.073771&$\mathcal{U}$  \\ 
$\log{(K)}$    &[m\,s$^{-1}$]  &     0.58&     0.58&     0.58&     0.02&     0.56&     0.61&     0.53&     0.63&$\mathcal{U}$  \\ 
$\sqrt{e}.\cos{\omega}$&               &    -0.021425&    -0.028442&    -0.023510&     0.077999&    -0.117180&     0.059704&    -0.209618&     0.142977&$\mathcal{U}$  \\ 
$\sqrt{e}.\sin{\omega}$&               &    -0.596953&    -0.593533&    -0.597847&     0.037382&    -0.633941&    -0.548640&    -0.670370&    -0.497859&$\mathcal{U}$  \\ 
$\lambda_{0}$  &[deg]          &   225.658516&   225.988820&   225.701948&     3.096407&   222.316287&   229.436249&   218.679062&   232.672381&$\mathcal{U}$  \\ 
\hline 
$\log{(P)}$    &[d]            &     1.530369&     1.530505&     1.530577&     0.000282&     1.530165&     1.530811&     1.529813&     1.531065&$\mathcal{U}$  \\ 
$\log{(K)}$    &[m\,s$^{-1}$]  &     0.49&     0.50&     0.50&     0.03&     0.47&     0.53&     0.44&     0.55&$\mathcal{U}$  \\ 
$\sqrt{e}.\cos{\omega}$&               &     0.089332&     0.163526&     0.213761&     0.146216&    -0.035613&     0.306443&    -0.203105&     0.412917&$\mathcal{U}$  \\ 
$\sqrt{e}.\sin{\omega}$&               &     0.339711&     0.219084&     0.272743&     0.133488&     0.030981&     0.334583&    -0.157897&     0.421322&$\mathcal{U}$  \\ 
$\lambda_{0}$  &[deg]          &   120.340507&   124.671746&   124.146671&     3.478923&   120.635953&   128.547476&   116.645239&   132.285391&$\mathcal{U}$  \\ 
\hline 
$\log{(P)}$    &[d]            &     2.262470&     2.263926&     2.263963&     0.002232&     2.261303&     2.266297&     2.258300&     2.268859&$\mathcal{U}$  \\ 
$\log{(K)}$    &[m\,s$^{-1}$]  &     0.23&     0.21&     0.20&     0.05&     0.15&     0.26&     0.08&     0.30&$\mathcal{U}$  \\ 
$\sqrt{e}.\cos{\omega}$&               &     0.191246&     0.167121&     0.209789&     0.189206&    -0.078520&     0.364574&    -0.282064&     0.506428&$\mathcal{U}$  \\ 
$\sqrt{e}.\sin{\omega}$&               &     0.419103&     0.213797&     0.320208&     0.195134&    -0.049677&     0.402924&    -0.274331&     0.537274&$\mathcal{U}$  \\ 
$\lambda_{0}$  &[deg]          &   197.675128&   196.751843&   196.887430&     6.258785&   189.645165&   203.799985&   182.342310&   211.192757&$\mathcal{U}$  \\ 
\hline 
$\log{(P)}$    &[d]            &     3.582659&     3.606144&     3.595842&     0.028188&     3.580559&     3.640255&     3.557643&     3.702132&$\mathcal{U}$  \\ 
$\log{(K)}$    &[m\,s$^{-1}$]  &     0.74&     0.76&     0.75&     0.02&     0.73&     0.78&     0.71&     0.80&$\mathcal{U}$  \\ 
$\sqrt{e}.\cos{\omega}$&               &     0.028669&    -0.069460&    -0.042689&     0.161723&    -0.257669&     0.126252&    -0.420089&     0.262372&$\mathcal{U}$  \\ 
$\sqrt{e}.\sin{\omega}$&               &    -0.205147&    -0.146710&    -0.208581&     0.122078&    -0.262944&     0.019978&    -0.347443&     0.162734&$\mathcal{U}$  \\ 
$\lambda_{0}$  &[deg]          &   137.392067&   139.338913&   137.305526&     3.530496&   135.946353&   143.669909&   132.569853&   149.602355&$\mathcal{U}$  \\ 
\hline 
\end{tabular}
\end{center}
\end{table*}

\begin{table*}
\scriptsize
\caption{Same as Table\,\ref{HD20003_tab-mcmc-Probed_params} for \object{HD\,20781}}
 \label{HD20781_tab-mcmc-Probed_params}
\def\arraystretch{1.5}
\begin{center}
\begin{tabular}{lcclcccccccc}
\hline
\hline
Param. & Units & Max(Like) & Med & Mod &Std & CI(15.85) & CI(84.15) &CI(2.275) & CI(97.725) & Prior\\
\hline
\multicolumn{11}{c}{ \bf Likelihood}\\
\hline
$\log{(\rm Post})$&               &  -388.542397&  -397.139540&  -396.801832&     3.086207&  -400.992042&  -394.045214&  -405.822507&  -391.632479&               \\ 
$\log{(\rm Like)}$&               &  -388.379068&  -396.816561&  -396.641259&     3.045776&  -400.618909&  -393.744957&  -405.246050&  -391.364866&               \\ 
$\log{(\rm Prior)}$&               &    -0.163328&    -0.271536&    -0.192668&     0.188786&    -0.532565&    -0.133491&    -0.959781&    -0.053132&               \\ 
\hline 
M$_{\star}$    &[M$_{\odot}$]  &     0.798309&     0.700500&     0.675260&     0.088466&     0.599877&     0.801716&     0.500287&     0.905060&$\mathcal{N}(0.7,0.1)$\\ 
\hline 
$\sigma_{JIT}$ &[m\,s$^{-1}$]  &     1.14&     1.27&     1.21&     0.19&     1.09&     1.01&     1.00&     0.69&$\mathcal{U}$  \\ 
\hline 
$\gamma_{HARPS}$&[m\,s$^{-1}$]  & 40369.22& 40369.21& 40369.17&     0.10& 40369.10& 40369.32& 40368.98& 40369.44&$\mathcal{U}$  \\ 
\hline 
$\log{(P)}$    &[d]            &     0.725379&     0.725377&     0.725315&     0.000076&     0.725292&     0.725459&     0.725196&     0.725567&$\mathcal{U}$  \\ 
$\log{(K)}$    &[m\,s$^{-1}$]  &     0.05&    -0.04&    -0.04&     0.07&    -0.12&     0.03&    -0.22&     0.09&$\mathcal{U}$  \\ 
$\sqrt{e}.\cos{\omega}$&               &    -0.061140&     0.043862&     0.039181&     0.230069&    -0.243492&     0.299379&    -0.460848&     0.499573&$\mathcal{U}$  \\ 
$\sqrt{e}.\sin{\omega}$&               &     0.216991&    -0.013862&    -0.028798&     0.215428&    -0.264618&     0.241274&    -0.455895&     0.436077&$\mathcal{U}$  \\ 
$\lambda_{0}$  &[deg]          &  -135.345360&  -134.251473&  -137.352873&     9.387901&  -144.854378&  -123.506806&  -156.133446&  -112.221519&$\mathcal{U}$  \\ 
\hline 
$\log{(P)}$    &[d]            &     1.142791&     1.142717&     1.142690&     0.000092&     1.142611&     1.142819&     1.142490&     1.142923&$\mathcal{U}$  \\ 
$\log{(K)}$    &[m\,s$^{-1}$]  &     0.29&     0.26&     0.24&     0.03&     0.22&     0.29&     0.18&     0.33&$\mathcal{U}$  \\ 
$\sqrt{e}.\cos{\omega}$&               &     0.327288&     0.203741&     0.321124&     0.172122&    -0.029464&     0.370760&    -0.219811&     0.484086&$\mathcal{U}$  \\ 
$\sqrt{e}.\sin{\omega}$&               &     0.021983&     0.026817&     0.016131&     0.159055&    -0.169199&     0.211170&    -0.318653&     0.343092&$\mathcal{U}$  \\ 
$\lambda_{0}$  &[deg]          &   -84.722610&   -83.737346&   -84.982562&     4.380473&   -88.530458&   -78.631060&   -93.607328&   -73.479892&$\mathcal{U}$  \\ 
\hline 
$\log{(P)}$    &[d]            &     1.464692&     1.464758&     1.464755&     0.000134&     1.464609&     1.464910&     1.464452&     1.465071&$\mathcal{U}$  \\ 
$\log{(K)}$    &[m\,s$^{-1}$]  &     0.45&     0.45&     0.45&     0.02&     0.42&     0.48&     0.40&     0.50&$\mathcal{U}$  \\ 
$\sqrt{e}.\cos{\omega}$&               &     0.139677&     0.145327&     0.162099&     0.120006&    -0.010957&     0.265346&    -0.162692&     0.358965&$\mathcal{U}$  \\ 
$\sqrt{e}.\sin{\omega}$&               &     0.246274&     0.263747&     0.306374&     0.113409&     0.114332&     0.358597&    -0.077325&     0.434823&$\mathcal{U}$  \\ 
$\lambda_{0}$  &[deg]          &   -77.213730&   -78.248282&   -78.600903&     2.996830&   -81.624509&   -74.896708&   -84.983023&   -71.294902&$\mathcal{U}$  \\ 
\hline 
$\log{(P)}$    &[d]            &     1.931848&     1.932003&     1.932093&     0.000430&     1.931522&     1.932502&     1.931041&     1.932995&$\mathcal{U}$  \\ 
$\log{(K)}$    &[m\,s$^{-1}$]  &     0.43&     0.41&     0.41&     0.02&     0.39&     0.44&     0.37&     0.46&$\mathcal{U}$  \\ 
$\sqrt{e}.\cos{\omega}$&               &     0.148239&     0.050362&     0.090292&     0.146393&    -0.126620&     0.219662&    -0.260249&     0.340229&$\mathcal{U}$  \\ 
$\sqrt{e}.\sin{\omega}$&               &     0.268142&     0.155008&     0.226421&     0.139655&    -0.035773&     0.293236&    -0.188818&     0.387469&$\mathcal{U}$  \\ 
$\lambda_{0}$  &[deg]          &    12.584852&    15.024637&    14.339843&     3.408448&    11.079349&    18.883891&     7.424984&    22.814245&$\mathcal{U}$  \\ 
\hline 
\end{tabular}
\end{center}
\end{table*}

\begin{table*}
\scriptsize
\caption{Same as Table\,\ref{HD20003_tab-mcmc-Probed_params} for \object{HD\,20782}}
 \label{HD20782_tab-mcmc-Probed_params}
\def\arraystretch{1.5}
\begin{tabular}{lcclcccccccc}
\hline
\hline
Param. & Units & Max(Like) & Med & Mod &Std & CI(15.85) & CI(84.15) &CI(2.275) & CI(97.725) & Prior\\
\hline
\multicolumn{11}{c}{ \bf Likelihood}\\
\hline
$\log{(\rm Post})$&               &  -372.933997&  -379.659609&  -379.184909&     2.652630&  -382.940155&  -377.007478&  -387.071255&  -375.024009&               \\ 
$\log{(\rm Like)}$&               &  -367.930148&  -374.651836&  -374.180899&     2.652279&  -377.934595&  -371.997235&  -382.071125&  -370.018358&               \\ 
$\log{(\rm Prior)}$&               &    -5.003849&    -5.008809&    -5.011567&     0.009534&    -5.019713&    -4.997733&    -5.030582&    -4.987338&               \\ 
\hline 
M$_{\star}$    &[M$_{\odot}$]  &     0.985227&     0.950073&     0.916830&     0.087670&     0.850317&     1.050787&     0.755244&     1.148009&$\mathcal{N}(0.95,0.1)$\\ 
\hline 
$\sigma_{COR07-DRS-3-4}$&[m\,s$^{-1}$]  &     5.33&     6.45&     5.20&     1.77&     4.84&     8.72&     3.61&    12.03&$\mathcal{U}$  \\ 
$\sigma_{COR14-DRS-3-8}$&[m\,s$^{-1}$]  &     4.32&     4.66&     4.01&     1.59&     3.08&     6.58&     1.49&     9.29&$\mathcal{U}$  \\ 
$\sigma_{COR98-DRS-3-3}$&[m\,s$^{-1}$]  &     5.26&     5.88&     5.33&     1.74&     4.14&     7.98&     2.33&    10.72&$\mathcal{U}$  \\ 
$\sigma_{HARPS}$&[m\,s$^{-1}$]  &     1.42&     1.49&     1.47&     0.21&     1.25&     1.71&     1.00&     1.95&$\mathcal{U}$  \\ 
$\sigma_{UCLES-PUB-2006}$&[m\,s$^{-1}$]  &     0.06&     1.60&     0.00&     1.05&     0.49&     2.95&     0.07&     4.41&$\mathcal{U}$  \\ 
$\sigma_{JIT}$ &[m\,s$^{-1}$]  &     1.29&     0.94&     1.00&     0.46&     0.37&     1.47&     0.06&     1.93&$\mathcal{U}$  \\ 
\hline 
$\gamma_{COR07-DRS-3-4}$&[m\,s$^{-1}$]  & 39931.00& 39930.61& 39930.34&     1.92& 39928.44& 39932.70& 39925.88& 39935.32&$\mathcal{U}$  \\ 
$\gamma_{COR14-DRS-3-8}$&[m\,s$^{-1}$]  & 39958.04& 39956.56& 39956.34&     1.46& 39954.89& 39958.14& 39952.82& 39959.96&$\mathcal{U}$  \\ 
$\gamma_{COR98-DRS-3-3}$&[m\,s$^{-1}$]  & 39929.00& 39928.10& 39927.80&     1.68& 39926.19& 39929.98& 39924.15& 39932.04&$\mathcal{U}$  \\ 
$\gamma_{HARPS}$&[m\,s$^{-1}$]  & 39964.88& 39964.81& 39964.82&     0.20& 39964.58& 39965.03& 39964.35& 39965.25&$\mathcal{U}$  \\ 
$\mathrm{offset}_{UCLES-PUB-2006}$&[m\,s$^{-1}$]  &     5.30&     5.44&     5.26&     0.83&     4.51&     6.37&     3.53&     7.36&$\mathcal{U}$  \\ 
\hline 
$\log{(P)}$    &[d]            &     2.776014&     2.776021&     2.776014&     0.000016&     2.776003&     2.776040&     2.775987&     2.776056&$\mathcal{U}$  \\ 
$\log{(K)}$    &[m\,s$^{-1}$]  &     2.07&     2.07&     2.07&     0.01&     2.07&     2.08&     2.06&     2.09&$\mathcal{U}$  \\ 
$\sqrt{e}.\cos{\omega}$&               &    -0.780714&    -0.784083&    -0.787192&     0.005216&    -0.789895&    -0.777061&    -0.793328&    -0.773612&$\mathcal{U}$  \\ 
$\sqrt{e}.\sin{\omega}$&               &     0.582695&     0.578516&     0.573850&     0.006592&     0.571141&     0.587284&     0.566660&     0.591712&$\mathcal{U}$  \\ 
$\lambda_{0}$  &[deg]          &   295.804494&   296.111799&   296.327571&     0.488659&   295.460223&   296.661315&   295.142950&   296.986011&$\mathcal{U}$  \\ 
\hline 
\end{tabular}
\end{table*}

\begin{table*}
\scriptsize
\caption{Same as Table\,\ref{HD20003_tab-mcmc-Probed_params} for \object{HD\,21693}}
 \label{HD21693_tab-mcmc-Probed_params}
\def\arraystretch{1.5}
\begin{center}
\begin{tabular}{lcclcccccccc}
\hline
\hline
Param. & Units & Max(Like) & Med & Mod &Std & CI(15.85) & CI(84.15) &CI(2.275) & CI(97.725) & Prior\\
\hline
\multicolumn{11}{c}{ \bf Likelihood}\\
\hline
$\log{(\rm Post})$&               &  -435.045817&  -440.915985&  -440.662807&     2.436011&  -444.034771&  -438.620178&  -447.941122&  -436.845354&               \\ 
$\log{(\rm Like)}$&               &  -434.994431&  -440.747390&  -440.523392&     2.426114&  -443.837350&  -438.463925&  -447.707065&  -436.680607&               \\ 
$\log{(\rm Prior)}$&               &    -0.051386&    -0.133026&    -0.077867&     0.126005&    -0.308420&    -0.041395&    -0.592687&    -0.007046&               \\ 
\hline 
M$_{\star}$    &[M$_{\odot}$]  &     0.627847&     0.799037&     0.801958&     0.087416&     0.699111&     0.898724&     0.600636&     0.995794&$\mathcal{N}(0.8,0.1)$\\ 
\hline 
$\sigma_{JIT\,LOW}$&[m\,s$^{-1}$]  &     1.27&     1.26&     1.25&     0.44&     0.73&     0.79&     0.14&     0.38&$\mathcal{U}$  \\ 
$\sigma_{JIT\,HIGH}$&[m\,s$^{-1}$]  &     2.69&     2.56&     2.70&     0.56&     1.86&     2.06&     0.85&     1.60&$\mathcal{U}$  \\ 
\hline 
$\gamma_{HARPS}$&[m\,s$^{-1}$]  & 39768.81& 39768.81& 39768.82&     0.13& 39768.66& 39768.95& 39768.52& 39769.09&$\mathcal{U}$  \\ 
\hline 
RHK$_{\rm index}$\,$lin$&[\ms]&    11.066005&    10.860734&    10.732702&     0.440301&    10.361899&    11.369837&     9.877045&    11.871009&$\mathcal{U}$  \\ 
\hline 
$\log{(P)}$    &[d]            &     1.355767&     1.355617&     1.355612&     0.000146&     1.355449&     1.355780&     1.355285&     1.355943&$\mathcal{U}$  \\ 
$\log{(K)}$    &[m\,s$^{-1}$]  &     0.35&     0.34&     0.34&     0.04&     0.30&     0.38&     0.24&     0.42&$\mathcal{U}$  \\ 
$\sqrt{e}.\cos{\omega}$&               &    -0.033137&    -0.003440&    -0.067300&     0.179167&    -0.215702&     0.205008&    -0.392233&     0.358369&$\mathcal{U}$  \\ 
$\sqrt{e}.\sin{\omega}$&               &    -0.247432&    -0.267481&    -0.370194&     0.162410&    -0.411532&    -0.042976&    -0.522567&     0.169768&$\mathcal{U}$  \\ 
$\lambda_{0}$  &[deg]          &    34.107432&    35.450889&    33.905458&     4.533112&    30.349272&    40.588204&    25.065784&    46.053381&$\mathcal{U}$  \\ 
\hline 
$\log{(P)}$    &[d]            &     1.730230&     1.730263&     1.730205&     0.000221&     1.730013&     1.730515&     1.729784&     1.730800&$\mathcal{U}$  \\ 
$\log{(K)}$    &[m\,s$^{-1}$]  &     0.54&     0.54&     0.54&     0.02&     0.51&     0.56&     0.48&     0.59&$\mathcal{U}$  \\ 
$\sqrt{e}.\cos{\omega}$&               &     0.269425&     0.196029&     0.243736&     0.136666&     0.005735&     0.324520&    -0.157091&     0.416385&$\mathcal{U}$  \\ 
$\sqrt{e}.\sin{\omega}$&               &    -0.025692&    -0.063953&    -0.075276&     0.137633&    -0.218240&     0.105252&    -0.335603&     0.237976&$\mathcal{U}$  \\ 
$\lambda_{0}$  &[deg]          &  -206.679521&  -207.980760&  -207.807573&     2.957609&  -211.285700&  -204.615563&  -214.868823&  -201.248502&$\mathcal{U}$  \\ 
\hline 
\end{tabular}
\end{center}
\end{table*}

\begin{table*}
\scriptsize
\caption{Same as Table\,\ref{HD20003_tab-mcmc-Probed_params} for \object{HD\,31527}}
 \label{HD31527_tab-mcmc-Probed_params}
\def\arraystretch{1.5}
\begin{center}
\begin{tabular}{lcclcccccccc}
\hline
\hline
Param. & Units & Max(Like) & Med & Mod &Std & CI(15.85) & CI(84.15) &CI(2.275) & CI(97.725) & Prior\\
\hline
\multicolumn{11}{c}{ \bf Likelihood}\\
\hline
$\log{(\rm Post})$&               &  -431.457393&  -439.413776&  -439.404479&     2.693424&  -442.806691&  -436.678805&  -446.790781&  -434.536716&               \\ 
$\log{(\rm Like)}$&               &  -430.869912&  -438.920421&  -438.066741&     2.744421&  -442.319982&  -436.104557&  -446.392727&  -433.960853&               \\ 
$\log{(\rm Prior)}$&               &    -0.587481&    -0.421477&    -0.228174&     0.322560&    -0.872877&    -0.155258&    -1.460968&    -0.044428&               \\ 
\hline 
M$_{\star}$    &[M$_{\odot}$]  &     0.964548&     0.960466&     0.935623&     0.087688&     0.861241&     1.058519&     0.761776&     1.159099&$\mathcal{N}(0.96,0.1)$\\ 
\hline 
$\sigma_{JIT}$ &[m\,s$^{-1}$]  &     1.24&     1.29&     1.29&     0.30&     0.95&     0.93&     0.77&     0.62&$\mathcal{U}$  \\ 
\hline 
$\gamma_{HARPS}$&[m\,s$^{-1}$]  & 25739.68& 25739.70& 25739.69&     0.09& 25739.61& 25739.80& 25739.51& 25739.90&$\mathcal{U}$  \\ 
\hline 
$\log{(P)}$    &[d]            &     1.218857&     1.218889&     1.218885&     0.000080&     1.218798&     1.218979&     1.218706&     1.219071&$\mathcal{U}$  \\ 
$\log{(K)}$    &[m\,s$^{-1}$]  &     0.43&     0.43&     0.43&     0.02&     0.41&     0.46&     0.39&     0.48&$\mathcal{U}$  \\ 
$\sqrt{e}.\cos{\omega}$&               &     0.242808&     0.210416&     0.235029&     0.104713&     0.075031&     0.311766&    -0.082660&     0.383529&$\mathcal{U}$  \\ 
$\sqrt{e}.\sin{\omega}$&               &     0.236848&     0.188237&     0.211870&     0.120112&     0.021413&     0.298912&    -0.126407&     0.384773&$\mathcal{U}$  \\ 
$\lambda_{0}$  &[deg]          &    52.547369&    51.262624&    51.323989&     2.622456&    48.258036&    54.220352&    45.382590&    57.230513&$\mathcal{U}$  \\ 
\hline 
$\log{(P)}$    &[d]            &     1.709426&     1.709315&     1.709215&     0.000282&     1.709003&     1.709631&     1.708664&     1.709963&$\mathcal{U}$  \\ 
$\log{(K)}$    &[m\,s$^{-1}$]  &     0.40&     0.40&     0.40&     0.02&     0.37&     0.42&     0.35&     0.45&$\mathcal{U}$  \\ 
$\sqrt{e}.\cos{\omega}$&               &     0.023117&     0.081468&     0.086880&     0.144497&    -0.097212&     0.242734&    -0.237589&     0.358056&$\mathcal{U}$  \\ 
$\sqrt{e}.\sin{\omega}$&               &    -0.164323&     0.014834&    -0.010018&     0.126248&    -0.134951&     0.164944&    -0.248677&     0.274415&$\mathcal{U}$  \\ 
$\lambda_{0}$  &[deg]          &  -208.406874&  -207.583870&  -207.914089&     2.785788&  -210.745542&  -204.465797&  -213.962526&  -201.198653&$\mathcal{U}$  \\ 
\hline 
$\log{(P)}$    &[d]            &     2.434889&     2.434048&     2.433252&     0.003107&     2.430440&     2.437413&     2.426290&     2.440704&$\mathcal{U}$  \\ 
$\log{(K)}$    &[m\,s$^{-1}$]  &     0.11&     0.10&     0.09&     0.05&     0.04&     0.15&    -0.03&     0.21&$\mathcal{U}$  \\ 
$\sqrt{e}.\cos{\omega}$&               &    -0.541593&    -0.450052&    -0.536450&     0.169210&    -0.588081&    -0.235463&    -0.692747&     0.104560&$\mathcal{U}$  \\ 
$\sqrt{e}.\sin{\omega}$&               &     0.097884&     0.006756&     0.058079&     0.173485&    -0.200976&     0.195484&    -0.403356&     0.363079&$\mathcal{U}$  \\ 
$\lambda_{0}$  &[deg]          &  -111.177767&  -110.997045&  -113.124819&     5.834754&  -117.548307&  -104.311167&  -124.016867&   -97.254697&$\mathcal{U}$  \\ 
\hline 
\end{tabular}
\end{center}
\end{table*}

\begin{table*}
\scriptsize
\caption{Same as Table\,\ref{HD20003_tab-mcmc-Probed_params} for \object{HD\,45184}}
 \label{HD45184_tab-mcmc-Probed_params}
\def\arraystretch{1.5}
\begin{center}
\begin{tabular}{lcclcccccccc}
\hline
\hline
Param. & Units & Max(Like) & Med & Mod &Std & CI(15.85) & CI(84.15) &CI(2.275) & CI(97.725) & Prior\\
\hline
\multicolumn{11}{c}{ \bf Likelihood}\\
\hline
$\log{(\rm Post})$&               &  -375.948188&  -382.533843&  -381.953342&     2.376383&  -385.497577&  -380.151077&  -389.208051&  -378.372729&               \\ 
$\log{(\rm Like)}$&               &  -375.907261&  -382.434308&  -381.860465&     2.358405&  -385.390539&  -380.066476&  -389.064124&  -378.293195&               \\ 
$\log{(\rm Prior)}$&               &    -0.040927&    -0.076679&    -0.029834&     0.074946&    -0.179107&    -0.022112&    -0.352708&    -0.003342&               \\ 
\hline 
M$_{\star}$    &[M$_{\odot}$]  &     1.058757&     1.029104&     1.010914&     0.087356&     0.930545&     1.127460&     0.830397&     1.231321&$\mathcal{N}(1.03,0.1)$\\ 
\hline 
$\sigma_{JIT\,LOW}$&[m\,s$^{-1}$]  &     1.38&     1.76&     1.90&     0.64&     0.73&     1.27&     0.11&     0.83&$\mathcal{U}$  \\ 
$\sigma_{JIT\,HIGH}$&[m\,s$^{-1}$]  &     2.33&     2.74&     2.77&     0.68&     1.70&     2.17&     0.93&     1.62&$\mathcal{U}$  \\ 
\hline 
$\gamma_{HARPS}$&[m\,s$^{-1}$]  & -3757.68& -3757.65& -3757.69&     0.14& -3757.81& -3757.49& -3757.98& -3757.34&$\mathcal{U}$  \\ 
\hline 
RHK$_{\rm index}$\,$lin$&[\ms]&    10.723380&    10.634926&    10.630724&     0.607624&     9.941268&    11.330657&     9.258677&    11.996205&$\mathcal{U}$  \\ 
\hline 
$\log{(P)}$    &[d]            &     0.769779&     0.769779&     0.769776&     0.000020&     0.769758&     0.769802&     0.769736&     0.769826&$\mathcal{U}$  \\ 
$\log{(K)}$    &[m\,s$^{-1}$]  &     0.63&     0.63&     0.63&     0.02&     0.61&     0.65&     0.58&     0.67&$\mathcal{U}$  \\ 
$\sqrt{e}.\cos{\omega}$&               &    -0.220553&    -0.170838&    -0.275286&     0.126595&    -0.294336&    -0.000876&    -0.378620&     0.156389&$\mathcal{U}$  \\ 
$\sqrt{e}.\sin{\omega}$&               &     0.182613&     0.117241&     0.149405&     0.131397&    -0.053453&     0.252324&    -0.200167&     0.349677&$\mathcal{U}$  \\ 
$\lambda_{0}$  &[deg]          &  -230.080798&  -229.352706&  -229.816814&     2.694368&  -232.391963&  -226.300698&  -235.479746&  -223.032360&$\mathcal{U}$  \\ 
\hline 
$\log{(P)}$    &[d]            &     1.118470&     1.118443&     1.118431&     0.000075&     1.118359&     1.118527&     1.118271&     1.118619&$\mathcal{U}$  \\ 
$\log{(K)}$    &[m\,s$^{-1}$]  &     0.40&     0.37&     0.37&     0.04&     0.33&     0.41&     0.28&     0.45&$\mathcal{U}$  \\ 
$\sqrt{e}.\cos{\omega}$&               &    -0.147627&    -0.073911&    -0.085546&     0.176247&    -0.277303&     0.140983&    -0.420074&     0.304579&$\mathcal{U}$  \\ 
$\sqrt{e}.\sin{\omega}$&               &    -0.059901&     0.065831&     0.087988&     0.173593&    -0.146692&     0.260735&    -0.312367&     0.404779&$\mathcal{U}$  \\ 
$\lambda_{0}$  &[deg]          &  -129.227327&  -127.324650&  -126.147264&     4.833219&  -132.789170&  -121.833139&  -138.259570&  -116.150675&$\mathcal{U}$  \\ 
\hline 
\end{tabular}
\end{center}
\end{table*}

\begin{table*}
\scriptsize
\caption{Same as Table\,\ref{HD20003_tab-mcmc-Probed_params} for \object{HD\,51608}}
 \label{HD51608_tab-mcmc-Probed_params}
\def\arraystretch{1.5}
\begin{center}
\begin{tabular}{lcclcccccccc}
\hline
\hline
Param. & Units & Max(Like) & Med & Mod &Std & CI(15.85) & CI(84.15) &CI(2.275) & CI(97.725) & Prior\\
\hline
\multicolumn{11}{c}{ \bf Likelihood}\\
\hline
$\log{(\rm Post})$&               &  -402.833168&  -409.139845&  -408.668989&     2.465262&  -412.297050&  -406.697560&  -415.940872&  -404.885802&               \\ 
$\log{(\rm Like)}$&               &  -402.629727&  -408.960474&  -408.454977&     2.467315&  -412.097485&  -406.520548&  -415.757993&  -404.678963&               \\ 
$\log{(\rm Prior)}$&               &    -0.203442&    -0.163596&    -0.147311&     0.100964&    -0.296183&    -0.073908&    -0.485685&    -0.023113&               \\ 
\hline 
M$_{\star}$    &[M$_{\odot}$]  &     0.980358&     0.800314&     0.788707&     0.088192&     0.701466&     0.900776&     0.600351&     0.999712&$\mathcal{N}(0.8,0.1)$\\ 
\hline 
$\sigma_{JIT\,LOW}$&[m\,s$^{-1}$]  &     1.20&     1.24&     1.21&     0.42&     0.76&     0.78&     0.16&     0.39&$\mathcal{U}$  \\ 
$\sigma_{JIT\,HIGH}$&[m\,s$^{-1}$]  &     2.04&     1.94&     1.99&     0.52&     1.30&     1.49&     0.31&     1.06&$\mathcal{U}$  \\ 
\hline 
$\gamma_{HARPS}$&[m\,s$^{-1}$]  & 39977.30& 39977.24& 39977.20&     0.10& 39977.12& 39977.35& 39977.00& 39977.47&$\mathcal{U}$  \\ 
\hline 
RHK$_{\rm index}$\,$lin$&[\ms]&     3.902459&     4.129760&     4.057943&     0.501237&     3.567302&     4.713965&     2.972341&     5.281426&$\mathcal{U}$  \\ 
\hline 
$\log{(P)}$    &[d]            &     1.148361&     1.148375&     1.148387&     0.000043&     1.148326&     1.148424&     1.148275&     1.148474&$\mathcal{U}$  \\ 
$\log{(K)}$    &[m\,s$^{-1}$]  &     0.61&     0.60&     0.60&     0.02&     0.58&     0.61&     0.56&     0.63&$\mathcal{U}$  \\ 
$\sqrt{e}.\cos{\omega}$&               &    -0.118480&    -0.123787&    -0.183326&     0.102911&    -0.229128&     0.007749&    -0.308200&     0.133668&$\mathcal{U}$  \\ 
$\sqrt{e}.\sin{\omega}$&               &     0.335597&     0.239601&     0.269015&     0.095793&     0.113091&     0.322123&    -0.059036&     0.387245&$\mathcal{U}$  \\ 
$\lambda_{0}$  &[deg]          &   257.734884&   257.274281&   256.797681&     2.077620&   254.933378&   259.655465&   252.551371&   262.119813&$\mathcal{U}$  \\ 
\hline 
$\log{(P)}$    &[d]            &     1.981979&     1.982021&     1.981746&     0.000578&     1.981402&     1.982724&     1.980805&     1.983426&$\mathcal{U}$  \\ 
$\log{(K)}$    &[m\,s$^{-1}$]  &     0.37&     0.37&     0.37&     0.03&     0.34&     0.40&     0.30&     0.43&$\mathcal{U}$  \\ 
$\sqrt{e}.\cos{\omega}$&               &    -0.370075&    -0.326983&    -0.359983&     0.111428&    -0.420448&    -0.181879&    -0.492776&     0.046683&$\mathcal{U}$  \\ 
$\sqrt{e}.\sin{\omega}$&               &    -0.080548&    -0.087455&    -0.136843&     0.149558&    -0.250395&     0.099871&    -0.366481&     0.262974&$\mathcal{U}$  \\ 
$\lambda_{0}$  &[deg]          &  -160.261307&  -158.719759&  -159.674199&     3.735633&  -162.883613&  -154.431466&  -167.054572&  -149.921315&$\mathcal{U}$  \\ 
\hline 
\end{tabular}
\end{center}
\end{table*}

\begin{table*}
\scriptsize
\caption{Same as Table\,\ref{HD20003_tab-mcmc-Probed_params} for \object{HD\,134060}}
 \label{HD134060_tab-mcmc-Probed_params}
\def\arraystretch{1.5}
\begin{tabular}{lcclcccccccc}
\hline
\hline
Param. & Units & Max(Like) & Med & Mod &Std & CI(15.85) & CI(84.15) &CI(2.275) & CI(97.725) & Prior\\
\hline
\multicolumn{11}{c}{ \bf Likelihood}\\
\hline
$\log{(\rm Post})$&               &  -298.850697&  -304.216566&  -304.307005&     2.241852&  -307.134697&  -302.087246&  -310.611617&  -300.536513&               \\ 
$\log{(\rm Like)}$&               &  -297.620659&  -302.893646&  -302.472853&     2.232698&  -305.807282&  -300.789692&  -309.374131&  -299.237149&               \\ 
$\log{(\rm Prior)}$&               &    -1.230038&    -1.253327&    -1.227765&     0.254368&    -1.556962&    -1.023282&    -2.140093&    -0.820972&               \\ 
\hline 
M$_{\star}$    &[M$_{\odot}$]  &     1.099460&     1.094932&     1.086339&     0.088166&     0.993140&     1.195814&     0.898980&     1.298334&$\mathcal{N}(1.095,0.1)$\\ 
\hline 
$\sigma_{JIT}$ &[m\,s$^{-1}$]  &     1.58&     1.65&     1.63&     0.22&     1.43&     1.35&     1.31&     1.04&$\mathcal{U}$  \\ 
\hline 
$\gamma_{HARPS}$&[m\,s$^{-1}$]  & 37987.94& 37987.95& 37987.91&     0.13& 37987.80& 37988.10& 37987.65& 37988.25&$\mathcal{U}$  \\ 
\hline 
$\log{(P)}$    &[d]            &     0.514500&     0.514500&     0.514499&     0.000011&     0.514488&     0.514512&     0.514478&     0.514526&$\mathcal{U}$  \\ 
$\log{(K)}$    &[m\,s$^{-1}$]  &     0.67&     0.66&     0.66&     0.02&     0.64&     0.68&     0.62&     0.70&$\mathcal{U}$  \\ 
$\sqrt{e}.\cos{\omega}$&               &    -0.125031&    -0.096070&    -0.087114&     0.067809&    -0.173750&    -0.018918&    -0.246029&     0.063097&$\mathcal{U}$  \\ 
$\sqrt{e}.\sin{\omega}$&               &    -0.673348&    -0.660554&    -0.664313&     0.027638&    -0.690593&    -0.628249&    -0.720202&    -0.591182&$\mathcal{U}$  \\ 
$\lambda_{0}$  &[deg]          &   -60.971221&   -60.219046&   -60.857964&     2.817351&   -63.378221&   -57.046679&   -66.686867&   -53.740675&$\mathcal{U}$  \\ 
\hline 
$\log{(P)}$    &[d]            &     3.115614&     3.111116&     3.110321&     0.013896&     3.095987&     3.126974&     3.080422&     3.146266&$\mathcal{U}$  \\ 
$\log{(K)}$    &[m\,s$^{-1}$]  &     0.19&     0.22&     0.22&     0.05&     0.15&     0.27&     0.08&     0.33&$\mathcal{U}$  \\ 
$\sqrt{e}.\cos{\omega}$&               &     0.018089&    -0.131288&    -0.033631&     0.237542&    -0.408103&     0.151476&    -0.601802&     0.365222&$\mathcal{U}$  \\ 
$\sqrt{e}.\sin{\omega}$&               &    -0.193191&    -0.102419&    -0.157397&     0.182951&    -0.296909&     0.130699&    -0.452167&     0.321316&$\mathcal{U}$  \\ 
$\lambda_{0}$  &[deg]          &   -57.886308&   -59.605407&   -62.436000&     6.753625&   -66.940199&   -51.615752&   -74.094300&   -42.949848&$\mathcal{U}$  \\ 
\hline 
\end{tabular}
\end{table*}

\begin{table*}
\scriptsize
\caption{Same as Table\,\ref{HD20003_tab-mcmc-Probed_params} for \object{HD\,136352}}
 \label{HD136352_tab-mcmc-Probed_params}
\def\arraystretch{1.5}
\begin{tabular}{lcclcccccccc}
\hline
\hline
Param. & Units & Max(Like) & Med & Mod &Std & CI(15.85) & CI(84.15) &CI(2.275) & CI(97.725) & Prior\\
\hline
\multicolumn{11}{c}{ \bf Likelihood}\\
\hline
$\log{(\rm Post})$&               &  -413.840756&  -420.573939&  -420.519231&     2.525979&  -423.752736&  -418.071226&  -427.592767&  -416.139887&               \\ 
$\log{(\rm Like)}$&               &  -413.418243&  -420.301368&  -420.080839&     2.517934&  -423.494074&  -417.796941&  -427.287283&  -415.883659&               \\ 
$\log{(\rm Prior)}$&               &    -0.422514&    -0.225282&    -0.181747&     0.168419&    -0.456156&    -0.087916&    -0.795628&    -0.025262&               \\ 
\hline 
M$_{\star}$    &[M$_{\odot}$]  &     0.836645&     0.811051&     0.810685&     0.088144&     0.710931&     0.910337&     0.609205&     1.008754&$\mathcal{N}(0.81,0.1)$\\ 
\hline 
$\sigma_{JIT}$ &[m\,s$^{-1}$]  &     1.43&     1.41&     1.41&     0.30&     1.02&     1.08&     0.76&     0.81&$\mathcal{U}$  \\ 
\hline 
$\gamma_{HARPS}$&[m\,s$^{-1}$]  &-68709.12&-68709.03&-68709.05&     0.08&-68709.13&-68708.94&-68709.22&-68708.84&$\mathcal{U}$  \\ 
\hline 
$\log{(P)}$    &[d]            &     1.063786&     1.063799&     1.063829&     0.000081&     1.063703&     1.063890&     1.063608&     1.063979&$\mathcal{U}$  \\ 
$\log{(K)}$    &[m\,s$^{-1}$]  &     0.23&     0.20&     0.19&     0.03&     0.16&     0.24&     0.12&     0.27&$\mathcal{U}$  \\ 
$\sqrt{e}.\cos{\omega}$&               &    -0.342695&    -0.314967&    -0.410810&     0.145521&    -0.437347&    -0.116554&    -0.524238&     0.125811&$\mathcal{U}$  \\ 
$\sqrt{e}.\sin{\omega}$&               &    -0.090658&     0.033773&     0.091648&     0.156779&    -0.152919&     0.214589&    -0.306061&     0.352353&$\mathcal{U}$  \\ 
$\lambda_{0}$  &[deg]          &   277.740369&   277.024300&   275.232882&     4.443846&   272.023532&   282.083700&   266.558482&   287.198688&$\mathcal{U}$  \\ 
\hline 
$\log{(P)}$    &[d]            &     1.440604&     1.440627&     1.440596&     0.000121&     1.440492&     1.440768&     1.440359&     1.440909&$\mathcal{U}$  \\ 
$\log{(K)}$    &[m\,s$^{-1}$]  &     0.43&     0.42&     0.42&     0.02&     0.40&     0.45&     0.37&     0.47&$\mathcal{U}$  \\ 
$\sqrt{e}.\cos{\omega}$&               &    -0.070645&    -0.042094&    -0.084481&     0.126899&    -0.182267&     0.113998&    -0.296099&     0.237968&$\mathcal{U}$  \\ 
$\sqrt{e}.\sin{\omega}$&               &    -0.212716&    -0.113587&    -0.193076&     0.134373&    -0.259032&     0.060329&    -0.348573&     0.197605&$\mathcal{U}$  \\ 
$\lambda_{0}$  &[deg]          &    19.705439&    19.919436&    19.941697&     2.590743&    16.987116&    22.886544&    14.008324&    25.935025&$\mathcal{U}$  \\ 
\hline 
$\log{(P)}$    &[d]            &     2.030837&     2.031806&     2.031583&     0.000968&     2.030727&     2.032933&     2.029701&     2.034155&$\mathcal{U}$  \\ 
$\log{(K)}$    &[m\,s$^{-1}$]  &     0.12&     0.13&     0.12&     0.04&     0.08&     0.17&     0.03&     0.22&$\mathcal{U}$  \\ 
$\sqrt{e}.\cos{\omega}$&               &    -0.489331&    -0.199704&    -0.273372&     0.203550&    -0.406230&     0.080327&    -0.542331&     0.280257&$\mathcal{U}$  \\ 
$\sqrt{e}.\sin{\omega}$&               &     0.029048&     0.037665&     0.048956&     0.162724&    -0.157033&     0.224967&    -0.309643&     0.370264&$\mathcal{U}$  \\ 
$\lambda_{0}$  &[deg]          &   185.975741&   188.581732&   187.252622&     5.246122&   182.652564&   194.619420&   176.751015&   200.688274&$\mathcal{U}$  \\ 
\hline 
\end{tabular}
\end{table*}

\end{appendix}

\end{document}